\documentclass[twoside]{IEEEtran}
\usepackage[cmex10]{amsmath}
\usepackage{amsfonts}
\usepackage{amssymb}
\usepackage{amsbsy}
\usepackage{amsthm}
\usepackage{mathrsfs}
\usepackage{graphicx}
\usepackage{times}
\usepackage{rotating}
\usepackage{bm}
\usepackage{bbm}
\usepackage[table]{xcolor}

\usepackage{enumitem} 
\usepackage{multirow}
\usepackage{subfigure}
\usepackage{tikz}
\usepackage{pgfplots}
\usepackage{centernot}

\allowdisplaybreaks[1]

\newtheoremstyle{unnumbered}
  {3pt plus 1pt minus 2pt} % Space above
  {3pt plus 1pt minus 2pt} % Space below
  {} % Body font
  {1em} % Indent amount
  {\itshape} % Theorem head font\itshape\bfseries
  {:} % Punctuation after theorem head
  {.3em} % Space after theorem head
  {#1} % Theorem head spec (can be left empty, meaning `normal')

\newtheoremstyle{numbered}
  {3pt plus 1pt minus 2pt} % Space above
  {3pt plus 1pt minus 2pt} % Space below
  {} % Body font
  {1em} % Indent amount
  {\itshape} % Theorem head font\itshape\bfseries
  {:} % Punctuation after theorem head
  {.3em} % Space after theorem head
  {\thmname{#1}\thmnumber{ #2}\thmnote{ (#3)}} % Theorem head spec (can be left empty, meaning `normal')

\theoremstyle{numbered}
\newtheorem{theorem}{Theorem}%[section]
\newtheorem{corollary}[theorem]{Corollary}%[section]
\newtheorem{lemma}[theorem]{Lemma}%[section]
\newtheorem{definition}[theorem]{Definition}%[section]
\newtheorem{remark}[theorem]{Remark}%[section]
\newtheorem{example}[theorem]{Example}%[section]

\theoremstyle{unnumbered}
\newcommand{\be}{\begin{equation}}
\newcommand{\ee}{\end{equation}}
\newcommand{\ist}{\hspace*{.3mm}}
\newcommand{\rmv}{\hspace*{-.3mm}}

\def\ba#1\ea{\begin{align}#1\end{align}}
\def\bas#1\eas{\begin{align*}#1\end{align*}}

\providecommand{\abs}[1]{\lvert#1\rvert}

\providecommand{\norm}[1]{\lVert#1\rVert}
\providecommand{\Hm}[1]{\mathscr{H}^{#1}}
\providecommand{\eh}[1]{\mathfrak{h}^{#1}}

\IEEEoverridecommandlockouts
\makeatletter
\renewcommand\section{\@startsection{section}{1}{\z@}{2ex plus 1.5ex minus 0.8ex}{1ex plus 1ex minus 0.5ex}{\normalfont\normalsize\centering\scshape}} %change spacing before and after sections
\renewcommand\subsection{\@startsection{subsection}{2}{\z@}{2ex plus 1.5ex minus 0.8ex}{1ex plus 1ex minus 0.5ex}{\normalfont\normalsize\itshape}}%change spacing before and after subsections
\renewcommand\@IEEEauthorblockconfadjspace{0.2em} % spacing between title and authors conference
\renewcommand\@IEEEauthorblockNtopspace{0.ex} 
\renewcommand\@IEEEauthorblockAtopspace{1ex}% spacing between authors and affiliations

\makeatother

\def\Dt{D^*}
\def\0v{\boldsymbol{0}}
\def\Iv{\mathbf{I}}

\def\ev{\boldsymbol{e}}

\def\kv{\boldsymbol{k}}

\def\sv{\boldsymbol{s}}

\def\yv{\boldsymbol{t}}

\def\xv{\boldsymbol{x}}
\def\xvt{\tilde{\xv}}

\def\yv{\boldsymbol{y}}
\def\yvb{\bar{\yv} }  %%%%%%%%%%%%%%%%%%
\def\yvt{\tilde{\yv}}
\def\zv{\boldsymbol{z}}

\def\Dm{\boldsymbol{D}}

\def\Jm{J}
\def\Sigm{\boldsymbol{\Sigma}}

\def\rx{\mathsf{x}}
\def\ry{\mathsf{y}}
\def\rz{\mathsf{z}}

\def\r0v{\boldsymbol{\mathsf{0}}}

\def\rwv{\boldsymbol{\mathsf{w}}}
\def\rxv{\boldsymbol{\mathsf{x}}}

\def\ryv{\boldsymbol{\mathsf{y}}}
\def\rzv{\boldsymbol{\mathsf{z}}}

\def\rWm{\boldsymbol{\mathsf{W}}}

\def\rXm{\boldsymbol{\mathsf{X}}}

\def\IN{\mathbb{N}}
\def\N{\mathbb{N}}

\def\R{\mathbb{R}}
\def\IZ{\mathbb{Z}}

\def\sA{\mathcal{A}}
\def\sB{\mathcal{B}}
\def\srB{\mathcal{B}}
\def\sC{\mathcal{C}}
\def\sD{\mathcal{D}}
\def\sDt{\widetilde{\sD}}
\def\sE{\mathcal{E}}
\def\sEt{\widetilde{\sE}}
\def\sF{\mathcal{F}}
\def\sFt{\widetilde{\sF}}
\def\sI{\mathcal{I}}

\def\sG{\mathcal{G}}

\def\sT{\mathcal{T}}
\def\sS{\mathcal{S}}

\def\Pr{\operatorname{Pr}}

\def\ld{\operatorname{ld}}
\DeclareMathOperator*{\argmax}{arg\,max}

\def\trans{{\operatorname{T}}}
\def\E{\mathbb{E}}
\def\proj{\mathfrak{p}}

\def\Leb{\mathscr{L}}
\def\condi{\,|\,}

\begin{document}

%\IEEEoverridecommandlockouts

\title{
Entropy and Source Coding for
Integer- \\ Dimensional Singular Random Variables
}

\author{
G\"unther~Koliander, %~\IEEEmembership{Member,~IEEE,} 
Georg~Pichler,~\IEEEmembership{Student~Member,~IEEE,} \\
Erwin~Riegler, %~\IEEEmembership{Member,~IEEE,}
and~Franz~Hlawatsch,~\IEEEmembership{Fellow,~IEEE}%
%\IEEEauthorblockA{
%$^1$Institute of Telecommunications, Vienna University of Technology, 1040 Vienna, Austria\\
%$^2$Department of Signals and Systems, Chalmers University of Technology, 41296 Gothenburg, Sweden\\
\thanks{This paper was presented in part at the IEEE International Symposium on Information Theory (ISIT),  Honolulu, HA, July 2014.
}%
\thanks{G. Koliander, G. Pichler, and F. Hlawatsch are with the Institute of Telecommunications, TU Wien, 1040 Vienna, Austria (e-mail: \{gkoliand, gpichler, fhlawats\}@nt.tuwien.ac.at).}%
\thanks{E.~Riegler is with the Department of Information Technology and Electrical Engineering, ETH Zurich, CH 8092 Zurich, Switzerland (e-mail: eriegler@nari.ee.ethz.ch).}%
\thanks{This work was supported by the WWTF under grants ICT10-066 (NO\-WIRE) and ICT12-054 (TINCOIN), and by the FWF under grant P27370-N30.}%
\thanks{Copyright (c) 2016 IEEE. Personal use of this material is permitted.  However, permission to use this material for any other purposes must be obtained from the IEEE by sending a request to pubs-permissions@ieee.org.}
}%

\maketitle

\begin{abstract}
%abstract here
Entropy and differential entropy are important quantities in information theory.
%% Unfortunately, a 
A tractable extension to singular random variables---which are neither discrete nor continuous---has not been available so far. 
Here, we present  such an extension for the practically relevant class of integer-dimensional singular random variables. 
The proposed entropy definition contains the entropy of discrete random variables and the differential entropy of continuous random variables as special cases.
We show that it transforms in a natural manner under Lipschitz functions, and that it is invariant under unitary transformations.
We define joint entropy and conditional entropy for integer-dimensional singular random variables, and we show that the proposed entropy conveys useful expressions of the mutual information. 
As first applications of our entropy definition, we present a result on the minimal expected codeword length of quantized integer-dimensional singular sources and a Shannon lower bound for integer-dimensional singular sources.
\end{abstract}

%%%%%%%%%%%%%%%%%%%%%%%%%%%%%%%
\section{Introduction} \label{sec:intro}   
%%%%%%%%%%%%%%%%%%%%%%%%%%%%%%%
%
%
%%%%%%%%%%%%%%%%%%%%%%
\subsection{Background and Motivation}
\label{sec:intro_backg}
%%%%%%%%%%%%%%%%%%%%%%
%
%
Entropy is one of the fundamental concepts in information theory. 
The classical definition of entropy for discrete random variables and 
its interpretation as information content go back to Shannon \cite{sh48} and were analyzed thoroughly from axiomatic \cite{cs08} and operational \cite{sh48} viewpoints.
A similar definition for continuous random variables, differential entropy, was also introduced by Shannon  \cite{sh48}, but its interpretation as  information content is  controversial \cite{Kol56}.  
Nonetheless,  information-theoretic derivations involving undisputed quantities like Kullback-Leibler divergence or mutual information between continuous random variables can often be simplified using differential entropy. 
Furthermore, in rate-distortion theory, a lower bound on the rate-distortion function known as the Shannon lower bound can be calculated using differential entropy \cite[Sec.~4.6]{gr90}. 
Finally, differential entropy arises in asymptotic expansions of the entropy of ever finer quantizations of a continuous random variable \cite[Sec.~IV]{Kol56}.
Hence, although the interpretation of differential entropy is disputed, its operational relevance renders it a useful quantity. 

The concepts of entropy and differential entropy thus simplify the understanding and information-theoretic treatment of discrete and continuous random variables. 
However, these two kinds of random variables do not cover all interesting information-theoretic problems. 
In fact, a number of information-theoretic problems involving \textit{singular} random variables, which are neither discrete nor continuous, have been described recently: 
\begin{itemize}
\item For the vector interference channel, a singular input distribution has to be used to fully utilize the available degrees of freedom \cite{stbo16}.
\item In a probabilistic formulation of analog compression, the underlying source distribution is singular \cite{wuve10}.
\item In block-fading channel models, two different kinds of singular distributions arise: 
the optimal input distribution 
is singular in some settings \cite[Ch.~6]{pa04},
and the noiseless output distribution is singular except for special cases \cite{koriduhl13a}.
\end{itemize}
Thus, a suitable generalization of (differential) entropy  to singular random variables has the potential to simplify theoretical work in these areas and to provide valuable insights.

Another field where singular random variables appear is source coding. 
In many high-dimensional problems, deterministic dependencies reduce the intrinsic dimension of a source. Thus, the random variable describing the source cannot be continuous but  often is not discrete either.
A basic example is a random variable $\rxv=(\rx_1 \; \rx_2)^{\trans}\in \R^2$ supported on the unit circle, i.e., exhibiting the deterministic dependence $\rx_1^2+\rx_2^2=1$. 
Although $\rxv$ is defined on $\R^2$ and both components $\rx_1, \rx_2$ are continuous random variables, $\rxv$ itself is intrinsically only one-dimensional.
The differential entropy of $\rxv$ is not defined and, in fact, classical information theory does not provide a rigorous definition of entropy for this random variable.
Another, less trivial, example of a singular random variable is a rank-one 
random matrix of the form $\rXm = \rzv\rzv^{\trans}$,
%%   \in \R^{m\times m}$, 
where $\rzv$ is a continuous random vector.
%% in $\R^m$ 

The case of arbitrary probability distributions is very hard to handle, and due to its generality even the mere definition of a meaningful entropy seems impossible.
Two existing approaches to defining (differential) entropy for more general distributions are based on quantizations of the random variable in question. 
Usually, the entropy of these discretizations converges to infinity and, thus, a normalization has to be employed to obtain a useful result. 
In \cite{Renyi59}, this approach is adopted for very specific quantizations of a random variable. 
Unfortunately, this does not always result in a well-defined entropy and sometimes even fails for continuous random variables of finite differential entropy \cite[pp.~197f]{Renyi59}. 
Moreover, the quantization process seems difficult to deal with analytically and no theory was built based on this definition of entropy.%
\footnote{This entropy should not be confused with the \textit{information dimension} defined in the same paper \cite{Renyi59}, which is indeed a very useful and widely used tool.}
A similar approach is to consider arbitrary quantizations that are constrained by some measure of fineness to enable a limit operation.
In \cite{Kol56} and \cite{PoRo71}, $\varepsilon$-entropy is introduced as the minimal entropy of all quantizations using sets of diameter less than $\varepsilon$. 
However, to specify a diameter, a distortion function has to be defined. 
Since all basic information-theoretic quantities (e.g., mutual information or Kullback-Leibler divergence) do not depend on a specific distortion function, it is hardly possible to embed  $\varepsilon$-entropy into a general information-theoretic framework.
Furthermore, once again the quantization process seems difficult to handle analytically.

Since the aforementioned approaches do not provide a satisfactory generalization of (differential) entropy, we follow a different approach, which is also motivated by ever finer quantizations of the random variable. 
However, in our approach, the order of the two steps ``taking the limit of quantizations'' and ``calculating the entropy as the expectation of the logarithm of a probability (mass) function'' is changed. 
More precisely, we first consider the probability mass functions of quantizations and take a normalized limit.
(In the special case of a continuous random variable, this results in the probability density function due to  Lebesgue's differentiation theorem.)
Then we take the expectation of the logarithm of the resulting density function. 
Due to fundamental results in geometric measure theory, this approach can result in a well-defined entropy only for integer-dimensional distributions, since otherwise the density function does not exist \cite[Th.~3.1]{DeLe08}.
In fact, the existence of the density function implies that the random variable is distributed according to a \textit{rectifiable measure} \cite[Th.~1.1]{DeLe08}.
%Thus,  the kind of random variables considered in the present paper.
Thus, the distributions considered in the present paper are rectifiable distributions on Euclidean space. 
Although this is still far from the generality of arbitrary probability distributions, it covers numerous interesting cases---including all the examples mentioned above---and gives valuable insights.

%Rev1.5: Add Csiszar paper
The density function of rectifiable measures can also be defined as a certain Radon-Nikodym derivative.
A generalized (differential) entropy based on a Radon-Nikodym derivative with respect to a ``measure of the  observer's interest'' was considered in \cite{cs73}.
Our entropy is consistent with this approach, and at a certain point we will use a result on quantization problems established in  \cite{cs73}.
However, because in our setting a concrete measure is considered,
%% in our setting, we can 
the results we obtain go beyond the basic properties derived in \cite{cs73} for general measures.
%In particular, it was shown that in  \cite{cs73} that the entropy of ``fine'' (based on the measure in the definition) quantizations can be characterized by this generalized entropy.

\subsection{Contributions}
\label{sec:intro_contr}
We provide a generalization of the classical concepts of entropy and differential entropy to integer-dimensional random variables.
Our entropy satisfies several well-known properties of differential entropy: it is invariant under unitary transformations, transforms as expected under Lipschitz mappings, and can be extended to joint and conditional entropy. 
We show that the entropy of discrete random variables and the differential entropy of continuous random variables are special cases of our entropy definition.
For joint entropy, we prove a chain rule which takes the geometry of the support set into account.
Furthermore, we discuss why in certain cases our entropy definition may violate the classical result that conditioning does not increase (differential) entropy. 
We provide expressions of the mutual information between integer-dimensional random variables in terms of our entropy. 
We also show that an asymptotic equipartition property  analogous to \cite[Sec.~8.2]{Cover91} holds for our entropy, but with the Lebesgue measure replaced by the Hausdorff measure of appropriate dimension. 

In our proofs, we exercise care to detail all assumptions and to obtain mathematically rigorous statements. 
Thus, although many of our results might seem obvious to the cursory reader because of their similarity to well-known results for (differential) entropy, we emphasize that they are not simply replicas or straightforward adaptations of known results. 
This becomes evident, e.g.,  for the chain rule (see Theorem~\ref{th:chainrule} in Section~\ref{sec:chainrule}), which might be expected to have the same form as the chain rule  for differential entropy. 
However, already a simple example will show that the geometry of the support set may lead to an additional term, which is not present in the special case of continuous random variables.

As a first application of the proposed entropy, we derive a  result on the minimal expected binary codeword length of quantized integer-dimensional singular sources. More specifically, we show that our entropy characterizes the rate at which an arbitrarily fine quantization of an integer-dimensional singular  source can be compressed.
Another application is 
%a new result in rate-distortion theory. Based on the characterization of mutual information in~\cite{csiszar74}, we prove 
a lower bound on the rate-distortion function of an integer-dimensional singular  source that resembles the Shannon lower bound for discrete \cite[Sec.~4.3]{gr90} and continuous \cite[Sec.~4.6]{gr90} random variables.
For the specific case of a singular source that is uniformly distributed  on the unit circle, we demonstrate that our bound  is within $0.2\,\textrm{nat}$ of the true rate-distortion function.

\subsection{Notation}
\label{sec:notation}
Sets are denoted by calligraphic letters (e.g., $\sA$). 
The complement of a set $\sA$ is denoted $\sA^c$.
Sets of sets are denoted by fraktur letters (e.g., $\mathfrak{M}$).
The set of natural numbers $\{1, 2, \dots\}$ is denoted as $\IN$.
The open ball with center $\xv\in \R^M$ and radius $r>0$ is denoted by $\srB_r(\xv)$, i.e., $\srB_r(\xv)\triangleq \{\yv\in \R^M: \norm{\yv-\xv}< r\}$.
The symbol $\omega(M)$ denotes the volume of the $M$-dimensional unit ball, i.e., $\omega(M)=\pi^{M/2}/\Gamma(1+M/2)$ where $\Gamma$ is the Gamma function.
Boldface uppercase and lowercase letters denote matrices and vectors, respectively. 
The $m\times m$ identity matrix is denoted by $\Iv_m$.
Sans serif letters denote random quantities,  e.g., 
$\rxv$ is a random vector and $\rx$ is a random scalar. 
The superscript ${}^{\operatorname{T}}$ stands for transposition. % and 
For $x \!\in\!\R$, $\lfloor x\rfloor\triangleq \max\{m \!\in\! \IZ : m \!\leq\! x\}$ and for  $\xv \!\in\!\R^M$, $\lfloor \xv\rfloor\triangleq (\lfloor x_1\rfloor\, \cdots \, \lfloor x_M\rfloor)^{\trans}$.
Similarly, $\lceil x\rceil\triangleq \min\{m \!\in\! \IZ : m \!\geq\! x\}$.
We write $\E_{\rxv}[\cdot]$ for the expectation operator with respect to the random variable $\rxv$.
$\Pr\{\rxv\in \sA\}$ denotes the probability that $\rxv\in \sA$.
For $\xv \in \R^{M_1}$ and $\yv\in \R^{M_2}$, we denote by $\proj_{\rxv}\colon \R^{M_1+M_2}\to \R^{M_1}$, 
$\proj_{\rxv}(\xv,\yv)=\xv$,
the projection of $\R^{M_1+M_2}$ to the first $M_1$ components.
Similarly, $\proj_{\ryv}\colon \R^{M_1+M_2}\to \R^{M_2}$, 
$\proj_{\ryv}(\xv,\yv)=\yv$,
denotes the projection of $\R^{M_1+M_2}$ to the last $M_2$ components.
The generalized Jacobian determinant of a Lipschitz function%
\footnote{By Rademacher's theorem \cite[Th.~2.14]{Ambrosio2000Functions}, a Lipschitz function is differentiable almost everywhere and, thus, the Jacobian determinant is well defined almost everywhere.}
 $\phi$ is written as $\Jm_{\phi}$.
For a function $\phi$ with domain $\sD$ and a subset $\sDt\subseteq \sD$, we denote by $\phi\big|_{\sDt}$ the restriction of $\phi$ to the domain $\sDt$.
$\Hm{m}$ denotes the $m$-dimensional Hausdorff measure.%
\footnote{Readers unfamiliar with this concept may think of it as a measure of an $m$-dimensional area in a higher-dimensional space (e.g., surfaces in $\R^3$).
%Rev2.1 add reference to definition of Hausdorff measures
An introduction and definition can be found in \cite[Sec.~2.8]{Ambrosio2000Functions}.}
$\Leb^M$ denotes the $M$-dimensional Lebesgue measure and
$\mathfrak{B}_M$ denotes the Borel $\sigma$-algebra on $\R^M$.
For a measure $\mu$ and a $\mu$-measurable function $f$, the induced measure is given as $\mu f^{-1}(\sA)\triangleq \mu(f^{-1}(\sA))$.
For two measures $\mu$ and $\nu$ on the same measurable space, we indicate by $\mu\ll\nu$ that $\mu$ is absolutely continuous with respect to $\nu$ (i.e., for any measurable set $\sA$, $\nu(\sA)=0$ implies $\mu(\sA)=0$). 
For a measure $\mu$ and a measurable set $\sE$, the measure $\mu|_{\sE}$ is the restriction of $\mu$ to $\sE$, i.e., $\mu|_{\sE}(\sA)=\mu(\sA\cap \sE)$.
The logarithm to the base $e$ is denoted $\log$ and the logarithm to the base $2$ is denoted $\ld$.
%Rev1.13: add notation stackrel=
In certain equations, we reference an equation number on top of the equality sign in order to indicate that the equality holds due 
to some previous equation: e.g., $\stackrel{(\text{42})}=$ indicates that the equality holds due to eq.\ (42).
%% In equations we point out that the equality holds due to some formerly shown equation by referencing the equation number above the equality sign, 
%% e.g., $\stackrel{(42)}=$ denotes that the equality holds due to $(42)$.}

\subsection{Organization of the Paper}
The rest of this paper is organized as follows. 
In Section \ref{sec:defsentropy}, we review the established definitions of entropy
%% concepts of R\'enyi entropy and $\varepsilon$ entropy 
and describe the intuitive idea behind our entropy definition.
Rectifiable sets, measures, and random variables are introduced  in Section~\ref{sec:rectifiablesets} as the basic setting for integer-dimensional distributions. 
In Section~\ref{sec:entropy}, we develop the theory of ``lower-dimensional entropy'': we define entropy for integer-dimensional random variables, 
%(Section~\ref{sec:MotDef}),
prove a transformation property and invariance under unitary transformations,
% (Section~\ref{sec:transformationproperties}),
demonstrate connections to classical entropy and differential entropy,
% (Section~\ref{sec:relationtoentropy}),
%Rev1.1: add Wishart matrices
and provide examples by calculating the entropy of random variables supported on the unit circle in $\R^2$ and of positive semidefinite rank-one random
%% Wishart 
matrices.
% (Section~\ref{sec:entropysingularwishart})}.
In Sections~\ref{sec:jointentropy} and \ref{sec:conditionalentropy}, we introduce and discuss joint entropy and conditional entropy, respectively.
Relations of our entropy to the mutual information between integer-dimensional random variables are demonstrated in Section~\ref{sec:mutualinformation}.
In Section~\ref{sec:aep}, we prove an asymptotic equipartition property for our entropy.
In Section~\ref{sec:sourcecoding}, we present a result on the minimal expected binary codeword length of quantized integer-dimensional sources.
In Section~\ref{sec:ratedistortion}, we derive a Shannon lower bound for integer-dimensional singular sources and evaluate it for a source that is uniformly distributed on the unit circle.

%%%%%%%%%%%%%%%%%%%%%%%%%%%%%%%
\section{Previous Work and Motivation} \label{sec:defsentropy}   
%%%%%%%%%%%%%%%%%%%%%%%%%%%%%%%
We first recall the definitions of entropy for discrete random variables \cite[Ch.~2]{Cover91} and differential entropy for continuous random variables \cite[Ch.~8]{Cover91}.
Let $\rxv$ be a discrete random variable with probability mass function  $p_{\rxv}(\xv_i)=\Pr\{\rxv=\xv_i\}$, $i\in \sI$, where $\sI$ is the finite or countably  infinite set indexing all possible realizations $\xv_i$ of $\rxv$. 
The entropy of $\rxv$ is
\be\label{eq:discentropy}
H(\rxv)\triangleq -\E_{\rxv}[\log p_{\rxv}(\rxv)] = -\sum_{i\in \sI} p_{\rxv}(\xv_i) \log p_{\rxv}(\xv_i)\,.
\ee
For a continuous random variable $\rxv$ on $\R^M$ with probability density function $f_{\rxv}$, the differential entropy is
\be
h(\rxv) \triangleq -\E_{\rxv}[\log f_{\rxv}(\rxv)]
%\notag \\ &  
= -\int_{\R^M} f_{\rxv}(\xv) \log f_{\rxv}(\xv)\, \mathrm{d}\Leb^M(\xv)\,.\label{eq:diffentropy}
\ee
%Rev1.3:add comment that h can be infty or may not exist
We note
%% recall 
that $h(\rxv)$ may be $\pm \infty$ or undefined.

\subsection{Entropy of Dimension $d(\rxv)$ and $\varepsilon$-Entropy}
\label{sec:renyiandepsent}
There exist two previously proposed generalizations of (differential) entropy to a larger set of probability distributions. 
The first generalization
is based on quantizations of the random variable to ever finer cubes \cite{Renyi59}.
More specifically, for a (possibly singular) random variable $\rxv\in \R^M$, the 
\textit{R\'enyi information dimension} of $\rxv$ is 
%If for some $n\in \N$ the discrete random variable $\frac{\lfloor n\rxv \rfloor}{n}$ has finite entropy then
%\be
%\overline{d}(\rxv)\triangleq \limsup_{n\to \infty} \frac{H\big(\frac{\lfloor n\rxv \rfloor}{n}\big)}{\log n}, 
%\quad \underline{d}(\rxv)\triangleq \liminf_{n\to \infty} \frac{H\big(\frac{\lfloor n\rxv \rfloor}{n}\big)}{\log n}
%\ee
%are the \textit{upper and lower information dimension of $\rxv$}, respectively.
%If $\overline{d}(\rxv)=\underline{d}(\rxv)$ the limit
\be\label{eq:Rinfdim}
d(\rxv)\triangleq \lim_{n\to \infty} \frac{H\big(\frac{\lfloor n\rxv \rfloor}{n}\big)}{\log n}
\ee
 and the \textit{entropy of dimension $d(\rxv)$} of $\rxv$ is defined as
\be\label{eq:renyientropy}
h_{d(\rxv)}^{\text{R}}(\rxv) \triangleq \lim_{n\to \infty} \bigg(H\bigg(\frac{\lfloor n\rxv \rfloor}{n}\bigg)-d(\rxv) \log n\bigg)
\ee
provided the limits in \eqref{eq:Rinfdim} and \eqref{eq:renyientropy} exist.
%\end{definition}

This definition of entropy of dimension $d(\rxv)$ corresponds to the following procedure:
\begin{enumerate}
\item Quantize  $\rxv$ using the cubes $\prod_{i=1}^M\big[\frac{k_i}{n}, \frac{k_i+1}{n}\big)$, with $k_i \in \IZ$, i.e., consider the discrete random variable with probabilities $p_{\kv}=\Pr\big\{\rxv\in \prod_{i=1}^M\big[\frac{k_i}{n}, \frac{k_i+1}{n}\big)\big\}$. 
%This results in the sequence $\Big\{\frac{\lfloor n\rxv \rfloor}{n}\Big\}_{n\in \N}$ of discrete random variables.
\item Calculate the entropy of the quantized random variable, i.e., the negative expectation of the logarithm of the probability mass function $p_{\kv}$. 
\item Subtract the correction term $d(\rxv) \log n$ to account for the dimension of the random variable $\rxv$.
\item Take the limit $n\to \infty$.
\end{enumerate}
Although this approach seems reasonable, there are several issues. 
First,  the definition of $h_{d(\rxv)}^{\text{R}}(\rxv)$ seems to be difficult to handle analytically, and connections to major information-theoretic concepts such as mutual information are not available.
%Rev1.5: Include reference to singular distributions
Furthermore, the quantization used is just one of many possible---we might, e.g.,  consider a shifted version of the set of cubes $\prod_{i=1}^M\big[\frac{k_i}{n}, \frac{k_i+1}{n}\big)$, which, for singular distributions, may result in a different value of the resulting entropy.

An approach that overcomes the latter issue is the concept of $\varepsilon$-entropy \cite{Kol56, PoRo71}. 
The definition of $\varepsilon$-entropy does not use a specific quantization but takes the infimum of the entropy over all possible (countable) quantizations under a constraint on the diameter of the quantization sets. 
This is  motivated by data compression: the quantization should be such that an error of maximally $\varepsilon$ is made (thus, the quantization sets have maximal diameter $\varepsilon$) and at the same time the minimal possible number of bits should be used to encode the data (thus, the entropy is minimized over all possible quantizations).
More specifically, for a random variable $\rxv\in \R^M$, let $\mathfrak{P}_{\varepsilon}$ denote the set of all countable partitions of $\R^M$ into mutually disjoint, measurable sets of diameter at most $\varepsilon$.
Furthermore,  for a partition $\mathfrak{Q}=\{\sA_i:i\in \N\}\in \mathfrak{P}_{\varepsilon}$, the quantization $[\rxv]_{\mathfrak{Q}}\in \N$ is the discrete random variable defined by $p_i=\Pr\{[\rxv]_{\mathfrak{Q}}=i\}=\Pr\{\rxv\in \sA_i\}$ for $i\in \N$.
Then the \textit{$\varepsilon$-entropy of $\rxv$} is defined as
\be\label{eq:epsentropy}
H_{\varepsilon}(\rxv)\triangleq \inf_{\mathfrak{Q}\in \mathfrak{P}_{\varepsilon}} H([\rxv]_{\mathfrak{Q}})\,.
\ee 
Here, a problem is that $H_{\varepsilon}(\rxv)$ is only defined for a fixed $\varepsilon>0$ and the limit $\varepsilon\to 0$ converges to $\infty$ for nondiscrete distributions.
However, as in the case of R\'enyi information dimension, a correction term can be obtained using the following seemingly new definition of information dimension:
%\be\label{eq:infdim}
%\overline{d^*}(\rxv)\triangleq \limsup_{\varepsilon\to 0} \frac{H_{\varepsilon}(\rxv)}{\log \frac{1}{\varepsilon}}, 
%\quad \underline{d^*}(\rxv)\triangleq \liminf_{\varepsilon\to 0} \frac{H_{\varepsilon}(\rxv)}{\log \frac{1}{\varepsilon}}
%\ee
\be \notag %\label{eq:infdim}
d^*(\rxv)\triangleq \lim_{\varepsilon\to 0} \frac{H_{\varepsilon}(\rxv)}{\log \frac{1}{\varepsilon}}\,.
\ee
By~\cite[Prop.~3.3]{KaDe94}, the definitions of information dimension using 
 R\'enyi's approach and the $\varepsilon$-entropy approach coincide, i.e., $d^*(\rxv)=d(\rxv)$.
%In particular, we  again obtain R\'enyi's information dimension $d(\rxv)$ if the limit in~\eqref{eq:infdim} exists.
This suggests the following new definition of a $d(\rxv)$-dimensional entropy.
\begin{definition}\label{def:asyepsentropy}
Let $\rxv\in \R^M$ be a random variable with existing information dimension $d(\rxv)$.
Then the \textit{asymptotic $\varepsilon$-entropy of dimension $d(\rxv)$} is defined as
\be\notag 
h^{*}_{d(\rxv)}(\rxv)\triangleq \lim_{\varepsilon\to 0} \big(H_{\varepsilon}(\rxv) + d(\rxv) \log \varepsilon\big)\,.
\ee
\end{definition}
This definition corresponds to the following procedure:
\begin{enumerate}
\item Quantize $\rxv$ using an entropy-min\-imizing quantization%
\footnote{We assume for simplicity that an entropy-minimizing quantization exists although in general the infimum in \eqref{eq:epsentropy} may not be attained.} 
$\mathfrak{Q}$ given a diameter constraint $\varepsilon$, i.e., consider the discrete random variable $[\rxv]_{\mathfrak{Q}}$ with probabilities $p_i = \Pr\{[\rxv]_{\mathfrak{Q}}=i\} = \Pr\{\rxv\in \sA_i\}$ for $\sA_i\in \mathfrak{Q}$, where the diameter of each $\sA_i$ is upper bounded by $\varepsilon$.
\item Calculate the entropy of the quantized random variable $[\rxv]_{\mathfrak{Q}}$, i.e.,  the negative expectation of the logarithm of the probability mass function $p_i$.
\item Add the correction term $d(\rxv) \log \varepsilon$ to account for the dimension of the random variable $\rxv$.
\item Take the limit $\varepsilon \to 0$.
\end{enumerate}

Although this entropy is more general than the entropy of dimension $d(\rxv)$ in \eqref{eq:renyientropy}, the fundamental problems persist: we are still restricted to the choice of  sets of small diameter (this is of course useful if we consider maximal distance as a measure of distortion but can yield unnecessarily many quantization points for  areas of almost zero probability), 
and the definition still seems to be difficult to handle analytically and lacks connections to established information-theoretic quantities  such as mutual information.

\subsection{An Alternative Approach}\label{sec:ourapproach}
Here, we propose a different approach, which is motivated by the definition of differential entropy.
The basic idea is to circumvent the quantization step and perform the entropy calculation at the end.
Assuming $\rxv\in \R^M$, this results in the  following procedure: 
\begin{enumerate}
\item For some $\xv\in \R^M$, divide the probability $\Pr\{\rxv\in \srB_{\varepsilon}(\xv)\}$ by the correction factor%
\footnote{The  constant factor $\omega(d(\rxv))$ is included to obtain equality with differential entropy in the special case $d(\rxv)=M$. 
A different factor would result in an additive constant in the entropy definition.} 
$\omega(d(\rxv))\, \varepsilon^{d(\rxv)}$. (Recall that $\omega(d(\rxv))$ denotes the volume of the $d(\rxv)$-dimensional unit ball.)
\item Take the limit $\varepsilon \to 0$.
%\item Take the information dimension $d(\rxv)$ into account and calculate locally a density function being the equivalent to the discrete probability mass function, i.e., a ``probability per $d(\rxv)$-dimensional area''.
\item Calculate the entropy as the negative expectation of the logarithm of the resulting density function.
\end{enumerate}
More specifically, steps 1--2 yield the density function%
\footnote{A mathematically rigorous definition will be provided in Section~\ref{sec:rectifiablemeasures}.}
\be\label{eq:introdensity}
\theta_{\rxv}(\xv)\triangleq \lim_{\varepsilon\to 0} \frac{\Pr\{\rxv\in \srB_{\varepsilon}(\xv)\}}{\omega(d(\rxv))\, \varepsilon^{d(\rxv)}}
\ee
and the entropy in step 3 is thus given by
\be\label{eq:intentropy} 
\eh{d(\rxv)}(\rxv)\triangleq -\E_{\rxv}[\log \theta_{\rxv}(\rxv)]\,.
\ee

We will show that this  definition of entropy will lead to definitions of joint and conditional entropy, various useful relations, connections to mutual information, an asymptotic equipartition property, and bounds relevant to source coding.
However, our definition does have one limitation:
as pointed out in \cite[Sec.~VII-A]{wuve10}, the existence of the limit in \eqref{eq:introdensity} for almost every $\xv\in \R^M$ is a much stronger assumption than the existence of the R\'enyi information dimension \eqref{eq:Rinfdim}. 
Loosely speaking, the existence of the limit in \eqref{eq:introdensity} requires that the random variable $\rxv$ is $d(\rxv)$-dimensional almost everywhere whereas the existence of the R\'enyi information dimension merely requires that the random variable is  $d(\rxv)$-dimensional ``on average.''
By Preiss' Theorem \cite[Th.~5.6]{preiss87}, convergence in \eqref{eq:introdensity} even implies that the probability measure induced by the random variable $\rxv$ is rectifiable (see Definition~\ref{def:rectmeasure} in Section~\ref{sec:rectifiablemeasures}), which means that our definition does not apply to, e.g., self-similar fractal distributions. 
However, we are not aware of any application or calculation of the $d(\rxv)$-dimensional entropy in \eqref{eq:renyientropy} (or the asymptotic version of $\varepsilon$-entropy) for fractal distributions, and it does not seem clear whether the $d(\rxv)$-dimensional entropy is well defined in that case (although the information dimension \eqref{eq:Rinfdim} exists).

The rectifiability also implies that the density function $\theta_{\rxv}(\rxv)$ is equal to a certain Radon-Nikodym derivative. 
Based on this equality, the entropy $\eh{d(\rxv)}(\rxv)$ defined in \eqref{eq:intentropy} and \eqref{eq:introdensity}
%% the approach in \eqref{eq:intentropy} 
can be interpreted as a \textit{generalized entropy} as defined in \cite[eq.~(1.5)]{cs73} by 
\be\label{eq:genent}
H_{\lambda}(\mu)\triangleq 
\begin{cases}
- \displaystyle\int_{\R^{M}}\log \bigg(\frac{\mathrm{d}\mu}{\mathrm{d}\lambda}(\xv)\bigg) \, \mathrm{d}\mu(\xv) & \text{if }\mu \ll  \lambda \\
\infty &  \text{else.}
\end{cases}
\ee
Here, $\lambda$ is a $\sigma$-finite measure on $\R^M$ and $\mu$ is a probability measure on $\R^M$.
While $\mu$ can be chosen as the measure of a given random variable, the generalized entropy \eqref{eq:genent} provides no intuition on how to choose the measure $\lambda$.
It is more similar to a divergence between measures and, in particular, reduces to the Kullback-Leibler divergence \cite{KL51} for a probability measure $\lambda$.
We will see (cf.~Remark~\ref{rem:genent}) that our entropy definition coincides with \eqref{eq:genent} for the choice $\lambda=\Hm{m}|_{\sE}$, where $m$ and $\sE$ depend on the given random variable.
This  interpretation will allow us to use basic results from \cite{cs73} for our entropy definition.

%An extension of our theory to mixtures of rectifiable measures of different dimensions may provide an interesting direction for future work. 

Motivated by the entropy expression in \eqref{eq:intentropy}, a formal definition of the entropy of an integer-dimensional  random variable will be given in Section~\ref{sec:MotDef}, based on the mathematical theory of rectifiable measures discussed next.

%%%%%%%%%%%%%%%%%%%%%%%%%%%%%%%
\section{Rectifiable Random Variables} \label{sec:rectifiablesets}   
%%%%%%%%%%%%%%%%%%%%%%%%%%%%%%%
%Here goes the definition of rectifiable sets, equivalent definitions and examples.
%Maybe also some theorems about approximation, tangent measures, etc.
As mentioned in Section~\ref{sec:ourapproach}, the existence of a $d(\rxv)$-dimensional density implies that the random variable $\rxv$ is rectifiable. 
In this section, we recall the definitions of rectifiable sets and measures and introduce rectifiable random variables as a straightforward extension.
Furthermore, we present some basic properties that will be used in subsequent sections.
For the convenience of readers who prefer to skip the mathematical details, we summarize the most important facts in Corollary~\ref{cor:propsrectrandvar}. 

\subsection{Rectifiable Sets}
Our basic geometric objects of interest are rectifiable sets \cite[Sec.~3.2.14]{fed69}.
As the definition of rectifiable sets is not consistent in the literature,  we provide the definition most convenient for our purpose.
We recall that $\Hm{m}$ denotes the $m$-dimensional Hausdorff measure.

%Rev1.1: Changed definition to avoid null sets we don't need
\begin{definition}[\mbox{\cite[Def.~2.57]{Ambrosio2000Functions}}]\label{def:rectset}
For $m\in \IN$, an $\Hm{m}$-mea\-surable 
set $\sE\subseteq \R^M$ (with $M\geq m$) is called \textit{$m$-rectifiable}%
\footnote{In \cite[Def.~2.57]{Ambrosio2000Functions}, these sets are called \textit{countably $\Hm{m}$-rectifiable}.} 
if there exist $\Leb^m$-measurable, bounded sets $\sA_k\subseteq \R^m$ and Lipschitz functions $f_k\colon \sA_k\to \R^M$, both for%
\footnote{This definition also encompasses finite index sets $k\in \{1, \dots, K\}$; it suffices to set $\sA_k=\emptyset$ for $k> K$.}
$k\in \N$, such that $\Hm{m}\left(\sE\setminus \bigcup_{k\in \IN}f_k(\sA_k)\right)=0$.
%Here $\sA_k\subseteq \R^m$ are bounded sets and $f_k\colon \sA_k\mapsto \R^M$ with $M>m\geq 1$ are Lipschitz functions. 
%Then $f(\sA)\triangleq\{f(\xv)\in \R^m: \xv\in \sA\}$ is called an $m$-rectifiable set. 
A set $\sE\subseteq \R^M$ is called $0$-rectifiable if it is finite or countably infinite.
\end{definition}
\begin{remark}
%Although Definition~\ref{def:rectset} holds for arbitrary $m,M\in \N$, the only interesting case is when $m\leq M$, because for $m>M$ $\Hm{m}(\R^M)=0$ and, thus, every subset of $\R^M$ is $m$-rectifiable. 
Hereafter, we will often consider the setting of $m$-rectifiable sets in $\R^M$ and tacitly assume $m\in \{0,\dots, M\}$.
\end{remark}

Rectifiable sets satisfy the following well-known basic properties.
\begin{lemma}\label{lem:proprectset} 
Let $\sE$ be an $m$-rectifiable subset of $\R^M$.
\begin{enumerate}
\item Any $\Hm{m}$-measurable 
subset $\sD\subseteq \sE$ is also $m$-rectifiable.\label{en:subsetrect}
\item The measure $\Hm{m}|_{\sE}$ is $\sigma$-finite. \label{en:hmesigmafinite}
\item Let $\phi\colon \R^M \to \R^N$ with $N\geq m$ be a Lipschitz function.
 %with approximate $m$-dimensional Jacobian $\Jm_{\phi}>0$ $\Hm{m}|_{\sE}$-almost everywhere.
If  $\phi(\sE)$ is $\Hm{m}$-measurable, 
then it is $m$-rectifiable. \label{en:conslipschitz}
\item For $n>m$, we have $\Hm{n}(\sE)=0$. \label{en:rectsetsdimm}
\item Let $\sE_i$ for $i\in \N$ be $m$-rectifiable sets. 
Then $\bigcup_{i\in \N}\sE_i$ is $m$-rectifiable. \label{en:unionrectsetsrect}
\item For $m\neq 0$, $\R^m$ is $m$-rectifiable. \label{en:rmmrectifiable}
%Rev1.1: Replace former Lemma 23
%\item \cha{If $\sE_1$ is $m_1$-rectifiable and $\sE_2$ is $m_2$-rectifiable, then $\sE_1\times \sE_2$ is $(m_1+m_2)-$rectifiable.}\label{en:prodrect}
\end{enumerate}
\end{lemma}
%Rev1.1: remove proof
%\begin{IEEEproof}
%Properties~\ref{en:subsetrect}--\ref{en:rmmrectifiable} are well known; however, their proofs are not always provided in the literature.
%Therefore, for the reader's convenience, we provide proofs in Appendix~\ref{app:proofproprectset}.
%\end{IEEEproof}

Intuitively, rectifiable sets are lower-dimensional subsets of Euclidean space. 
Examples   include affine subspaces, algebraic varieties, differentiable manifolds, and graphs of Lipschitz functions.
As   countable unions of  rectifiable sets are again rectifiable, further examples are countable unions of any of the aforementioned sets.

\begin{remark}\label{rem:charrect}
There are various characterizations of $m$-rec\-tifiable sets that provide connections to other mathematical disciplines. 
For example, an $\Hm{m}$-measurable set $\sE\subseteq \R^M$ is $m$-rectifiable if and only if there exist $\sT_k\subseteq \R^M$ such that
$\sE \subseteq \sT_0\cup \bigcup_{k\in \IN} \sT_k$, 
where $\Hm{m}(\sT_0)=0$ and each $\sT_k$ is an $m$-dimensional, embedded $C^1$ submanifold of $\R^M$ \cite[Lem.~5.4.2]{krpa09}. 
Another characterization, based on \cite[Cor.~3.2.4]{fed69}, is that $\sE\subseteq \R^M$ is $m$-rectifiable if and only if 
\be\label{eq:rectascupborel}
\sE\subseteq\sE_0 \cup \bigcup_{k\in \IN}f_k(\sA_k)
\ee
where $\Hm{m}(\sE_0)=0$, $\sA_k$ are bounded Borel sets, and $f_k\colon \R^m\to \R^M$ are  Lipschitz functions that are one-to-one on $\sA_k$.
Due to \cite[Th.~15.1]{ke95}, this implies that $f_k(\sA_k)$ are also Borel sets.
\end{remark}

%%%%%%%%%%%%%%%%%%%%%%%%%%%%%%%
\subsection{Rectifiable Measures} \label{sec:rectifiablemeasures}   
%%%%%%%%%%%%%%%%%%%%%%%%%%%%%%%
%In this section we define rectifiable measures, present Preiss' Theorem and talk about Hausdorff density and Radon Nikodym derivatives.
Loosely speaking, rectifiable measures are measures that are concentrated on a rectifiable set. 
The most convenient way to define ``concentrated on''  mathematically is in terms of absolute continuity with respect to a specific Hausdorff measure. 
\begin{definition}[\mbox{\cite[Def.~2.59]{Ambrosio2000Functions}}]
\label{def:rectmeasure}
A Borel measure $\mu$ on $\R^M$ is called \textit{$m$-rectifiable} if there exists an $m$-rectifiable set $\sE\subseteq \R^M$ such that $\mu\ll \Hm{m}|_{\sE}$. 
%The set $\sE$ is called a support of \mu.
\end{definition}
%Rev1.2:change definition of theta to Radon Nikodym derivative
For an $m$-rectifiable measure $\mu$, i.e., $\mu\ll \Hm{m}|_{\sE}$ for an $m$-rectifiable set $\sE\subseteq \R^{M}$, we have 
by  Property~\ref{en:hmesigmafinite} in Lemma~\ref{lem:proprectset} that $\Hm{m}|_{\sE}$ is $\sigma$-finite.
Thus, by the Radon-Nikodym theorem \cite[Th.~1.28]{Ambrosio2000Functions}, there exists the Radon-Nikodym derivative 
\be \label{eq:Hausdorff_RN}
\theta_{\mu}^{m}(\xv)\triangleq \frac{\mathrm{d}\mu}{\mathrm{d}\Hm{m}|_{\sE}}(\xv)
\ee
satisfying
$\mathrm{d}\mu=\theta_{\mu}^{m}\ \mathrm{d}\Hm{m}|_{\sE}$.
We will refer to $\theta_{\mu}^{m}(\xv)$ as the \textit{$m$-dimensional Hausdorff density of $\mu$.}
%Rev1.2: remove Lemma
%We next construct a rectifiable set  $\sEt\subseteq \sE$ such that $\frac{\mathrm{d}\mu}{\mathrm{d}\Hm{m}|_{\sEt}}>0$ almost everywhere.
%\begin{lemma}\label{lem:support}
%Let $\mu$ be an $m$-rectifiable measure on $\R^M$, i.e., $\mu\ll \Hm{m}|_{\sE}$ for an $m$-rectifiable set $\sE\subseteq \R^{M}$. 
%Then there exists an $m$-rectifiable
%set $\sEt\subseteq \sE$ such that 
%\begin{enumerate}
%\item \label{en:abscontwrset}$\mu\ll \Hm{m}|_{\sEt}$;
%\item \label{en:aeequalrdse}$\frac{\mathrm{d}\mu}{\mathrm{d}\Hm{m}|_{\sEt}}=\frac{\mathrm{d}\mu}{\mathrm{d}\Hm{m}|_{\sE}}$ $\Hm{m}|_{\sEt}$-almost everywhere;
%\item \label{en:aepositive}$\frac{\mathrm{d}\mu}{\mathrm{d}\Hm{m}|_{\sEt}}>0$ $\Hm{m}|_{\sEt}$-almost everywhere. 
%\end{enumerate}
%%In addition $\frac{\mathrm{d}\mu}{\mathrm{d}\Hm{m}|_{\sEt}}>0$ holds also $\mu$-almost everywhere.
%\end{lemma}
%\begin{IEEEproof}[\hspace{-1em}Proof]
%See Appendix~\ref{app:proofsupport}.
%\end{IEEEproof}

%By Lemma~\ref{lem:support}, there exists for an $m$-rectifiable measure $\mu$ a set $\sEt$ such that $\mu\ll \Hm{m}|_{\sEt}$ (as required by Definition~\ref{def:rectmeasure}) and, additionally, $\frac{\mathrm{d}\mu}{\mathrm{d}\Hm{m}|_{\sEt}}>0$ $\Hm{m}|_{\sEt}$-almost everywhere.
\begin{remark}\label{rem:uniquem}
If $\mu$ is an $m$-rectifiable probability measure, it  cannot be $n$-rectifiable for $n\neq m$. 
Indeed, suppose that $\mu$ is both $m$-rectifiable and $n$-rectifiable where, without loss of generality, $n>m$.
Then there exists an $m$-rectifiable set $\sE$ such that  $\mu\ll \Hm{m}|_{\sE}$, which implies  $\mu(\sE^c)=0$. 
There also exists an $n$-rectifiable set $\sF$ such that $\mu\ll \Hm{n}|_{\sF}$.
By Property~\ref{en:rectsetsdimm} in Lemma~\ref{lem:proprectset},  the $m$-rectifiable set $\sE$ satisfies $\Hm{n}(\sE)=0$ and, in particular, $\Hm{n}|_{\sF}(\sE)=0$. 
Because $\mu\ll \Hm{n}|_{\sF}$, this implies $\mu(\sE)=0$.
Hence, $\mu(\R^M)=\mu(\sE^c)+\mu(\sE)=0$, which is a contradiction to the assumption that $\mu$ is a probability measure.
%If $n<m$ we can simply change the roles of $m$ and $n$.
\end{remark}

%Rev1.2: add reason for support
To avoid the nuisance of separately considering the case $\frac{\mathrm{d}\mu}{\mathrm{d}\Hm{m}|_{\sE}}=0$ in many proofs and to reduce the class of $m$-rectifiable sets of interest, we define the following notion of a support of an $m$-rectifiable measure.

\begin{definition}\label{def:support}
For an $m$-rectifiable measure $\mu$ on $\R^M$, an $m$-rectifiable set $\sE\subseteq\R^M$ is called \textit{a support of $\mu$} if $\mu\ll \Hm{m}|_{\sE}$,
$\frac{\mathrm{d}\mu}{\mathrm{d}\Hm{m}|_{\sE}}>0$ $\Hm{m}|_{\sE}$-almost everywhere, 
and $
\sE=\bigcup_{k\in \IN}f_k(\sA_k)
$
where, for $k\in \N$, $\sA_k$ is a bounded Borel set and $f_k\colon \R^m\to \R^M$ is a Lipschitz function that is one-to-one on $\sA_k$.
\end{definition}

\begin{lemma}\label{lem:support}
Let $\mu$ be an $m$-rectifiable measure, i.e., $\mu\ll \Hm{m}|_{\sE}$ for an $m$-rectifiable set $\sE\subseteq \R^{M}$. 
Then there exists a support $\sEt\subseteq \sE$.
Furthermore, the support is unique up to  sets of $\Hm{m}$-measure zero.
\end{lemma}
\begin{IEEEproof}
See Appendix~\ref{app:proofsupport}.
\end{IEEEproof}

%Rev1.2:Change definition of hausdorff density
\begin{remark}
\label{rem:hausdorffdensity}
For $m$-rectifiable measures, it is possible to interpret the Hausdorff density $\theta_{\mu}^{m}(\xv)$ as a measure of ``local probability per area.''
Indeed, for an $m$-rectifiable measure   $\mu$, i.e., $\mu \ll \Hm{m}|_{\sE}$ for an $m$-rectifiable set $\sE$, we can write $\theta_{\mu}^{m}(\xv)$ in \eqref{eq:Hausdorff_RN} as
\be \label{eq:hausdorffdens}
\theta_{\mu}^{m}(\xv)= \lim_{r\to 0} \frac{\mu(\srB_r(\xv))}{\omega(m) r^m}
\ee
 $\Hm{m}|_{\sE}$-almost everywhere (for a proof see \cite[Th.~2.83 and eq.~(2.42)]{Ambrosio2000Functions}).
Furthermore, the right-hand side in \eqref{eq:hausdorffdens} vanishes for $\Hm{m}$-almost all points not in $\sE$.
Note the similarity of \eqref{eq:hausdorffdens} with the ad-hoc construction in Section~\ref{sec:ourapproach}.
Indeed, \eqref{eq:hausdorffdens} is the mathematically rigorous formulation of \eqref{eq:introdensity}. This formulation also provides details regarding 
the probability measures for which it
results in a well-defined quantity.
\end{remark}

%%%%%%%%%%%%%%%%%%%%%%%%%%%%%%%%%
\subsection{Rectifiable Random Variables}
\label{sec:rectifiable-rvs}
%%%%%%%%%%%%%%%%%%%%%%%%%%%%%%%%

As we are only interested in probability measures and because information theory is often formulated for random variables, we define $m$-rectifiable random variables.
In what follows, we consider a random variable $\rxv\colon (\Omega, \mathfrak{S}) \to (\R^M, \mathfrak{B}_M)$ on a probability space $(\Omega, \mathfrak{S}, \mu)$, i.e., $\Omega$ is a set, $\mathfrak{S}$ is a $\sigma$-algebra on $\Omega$, and $\mu$ is a probability measure on $(\Omega, \mathfrak{S})$.
The probability measure induced by the random variable $\rxv$ is denoted by $\mu\rxv^{-1}$.
For  $\sA\in \mathfrak{B}_M$, $\mu\rxv^{-1}(\sA)$ equals the probability that $\rxv\in \sA$, i.e., 
\be\label{eq:definvmeasure}
\mu\rxv^{-1}(\sA)=\mu(\rxv^{-1}(\sA))=\Pr\{\rxv\in \sA\}\,.
\ee
%Here, the intuitive notation $\Pr\{\rxv\in \sA\}$ is introduced for the convenience of readers not familiar with measure theory.
%We will state most of our results for random variables but have to use measures in the proofs. 

\begin{definition}\label{def:rectrandomvar}
A random variable $
\rxv\colon  \Omega \to  \R^M 
$
 on a probability space $(\Omega, \mathfrak{S}, \mu)$ is called \textit{$m$-rectifiable} if the induced probability measure $\mu\rxv^{-1}$ on $\R^M$ is $m$-rectifiable, i.e., there exists an $m$-rectifiable set $\sE\subseteq \R^M$ such that $\mu\rxv^{-1}\ll \Hm{m}|_{\sE}$.
The $m$-dimensional Hausdorff density of an $m$-rectifiable random variable $\rxv$ is defined as (cf.\ \eqref{eq:Hausdorff_RN})
\be\label{eq:defrvhddensity}
%Rev1.2:change definition of Hausdorff density
\theta_{\rxv}^m(\xv)\triangleq \theta_{\mu\rxv^{-1}}^m(\xv) 
= \frac{\mathrm{d}\mu\rxv^{-1}}{\mathrm{d}\Hm{m}|_{\sE}}(\xv) \,.
%= \lim_{r\to 0} \frac{\Pr\{\rxv\in \srB_r(\xv)\}}{\omega(m) r^m}\,.
\ee
Furthermore, a support of the measure  $\mu\rxv^{-1}$ is called a \textit{support of $\rxv$}, i.e., $\sE$ is a support of $\rxv$ if $\mu\rxv^{-1}\ll \Hm{m}|_{\sE}$, $\frac{\mathrm{d}\mu\rxv^{-1}}{\mathrm{d}\Hm{m}|_{\sE}}(\xv)>0$ $\Hm{m}|_{\sE}$-almost everywhere, 
and $
\sE=\bigcup_{k\in \IN}f_k(\sA_k)
$
where, for $k\in \N$, $\sA_k$ is a bounded Borel set and $f_k\colon \R^m\to \R^M$ is a Lipschitz function that is one-to-one on $\sA_k$.
\end{definition}
%Furthermore, we will use the notation $\theta_{\rxv}^m$ for the Hausdorff density of the induced probability measure $\mu\rxv^{-1}$. 
Note that due to Remark~\ref{rem:uniquem}, an $m$-rectifiable random variable cannot be $n$-rectifiable for $n\neq m$.

In the nontrivial case $m< M$, the $m$-dimensional Hausdorff density $\theta_{\rxv}^m(\xv)$ is not a probability density function in the classical sense and is nonzero only on an $m$-dimensional set $\sE$. 
Indeed, the random variable $\rxv$ will vanish everywhere except on a set of Lebesgue measure zero, and thus a probability density function cannot exist.
%% thus, not allowing for a probability density function. 
However, the $m$-dimensional Hausdorff measure of the support set does not vanish, and one can think of $\theta_{\rxv}^m$ as an $m$-dimensional probability density function 
of the random variable $\rxv$ on $\R^M$.

Based on  our discussion of rectifiable measures in Section~\ref{sec:rectifiablemeasures},  we can find a characterization of $m$-rectifiable random variables that resembles well-known properties of continuous random variables. 
This characterization is stated in the next corollary.
Note, however, that although everything seems to be similar to the continuous case, Hausdorff measures lack substantial properties of the Lebesgue measure, e.g., the product measure is not always again a Hausdorff measure. % which most readers will be familiar with. 
\begin{corollary}\label{cor:propsrectrandvar} 
Let $\rxv$ be an $m$-rectifiable random variable on $\R^M$, i.e., $\mu\rxv^{-1} \ll \Hm{m}|_{\sE}$ for an $m$-rectifiable set $\sE\subseteq \R^M$. 
Then there exists the $m$-dimensional Hausdorff density $\theta_{\rxv}^m$,  and  the following properties hold:
\begin{enumerate}
%\item \label{en:corhdenseqrn}The $m$-dimensional Hausdorff density $\theta_{\rxv}^m$ coincides with the Radon-Nikodym derivative $\frac{\mathrm{d}\mu\rxvm^{-1}}{\mathrm{d}\Hm{m}|_{\sE}}$ $\Hm{m}|_{\sE}$-almost everywhere, i.e.,
%\be\label{eq:corhdenseqrn}
%\theta_{\rxv}^m(\xv)=\frac{\mathrm{d}\mu\rxv^{-1}}{\mathrm{d}\Hm{m}|_{\sE}}(\xv)  \quad \Hm{m}|_{\sE}\text{-almost everywhere.}
%\ee
\item \label{en:cordens}The probability $\Pr\{\rxv\in \sA\}$ for a measurable set $\sA\subseteq \R^M$ can be calculated as the integral of $\theta_{\rxv}^m$ over $\sA$ with respect to the $m$-dimensional Hausdorff measure restricted to $\sE$, i.e.,
\be\label{eq:calcprobina}
\Pr\{\rxv\in \sA\}=\mu\rxv^{-1}(\sA)= \int_{\sA} \theta_{\rxv}^m(\xv) \, \mathrm{d}\Hm{m}|_{\sE}(\xv)\,.\hspace{-.5mm}
\ee
\item \label{en:corexpectation}The expectation of a measurable function $f\colon \R^M\to\R$ with respect to the random variable $\rxv$ can be expressed as
\be\label{eq:corexpectation}
\E_{\rxv}[f(\rxv)]= \int_{\R^M} f(\xv)\, \theta_{\rxv}^m(\xv) \, \mathrm{d}\Hm{m}|_{\sE}(\xv)\,.
\ee
%\item\label{en:rvsuppisrect}The support $\sE$ is $m$-rectifiable.
%\item\label{en:cordensnull}The Hausdorff density $\theta_{\rxv}^m$ is zero $\Hm{m}$-almost everywhere on $\sE^c$.
\item \label{en:ehasprob1}The random variable $\rxv$ is in $\sE$ with probability one, i.e.,
\be
\Pr\{\rxv\in \sE\}
 = \mu\rxv^{-1}(\sE) 
=\int_{\sE} \theta_{\rxv}^m(\xv) \, \mathrm{d}\Hm{m}|_{\sE}(\xv)
 =1\,.\label{eq:probineisone}
\ee
\item \label{en:corexistsupport}There exists a support $\sEt\subseteq \sE$ of $\rxv$.
%\item \label{en:cordenspos} $\sE$ is a support of $\rxv$ if and only if the Hausdorff density $\theta_{\rxv}^m$ is positive $\Hm{m}$-almost everywhere on $\sE$.
\end{enumerate} 
\end{corollary}

The special cases $m=0$ and $m=M$ reduce to well-known concepts.

\begin{theorem}\label{th:specialcaserect}
Let $\rxv$ be a random variable on $\R^M$. Then:
\begin{enumerate}
\item \label{en:specialcasedisc}$\rxv$ is $0$-rectifiable if and only if it is a discrete random variable, i.e., there exists a probability mass function $p_{\rxv}(\xv_i)=\Pr\{\rxv=\xv_i\}>0$, $i\in \sI$, where $\sI$ is a finite or countably infinite index set indicating all possible  realizations $\xv_i$ of $\rxv$.
In this case, $\theta_{\rxv}^0=p_{\rxv}$ and $\sE=\{\xv_i:i\in \sI\}$ is a support of $\rxv$.
\item \label{en:specialcasecont}$\rxv$ is $M$-rectifiable if and only if it is a continuous random variable, i.e., there exists a probability density function $f_{\rxv}$ such that $\Pr\{\rxv\in \sA\}=\int_{\sA}f_{\rxv}(\xv)\, \mathrm{d}\Leb^M(\xv)$.
In this case, $\theta_{\rxv}^M=f_{\rxv}$  $\Leb^M$-almost everywhere.
\end{enumerate}
\end{theorem}
\begin{IEEEproof}
See Appendix~\ref{app:proofspecialcaserect}.
\end{IEEEproof}

The following theorem introduces a nontrivial class of $m$-rectifiable random variables.
%Rev1.main:changed to LOCALLY Lipschitz to include Wishart matrices
\begin{theorem}\label{th:exrectmeasure}
Let $\rxv$ be  a continuous random variable on $\R^m$.
Furthermore, let $\phi\colon \R^m\to \R^M$ with $M\geq m$ be a locally Lipschitz mapping whose $m$-dimensional  Jacobian determinant%
\footnote{The $m$-dimensional  Jacobian determinant of $\phi$ is defined as $\Jm_{\phi}(\xv)=\sqrt{\det(\Dm_{\phi}^{\trans}(\xv)\,\Dm_{\phi}(\xv))}$, where $\Dm_{\phi}(\xv) \in \R^{M\times m}$ denotes the Jacobian matrix of $\phi$, which is guaranteed to exist almost everywhere. 
Note in particular that $\Jm_{\phi}(\xv)$ is nonnegative.} 
satisfies
$\Jm_{\phi}(\xv)> 0$ $\Leb^m$-almost everywhere, and assume that $\phi(\R^m)$ is $\Hm{m}$-measurable.
Then $\ryv \triangleq\phi(\rxv)$ is an $m$-rectifiable  random variable on $\R^M$.
\end{theorem}
\begin{IEEEproof}
According to Definition~\ref{def:rectrandomvar}, we have to show that $\mu\ryv^{-1}\ll \Hm{m}|_{\sE}$ for an $m$-rectifiable set $\sE\subseteq \R^M$.
By Properties~\ref{en:subsetrect}, \ref{en:conslipschitz}, and  \ref{en:rmmrectifiable} in Lemma~\ref{lem:proprectset}, the set $\sE\triangleq\phi(\sB_{r}(\0v))$ is $m$-rectifiable 
($\phi$ is Lipschitz on $\sB_{r}(\0v)$ for all $r>0$).
%Rev1.main: Changed proof for locally Lipschitz functions
Hence, by Property~\ref{en:unionrectsetsrect}  in Lemma~\ref{lem:proprectset}, the set $\sE\triangleq\phi(\R^m)=\bigcup_{r\in \N}\phi(\sB_{r}(\0v))$ is $m$-rectifiable.
Thus, it suffices to show that $\mu\ryv^{-1}\ll \Hm{m}|_{\phi(\R^m)}$, i.e., that for any $\Hm{m}$-measurable set $\sA  \subseteq \R^{M}$, $\Hm{m}|_{\phi(\R^m)}(\sA)=0$ implies $\mu\ryv^{-1}(\sA)=0$.
To this end, assume 
first
that $\Hm{m}|_{\phi(\R^m)}(\sA)=0$ for a 
\textit{bounded}
$\Hm{m}$-measurable set $\sA  \subseteq \R^{M}$. 
Let $f$ denote the probability density function of $\rxv$.
By the generalized change of variables formula \cite[eq.~(2.47)]{Ambrosio2000Functions}, we have
\ba
& \int_{\phi^{-1}(\sA)} f(\xv)\Jm_{\phi}(\xv)\, \mathrm{d}\Leb^m(\xv) \notag \\
& \rule{13mm}{0mm}=
\int_{\phi(\phi^{-1}(\sA))} \sum_{\xv\in \phi^{-1}(\sA)\cap \phi^{-1}(\{\yv\})}f(\xv)\, \mathrm{d}\Hm{m}(\yv) \notag \\
& \rule{13mm}{0mm}=\int_{\sA\cap \phi(\R^m)} \sum_{\xv\in \phi^{-1}(\sA)\cap \phi^{-1}(\{\yv\})}f(\xv)\, \mathrm{d}\Hm{m}(\yv) \notag \\
& \rule{13mm}{0mm}\stackrel{(a)}= 0 \label{eq:intdensiszero}
\ea
where $(a)$ holds because $\Hm{m}(\sA\cap \phi(\R^m))=0$.
%Rev1.6:Correctly added $\Leb^m$-almost everywhere on \phi^{-1}(\sA)
Because $\Jm_{\phi}(\xv)> 0$ $\Leb^{m}$-almost everywhere, \eqref{eq:intdensiszero} implies $f(\xv)=0$ $\Leb^m$-almost everywhere 
on $\phi^{-1}(\sA)$, and hence $\int_{\phi^{-1}(\sA)} f(\xv)\, \mathrm{d}\Leb^m(\xv)=0$.
Thus, we have
\be
\mu\ryv^{-1}(\sA)
 = \mu\rxv^{-1}(\phi^{-1}(\sA))
 =\int_{\phi^{-1}(\sA)} f(\xv)\, \mathrm{d}\Leb^m(\xv)
 =0\,.\notag
\ee
%Rev1.main: Changed proof for locally Lipschitz functions
For an \textit{unbounded} $\Hm{m}$-measurable set $\sA  \subseteq \R^{M}$ satisfying  $\Hm{m}|_{\phi(\R^m)}(\sA)=0$, following the arguments above, we  obtain $\mu\ryv^{-1}(\sA \cap \sB_{r}(\0v))=0$ for the bounded sets $\sA \cap \sB_{r}(\0v)$, $r\in \N$.
This implies $\mu\ryv^{-1}(\sA)\leq \sum_{r\in \N}\mu\ryv^{-1}(\sA \cap \sB_{r}(\0v))=0$.
\end{IEEEproof}

%%%%%%%%%%%%%%%%%%%%%%%%%%%%%%%%%
\subsection{Example: Distributions on the Unit Circle}
\label{sec:unitcircle}
%%%%%%%%%%%%%%%%%%%%%%%%%%%%%%%%
%
As a basic example of $1$-rectifiable singular random variables, we consider distributions on the unit circle in $\R^2$, i.e., on $\sS_{1}\triangleq \{\xv\in \R^2:\norm{\xv}=1\}$.
\begin{corollary}
Let $\rz$ be a continuous random variable on $\R$. 
Then $\rxv = (\rx_1 \; \rx_2)^{\trans} \triangleq (\cos \rz \;\ist \sin \rz)^{\trans}$ is a $1$-rectifiable random variable. 
\end{corollary}
\begin{IEEEproof}
The mapping $\phi\colon z\mapsto (\cos z \;\ist \sin z)^{\trans}$ is Lipschitz and its Jacobian determinant is identically one.
Thus, we can directly apply Theorem~\ref{th:exrectmeasure}.
\end{IEEEproof}
This toy example is intuitive and illustrates the concept of $m$-rectifiable singular random variables in a very simple setup.
In a similar way, one can analyze the rectifiability of distributions on various other geometric structures.

%
%\vspace{-4mm}
%
%%%%%%%%%%%%%%%%%%%%%%%%%%%%%%%%%
\subsection{Example: Positive Semidefinite Rank-One Random Matrices}
\label{sec:wishart}
%%%%%%%%%%%%%%%%%%%%%%%%%%%%%%%%
%
%Rev1.main: proof that rank one Wishart matrices are rectifiable
A less obvious example of an $m$-rectifiable singular random variable are
%% the distribution of 
positive semidefinite rank-one 
%% Wishart 
random matrices, i.e., matrices of the form $\rXm = \rzv\rzv^{\trans} \in \R^{m\times m}$, where $\rzv$ is a continuous random variable on $\R^m$. 
%% However, our 
%
\begin{corollary}\label{th:wishartrectifiable}
Let $\rzv$ be a continuous random variable on $\R^m$.
Then the random  matrix $\rXm\triangleq \rzv\rzv^{\trans}$ is $m$-rectifiable on $\R^{m^2}$.
%% In particular, rank-one Wishart matrices $\rWm_{1, \Sigma}$ are $m$-rectifiable.
\end{corollary}
\begin{IEEEproof}
The mapping $\phi\colon \zv\mapsto \zv\zv^{\trans}$ is locally Lipschitz.
Thus, in order to apply Theorem~\ref{th:exrectmeasure},
it remains to show that $\Jm_{\phi}(\zv)> 0$   $\Leb^{m}$-almost everywhere.
To calculate the Jacobian matrix $\Dm_{\phi}(\zv)$,  we stack the columns of the matrix $\zv\zv^{\trans}$ and differentiate the resulting vector with respect to each element $z_i$.
It is easily seen that the resulting Jacobian matrix is given by
\be\label{eq:diffwishart}
\Dm_{\phi}(\zv)=
\begin{pmatrix}
\zv \ev_1^{\trans} + z_1 \Iv_m \\[.5mm]  
\zv \ev_2^{\trans} + z_2 \Iv_m \\ 
\vdots \\ 
\zv \ev_m^{\trans} + z_m \Iv_m
\end{pmatrix}
\ee
where $\ev_i$ denotes the $i$th unit vector.
As long as at least one element $z_i$ is nonzero, $\Dm_{\phi}(\zv)$ has full rank. Thus, $\Jm_{\phi}(\zv)> 0$ $\Leb^{m}$-almost everywhere.
\end{IEEEproof}
%
%% \begin{remark}\label{rem:lowrankwishart}
\begin{remark}\label{rem:lowrankwishart}
For the case of positive definite random matrices, i.e., $\rXm_m = \sum_{i=1}^m\rzv_i\rzv_i^{\trans}$ with independent continuous $\rzv_i$, 
it is easy to see that the measures induced by these random matrices are absolutely continuous with respect to the $m(m+1)/2$-dimensional Lebesgue measure on the space of all symmetric matrices.
The intermediate case of  positive semidefinite rank-deficient
%% rank-$n$ 
random matrices $\rXm_n = \sum_{i=1}^n\rzv_i\rzv_i^{\trans}$ for $n\in \{2, \dots, m-1\}$,
where the $\rzv_i \in \R^m$, $i\in \{1, \dots, n\}$ are independent continuous random variables, is considerably more involved because the mapping 
$(\zv_1, \dots, \zv_n)\mapsto  \sum_{i=1}^n\zv_i\zv_i^{\trans}$ has a vanishing Jacobian determinant almost everywhere.
We conjecture that $\rXm_n$ is $(mn-n(n-1)/2)$-rectifiable, conforming to the dimension of the manifold of all positive semidefinite rank-$n$ matrices with $n$ distinct eigenvalues.
\end{remark}

%\vspace{1mm}

%%%%%%%%%%%%%%%%%%%%%%%%%%%%%%%
\section{Entropy of Rectifiable Random Variables} \label{sec:entropy}   
%%%%%%%%%%%%%%%%%%%%%%%%%%%%%%%
%We define entropy for rectifiable measures, present theorems that are similar to the ones known for differential entropy and connections to mutual information.
%Entropy maximizing measures will also be part of this section.

\subsection{Definition}\label{sec:MotDef}
The $m$-rectifiable random variables introduced in Definition~\ref{def:rectrandomvar} will be the objects  considered in our entropy definition.
Due to the existence of the $m$-dimensional Hausdorff density $\theta_{\rxv}^{m}$ for these random variables (see~\eqref{eq:hausdorffdens} and \eqref{eq:defrvhddensity}), the heuristic approach described in Section~\ref{sec:ourapproach} (see~\eqref{eq:introdensity} and \eqref{eq:intentropy}) can be made rigorous.
%Following the motivation in Section~\ref{sec:defsentropy} and using the definitions presented in Section~\ref{sec:rectifiablesets}, we can finally give a rigorous definition of $m$-dimensional entropy.
\begin{definition}\label{def:mdimentropy}
Let $\rxv$ be an $m$-rectifiable random variable on $\R^M$. 
The \textit{$m$-dimensional entropy of $\rxv$} is defined as
\be
\eh{m}(\rxv) \triangleq - \E_{\rxv}\big[\log  \theta_{\rxv}^m(\rxv) \big] 
%\notag \\ & 
=  - \int_{\R^M} \log  \theta_{\rxv}^{m}(\xv)  \, \mathrm{d}\mu\rxv^{-1}(\xv)\label{eq:defmdimentropy}
\ee
%Rev1.3:add possibility that the integral may not exist or be +-infty
provided the integral on the right-hand side exists in $\R\cup \{\pm\infty\}$.
\end{definition}
By~\eqref{eq:corexpectation}, we obtain
\ba
\eh{m}(\rxv)& = - \int_{\R^M} \theta_{\rxv}^{m}(\xv) \log  \theta_{\rxv}^{m}(\xv)  \, \mathrm{d}\Hm{m}|_{\sE}(\xv) \label{eq:entwithsupport1} \\[1mm]
& = - \int_{\sE} \theta_{\rxv}^{m}(\xv) \log  \theta_{\rxv}^{m}(\xv)  \, \mathrm{d}\Hm{m} (\xv)\label{eq:entwithsupport2}
\ea 
where $\sE\subseteq \R^M$ is an arbitrary $m$-rectifiable set satisfying $\mu\rxv^{-1} \ll \Hm{m}|_{\sE}$ (in particular, $\sE$ may be a support of $\rxv$). 
%Thus, the expression \eqref{eq:entwithsupport} is independent of the specific choice of the support $\sE$.

\begin{remark}\label{rem:genent}
For a fixed $m$-rectifiable measure $\mu$, our entropy definition~\eqref{eq:defmdimentropy} can be interpreted as a generalized entropy \eqref{eq:genent} with  $\lambda=\Hm{m}|_{\sE}$.
This will allow us to use basic results from \cite{cs73} for our entropy definition.
However, our definition changes the measure $\lambda$ based on the choice of $\mu$ and thus is not simply a special case of \eqref{eq:genent}.
\end{remark}

%Rev2.5:changed order of subsections, first transformation properties then relation to entropy, diff entropy
%%%%%%%%%%%%%%%%%%%%%%%%%%%%%%%%%%%
\subsection{Transformation Property}\label{sec:transformationproperties}
%%%%%%%%%%%%%%%%%%%%%%%%%%%%%%%%%%%%%

One important property of differential entropy is   its  invariance under unitary transformations.
A similar result holds for $m$-dimensional entropy.
We can even give a more general result for arbitrary one-to-one Lipschitz mappings.
\begin{theorem} \label{th:gentransformation}
Let $\rxv$ be an $m$-rectifiable random variable on $\R^N$ with $1\leq m\leq N$, finite $m$-dimensional entropy $\eh{m}(\rxv)$, support $\sE$, and $m$-dimensional Hausdorff density $\theta_{\rxv}^{m}$. 
Furthermore, let $\phi\colon \R^N\to \R^M$ with $M\geq m$ be a Lipschitz mapping such that%
\footnote{Here $\Jm^{\sE}_{\phi}$ denotes the Jacobian determinant of the tangential differential of $\phi$ in $\sE$. For details see \cite[Sec.~3.2.16]{fed69}.}
 $\Jm^{\sE}_{\phi}> 0$  $\Hm{m}|_{\sE}$-almost everywhere, $\phi(\sE)$ is $\Hm{m}$-measurable, and $\E_{\rxv}[\log \Jm^{\sE}_{\phi}(\rxv)]$ exists and is finite. 
If the restriction of $\phi$ to $\sE$ is  one-to-one, 
then $\ryv\triangleq\phi(\rxv)$ is an $m$-rectifiable random variable with $m$-dimensional Hausdorff density 
\be\notag %\label{eq:transformeddensth}
\theta_{\ryv}^m(\yv)=\frac{\theta_{\rxv}^m(\phi^{-1}(\yv))}{\Jm^{\sE}_{\phi}(\phi^{-1}(\yv))}
\ee
$\Hm{m}|_{\phi(\sE)}$-almost everywhere, and
its $m$-dimensional entropy is
\be \notag 
\eh{m}(\ryv)= \eh{m}(\rxv)+ \E_{\rxv}[\log \Jm^{\sE}_{\phi}(\rxv)]\,.
\ee
\end{theorem}
 \begin{IEEEproof}
See Appendix~\ref{app:proofgentransformation}.
\end{IEEEproof}

\begin{remark}
Theorem~\ref{th:gentransformation} shows that for the special case of a unitary transformation $\phi$ (e.g., a translation), 
\be 
\eh{m}(\phi(\rxv))= \eh{m}(\rxv) \notag 
\ee
 because $\Jm^{\sE}_{\phi}(\rxv)$ is identically one in that case.
\end{remark}

\begin{remark}\label{rem:gentrafo}
In general, no result resembling Theorem~\ref{th:gentransformation} holds for Lipschitz functions $\phi\colon \R^N \to \R^M$  that are not one-to-one on $\sE$. 
We can argue as in the proof of Theorem~\ref{th:gentransformation} and obtain that $\ryv=\phi(\rxv)$ is $m$-rectifiable and that the $m$-dimensional Hausdorff density is
\be \notag 
\theta_{\ryv}^m(\yv)=\sum_{\xv\in\phi^{-1}(\{\yv\})}\frac{\theta_{\rxv}^m(\xv)}{\Jm^{\sE}_{\phi}(\xv)}
\ee
$\Hm{m}|_{\phi(\sE)}$-almost everywhere.
We then obtain for  the $m$-dimensional entropy 
\ba
\eh{m}(\ryv)
& =
- \int_{\phi(\sE)}\Bigg(\sum_{\xv\in  \phi^{-1}(\{\yv\})}\frac{\theta_{\rxv}^m(\xv)}{\Jm^{\sE}_{\phi}(\xv)}\Bigg) \notag \\
& \rule{18mm}{0mm} \times\log \Bigg(\sum_{\xv\in  \phi^{-1}(\{\yv\})}\frac{\theta_{\rxv}^m(\xv)}{\Jm^{\sE}_{\phi}(\xv)}\Bigg) \, \mathrm{d}\Hm{m}(\yv) \notag \\ 
& \stackrel{\hidewidth (a) \hidewidth}= - \int_{\sE}\theta_{\rxv}^m(\xv) \notag \\
& \rule{13mm}{0mm} \times \log \Bigg(\sum_{\xv'\in  \phi^{-1}(\{\phi(\xv)\})}\frac{\theta_{\rxv}^m(\xv')}{\Jm^{\sE}_{\phi}(\xv')}\Bigg)  \, \mathrm{d}\Hm{m}(\xv) \notag%\label{eq:mostgentrafo}
\ea
where $(a)$ holds because of the generalized area formula
 \cite[Th.~2.91]{Ambrosio2000Functions}.
%Rev1.main: add information on finite-to-one mappings to use with Wishart matrices
In most cases, this cannot be easily expressed in terms of a differential entropy due to the sum in the logarithm.
However, in the special case of a Jacobian determinant $\Jm^{\sE}_{\phi}$ and a Hausdorff density  $\theta_{\rxv}^m$ that are symmetric in the sense that $\theta_{\rxv}^m(\xv')$ and $\Jm^{\sE}_{\phi}(\xv')$ are constant on $\phi^{-1}(\{\phi(\xv)\})$ for all $\xv\in \sE$, the summation reduces to a multiplication by the cardinality of $\phi^{-1}(\{\phi(\xv)\})$.
%An example of this setup will be given in Section~\ref{sec:entropysingularwishart} below.}
\end{remark}

\subsection{Relation to Entropy and Differential Entropy}\label{sec:relationtoentropy}
In the special cases $m=0$ and $m=M$, our entropy definition reduces to classical entropy \eqref{eq:discentropy} and differential entropy \eqref{eq:diffentropy}, respectively.
\begin{theorem}\label{th:specialcaseentropy}
Let $\rxv$ be a random variable on $\R^M$.
If $\rxv$ is a $0$-rectifiable (i.e., discrete) random variable, then the $0$-dimensional entropy of $\rxv$ coincides with the classical entropy, i.e., $\eh{0}(\rxv)=H(\rxv)$. 
If $\rxv$ is an $M$-rectifiable (i.e., continuous) random variable,  then the $M$-dimensional entropy of $\rxv$ coincides with the differential entropy, i.e., $\eh{M}(\rxv)= h(\rxv)$.
\end{theorem}
\begin{IEEEproof}
Let $\rxv$ be a $0$-rectifiable random variable.
By Theorem~\ref{th:specialcaserect}, $\rxv$ is a discrete random variable  with possible realizations $\xv_i$, $i\in \sI$,  
 the $0$-dimensional Hausdorff density $\theta_{\rxv}^0$ is the probability mass function of $\rxv$, and a support is given by $\sE=\{\xv_i:i\in \sI\}$. 
Thus, \eqref{eq:entwithsupport2} yields
\ba 
\eh{0}(\rxv) 
& = -\int_{\sE} \theta_{\rxv}^0(\xv) \log \theta_{\rxv}^0(\xv) \, \mathrm{d}\Hm{0}(\xv)\notag \\
& \stackrel{(a)}=-\sum_{i\in \sI}\Pr\{\rxv=\xv_i\}\log \Pr\{\rxv=\xv_i\} \notag \\
& \stackrel{\hidewidth\eqref{eq:discentropy} \hidewidth}=H(\rxv) \notag 
\ea
where $(a)$ holds because $\Hm{0}$ is the counting measure.

Let $\rxv$ be an $M$-rectifiable random variable.
By Theorem~\ref{th:specialcaserect}, $\rxv$ is a  continuous random variable   and the $M$-dimensional Hausdorff density $\theta_{\rxv}^M$ is equal to the probability density function $f_{\rxv}$. Thus, \eqref{eq:defmdimentropy} yields
\be  \notag 
\eh{M}(\rxv)
= - \E_{\rxv}\big[\log  \theta_{\rxv}^M(\rxv) \big]
= -\E_{\rxv}[\log f_{\rxv}(\rxv)]
\stackrel{\eqref{eq:diffentropy}}= h(\rxv)\,. %\IEEEQEDhereeqn
%\vspace{-5mm}
\ee
%The theorem  follows from Theorem~\ref{th:specialcaserect} and Definition~\ref{def:mdimentropy}.
\end{IEEEproof}

To get an idea of the $m$-dimensional entropy of random variables in between the discrete and continuous cases, we can  use Theorem~\ref{th:exrectmeasure} to construct $m$-rectifiable random variables.
More specifically, we consider a continuous random variable $\rxv$ on $\R^m$ and a one-to-one Lipschitz mapping $\phi\colon \R^m\to \R^M$ ($M\geq m$) whose  generalized Jacobian determinant satisfies
%\footnote{We use here and in what follows the term ``Jacobian'' or $\Jm_{\phi}$ to refer to the approximate $m$-dimensional Jacobian of a Lipschitz mapping $\phi$.} 
$\Jm_{\phi}>0$ $\Leb^m$-almost everywhere. 
Intuitively, we should see a connection between the differential entropy of $\rxv$ and the $m$-dimensional entropy of $\ryv\triangleq\phi(\rxv)$.
By Theorem~\ref{th:exrectmeasure}, the random variable $\ryv$ is $m$-rectifiable and, because $\phi$ is one-to-one, we can indeed calculate the $m$-dimensional entropy.
\begin{corollary}\label{th:diffentoto}
Let $\rxv$ be a continuous random variable on $\R^m$ with finite differential entropy $h(\rxv)$ and probability density function $f_{\rxv}$. 
Furthermore, let $\phi\colon \R^m\to \R^M$ ($M\geq m$) be a one-to-one Lipschitz mapping such that  $\Jm_{\phi}> 0$  $\Leb^m$-almost everywhere and $\E_{\rxv}[\log \Jm_{\phi}(\rxv)]$ exists and is finite. 
Then the $m$-dimensional Hausdorff density of the $m$-rectifiable random variable $\ryv\triangleq\phi(\rxv)$ is
\be \notag % \label{eq:transformdensac}
\theta_{\ryv}^m(\yv)=\frac{f_{\rxv}(\phi^{-1}(\yv))}{\Jm_{\phi}(\phi^{-1}(\yv))}
\ee
$\Hm{m}|_{\phi(\R^m)}$-almost everywhere,
and the $m$-dimensional entropy of $\ryv$ is
\be \notag 
\eh{m}(\ryv)= h(\rxv)+ \E_{\rxv}[\log \Jm_{\phi}(\rxv)]\,.
\ee
For the special case of the embedding $\phi\colon \R^m\to \R^M$, $\phi(x_1, \dots, x_m)=(x_1\, \cdots \, x_m \, \, 0\, \cdots \, 0)^{\trans}$, this results in 
\be\label{eq:transembedding}
\eh{m}(\rx_1, \dots, \rx_m, 0, \dots, 0)= h(\rxv)\,.
\ee
\end{corollary}
 \begin{IEEEproof}
The first part is the special case $N=m$ and $\sE=\R^m$ of Theorem~\ref{th:gentransformation}.
The result~\eqref{eq:transembedding} then follows from the fact that, for the considered embedding, $\Jm_{\phi}(\xv)$ is identically $1$.
\end{IEEEproof}

%COUNTABLE NUMBER ONE-TO-ONE LIPSCHITZ MAPPINGS?

%\vspace{-4mm}

%
%\subsection{Example: Entropy of Distributions on the Unit Circle}
%\label{sec:entropyunitcircle}
%
%
\subsection{Example: Entropy of Distributions on the Unit Circle}
\label{sec:entropyunitcircle}
%% \begin{example}\label{ex:entunitcircle}
It is now easy to calculate the entropy of the $1$-rectifiable singular random variables on the unit circle 
previously considered in Section \ref{sec:unitcircle}. 
%\begin{theorem}\label{th:entropyunitcircle}
%% To this end, 
Let $\rz$ be a continuous random variable on $\R$ with probability density function
%% density 
$f_{\rz}$ supported  on $[0,2 \pi)$, i.e., $f_{\rz}(z)=0$ for $z\notin [0,2 \pi)$.
By Corollary~\ref{th:diffentoto}, the $1$-dimensional Hausdorff density of the random variable $\rxv= \phi(\rz)= (\cos \rz \; \sin \rz)^{\trans}$ is given by (recall that the Jacobian determinant is identically one)
\be\label{eq:hddunitcircle}
\theta_{\rxv}^1(\xv)=f_{\rz}(\phi^{-1}(\xv))
\ee
$\Hm{1}|_{\sS_1}$-almost everywhere,
and the entropy of $\rxv$ is given by
\be\label{eq:finalentunitcircle}
\eh{1}(\rxv)= h(\rz)\,.
\ee
Of course, this result for $\eh{1}(\rxv)$ may have been conjectured by heuristic reasoning. Next, we consider a case where heuristic reasoning does not help.  
%% \end{example}
%\end{theorem}
%
%\begin{IEEEproof}
%Because 
%\end{IEEEproof}
%
%\vspace{-4mm}
%
%%Rev1.main:Example Wishart Matrices
%
%%% \newpage %%%%%%%%
%%
%
\subsection{Example: Entropy of Positive Semidefinite Rank-One Random Matrices}
\label{sec:entropysingularwishart}
As a more challenging example, we calculate the entropy of a specific type of $m$-rectifiable singular random variables, namely, the positive semidefinite rank-one random matrices
previously considered in Section \ref{sec:wishart}. 
\begin{theorem}\label{th:entropywishart}
Let $\rzv$ be a continuous random variable on $\R^m$ with probability density function
%% density 
$f_{\rzv}$, and let $\bar{\rzv}$ denote the random variable with probability density function
 $f_{\bar{\rzv}}(\zv)=(f_{\rzv}(\zv) + f_{\rzv}(-\zv))/2$.
 %%  for $\zv\in \R^{m}$.
Then the $m$-dimensional entropy of the random matrix $\rXm =\rzv\rzv^{\trans}$ is given by
\be\label{eq:finalentwishart}
\eh{m}(\rXm)= h(\bar{\rzv})+ \frac{m-1}{2} \log 2 + \frac{m}{2}\, \E_{\rzv}[\log \norm{\rzv}^2]\,.
\ee
\end{theorem}
\begin{IEEEproof}
We first calculate the Jacobian determinant of the mapping $\phi\colon \zv\mapsto \zv\zv^{\trans}$, which is given by
$\Jm_{\phi}(\zv) = \sqrt{\det(\Dm_{\phi}^{\trans}(\zv)\ist\Dm_{\phi}(\zv))}$. By \eqref{eq:diffwishart} and  some simple algebraic manipulations, one obtains
$\Jm_{\phi}(\zv)  = \sqrt{\det(2\norm{\zv}^2 \Iv_m+ 2 \zv\zv^{\trans})}$, and further
\ba
\Jm_{\phi}(\zv) & = 
\sqrt{2^m\norm{\zv}^{2m} \det\rmv\bigg(\Iv_m+ \frac{1}{\norm{\zv}^2}\zv\zv^{\trans}\bigg)} \notag \\
& \stackrel{(a)}= \sqrt{2^m\norm{\zv}^{2m} \bigg(1+ \frac{\zv^{\trans}\zv}{\norm{\zv}^2}\bigg)} \notag \\
& = \sqrt{2^{m+1}\norm{\zv}^{2m}} \notag \\
& = 2^{\frac{m+1}{2}}\norm{\zv}^{m} \label{eq:jacobianwishart}
\ea
where $(a)$ holds due to \cite[Example~1.3.24]{hojo13}.
Because the mapping $\phi \colon\zv\mapsto \zv\zv^{\trans}$ is not one-to-one, we cannot directly use Corollary~\ref{th:diffentoto}. 
However,  along the lines of Remark~\ref{rem:gentrafo}, we obtain
\ba
& \eh{m}(\rXm)  \notag \\
& \rule{5mm}{0mm} = - \int_{\R^m}f_{\rzv}(\zv) \log \rmv\Bigg(\sum_{\zv'\in  \phi^{-1}(\{\phi(\zv)\})}\frac{f_{\rzv}(\zv')}{\Jm_{\phi}(\zv')}\Bigg) \, \mathrm{d}\Leb^m(\zv)\,. \notag \\[-3mm]
\label{eq:firstentwishart}
\ea
Because the 
%% elements 
$\zv'\in \phi^{-1}(\{\phi(\zv)\})$ are given by $\pm \zv$, and because 
$f_{\rzv}(\zv)+f_{\rzv}(-\zv)=2 f_{\bar{\rzv}}(\zv)$ and 
$\Jm_{\phi}(\zv)= \Jm_{\phi}(-\zv)$ (see~\eqref{eq:jacobianwishart}), 
eq.\ \eqref{eq:firstentwishart} implies 
\ba
&  \eh{m}(\rXm) \notag \\
& \rule{3mm}{0mm}= - \int_{\R^m}f_{\rzv}(\zv)  \log \Bigg(2\frac{f_{\bar{\rzv}}(\zv)}{\Jm_{\phi}(\zv)}\Bigg)  \, \mathrm{d}\Leb^m(\zv) \notag \\
& \rule{3mm}{0mm}= - \int_{\R^m}f_{\rzv}(\zv)  \big( \log 2 + \log f_{\bar{\rzv}}(\zv)   - \log \Jm_{\phi}(\zv) \big)  \, \mathrm{d}\Leb^m(\zv) \notag \\
& \rule{3mm}{0mm}= -  \log 2 - \int_{\R^m}f_{\rzv}(\zv) \log f_{\bar{\rzv}}(\zv)  \, \mathrm{d}\Leb^m(\zv)   + \E_{\rzv}[\log \Jm_{\phi}(\rzv)]\notag \\
& \rule{3mm}{0mm}\stackrel{(a)}= -  \log 2 - \frac{1}{2}\int_{\R^m}f_{\rzv}(\zv) \log f_{\bar{\rzv}}(\zv)  \, \mathrm{d}\Leb^m(\zv) \notag \\
& \rule{12mm}{0mm} - \frac{1}{2}\int_{\R^m}f_{\rzv}(-\zv) \log f_{\bar{\rzv}}(\zv)  \, \mathrm{d}\Leb^m(\zv)  + \E_{\rzv}[\log \Jm_{\phi}(\rzv)]
\notag \\
& \rule{3mm}{0mm}= -  \log 2 - \int_{\R^m}f_{\bar{\rzv}}(\zv) \log f_{\bar{\rzv}}(\zv)  \, \mathrm{d}\Leb^m(\zv)  + \E_{\rzv}[\log \Jm_{\phi}(\rzv)]
\notag \\
& \rule{3mm}{0mm}= -  \log 2 + h(\bar{\rzv}) + \E_{\rzv}[\log \Jm_{\phi}(\rzv)]
\label{eq:secondentwishart}
\ea
where $(a)$ holds because $f_{\bar{\rzv}}(-\zv)=f_{\bar{\rzv}}(\zv)$.
Inserting \eqref{eq:jacobianwishart} into \eqref{eq:secondentwishart} gives \eqref{eq:finalentwishart}.
\end{IEEEproof}
%
%% ?Wishart matrices $\rWm_{\rzv,n}?\rWm_{n, \Sigma}$ for $n\in \{2, \dots, m-1\}$ is considerably more involved because the mapping $(\zv_1, \dots, \zv_n)\mapsto  \sum_{i=1}^n\zv_i\zv_i^{\trans}$ has a vanishing Jacobian determinant almost everywhere.
%
A practically interesting special case of symmetric random matrices is constituted by the class of Wishart matrices \cite{uh94}.
A rank-$n$ Wishart matrix is given by $\rWm_{n, \Sigm}\triangleq \sum_{i=1}^n\rzv_i\rzv_i^{\trans} \in \R^{m\times m}$, %with ?$n \ge 1$,
where the $\rzv_i$, $i\in \{1, \dots, n\}$ are independent and identically distributed  (i.i.d.)
Gaussian random variables on $\R^m$ with mean $\0v$ and some nonsingular covariance matrix $\Sigm$.
The differential entropy of a \emph{full-rank} Wishart matrix (i.e., $n\geq m$), considered as 
a random variable in the $m(m+1)/2$-dimensional space of symmetric matrices, is given by \cite[eq.~(B.82)]{bi06}
\ba 
h(\rWm_{n, \Sigm}) & = \log \rmv\bigg(2^{mn/2} \, \Gamma_m\bigg(\frac{n}{2}\bigg)( \det \Sigm)^{n/2} \bigg)\notag \\
& \rule{5mm}{0mm}+ \frac{mn}{2} + \frac{m-n+1}{2}\, \E_{\rzv}[\log \det(\rWm_{n, \Sigm})]\label{eq:entwishartfull}
\ea
where $\Gamma_m(\cdot)$ denotes the multivariate gamma function.
In our setting, full-rank Wishart matrices can be interpreted as $m(m+1)/2$-rectifiable random variables in the $m^2$-dimensional  
space of all $m\times m$ matrices by considering the  embedding of symmetric matrices into the space of all matrices and using Theorem~\ref{th:exrectmeasure}.
Using this interpretation, we can use Corollary~\ref{th:diffentoto} and obtain $h(\rWm_{n, \Sigm})=\eh{m(m+1)/2}(\rWm_{n, \Sigm})$.

The case of rank-deficient Wishart matrices, i.e., $n\in \{1, \dots, m-1\}$, has not been analyzed information-theoret\-ically so far.
For simplicity, we will consider the case of rank-one Wishart matrices, i.e., $\rWm_{1, \Sigm}= \rzv\rzv^{\trans} \in \R^{m\times m}$.
The $m$-dimensional entropy of $\rWm_{1, \Sigm}$ is given by \eqref{eq:finalentwishart} in
Theorem~\ref{th:entropywishart}. Because $\rzv$ is Gaussian with mean $\0v$, we have $\bar{\rzv} = \rzv$ in Theorem~\ref{th:entropywishart}, so that \eqref{eq:finalentwishart}
simplifies to
\be \notag %\label{eq:finalentwishart_wish}
\eh{m}(\rWm_{1, \Sigm})= h(\rzv)+ \frac{m-1}{2} \log 2 + \frac{m}{2}\, \E_{\rzv}[\log \norm{\rzv}^2]\,.
\ee
Again using the Gaussianity of $\rzv$, 
%% Inserting $h(\rzv) = \log \big((2\pi e)^{m/2}( \det \bm{\Sigma})^{1/2}\big)$, 
we obtain further
\ba 
\eh{m}(\rWm_{1, \Sigm})& = \log\rmv \big((2\pi e)^{m/2}( \det \Sigm)^{1/2}\big) \notag \\
& \rule{15mm}{0mm} + \frac{m-1}{2} \log 2 + \frac{m}{2}\, \E_{\rzv}[\log \norm{\rzv}^2]\notag \\[1mm]
& = \log\rmv \big(2^{m-1/2}\pi^{m/2}( \det \Sigm)^{1/2}\big) 
\notag \\ & \rule{25mm}{0mm}
+ \frac{m}{2} + \frac{m}{2}\, \E_{\rzv}[\log \norm{\rzv}^2] \,.
\label{eq:entwishartone}
\ea
If $\rzv$ contains independent standard normal entries, then $\norm{\rzv}^2$ is $\chi^2_m$ distributed and $\E_{\rzv}[\log \norm{\rzv}^2]=\psi(m/2)+\log 2$, where $\psi(\cdot)$ denotes the digamma function \cite[eq.~(B.81)]{bi06}.
It is interesting to compare \eqref{eq:entwishartone} with the differential entropy of the full-rank Wishart matrix as given by \eqref{eq:entwishartfull}.
Although there is a formal similarity,
%% striking similarity between \eqref{eq:entwishartone} and \eqref{eq:entwishartfull}, 
we emphasize that the differential entropy in \eqref{eq:entwishartfull} cannot be
%% have been
trivially extended to the setting $n<m$ because neither $\Gamma_m (\frac{n}{2})$ nor $\log \det(\rWm_{n, \Sigm})$ is defined in this case.
%% these cases. 
We conjecture
%% expect 
that an expression similar to \eqref{eq:entwishartone} can be derived for other rank-deficient Wishart matrices. 
However, as 
%% already 
mentioned in Section \ref{sec:wishart},
%% Remark~\ref{rem:lowrankwishart} 
the analysis of these matrices is significantly more involved and, thus, beyond the scope of this
%% the present 
paper.
\begin{remark}
A different approach to defining an entropy for rank-deficient Wishart matrices would be to use a coordinate system 
on the manifold of all positive semidefinite matrices of rank $n$ and calculate a probability density function with respect to volume elements of this manifold. 
Such a density was calculated for Wishart matrices in \cite{uh94}, and could be used for an alternative entropy definition. 
%We conjecture that, up to an additive constant, this would result in the same entropy as the more general measure-theoretic approach adopted here.
%% in the present paper.
\end{remark}

%%%%%%%%%%%%%%%%%%%%%%%%%%%%%%%%%%%%%%%%%%%%%%

\section{Joint Entropy}\label{sec:jointentropy}
%%%%%%%%%%%%%%%%%%%%%%%%%%%%%%%%%%%%%%%%%%%%%%

Joint entropy is a widely used concept although it can  be covered by the general concept of higher-dimensional entropy, because a pair of random variables $(\rxv, \ryv)$ with $\rxv\in \R^{M_1}$ and $\ryv\in \R^{M_2}$ can also be interpreted as a single random variable on $\R^{M_1+M_2}$.
Thus, our concept of entropy automatically generalizes to more than one random variable.
Using this interpretation, we obtain from~\eqref{eq:defmdimentropy} and \eqref{eq:entwithsupport1} for an $m$-rectifiable pair of random variables $(\rxv, \ryv)$ (i.e., $\mu(\rxv,\ryv)^{-1}\ll\Hm{m}|_{\sE}$ for an $m$-rectifiable set $\sE$)
\ba
\eh{m}(\rxv, \ryv) 
& \triangleq - \E_{(\rxv, \ryv)}\big[\log \theta_{(\rxv, \ryv)}^m(\rxv, \ryv)\big]  \label{eq:defjointentropyexp} \\[1mm]
&   =  - \int_{\R^M} \log \theta_{(\rxv, \ryv)}^{m}(\xv, \yv) \, \mathrm{d}\mu(\rxv, \ryv)^{-1}(\xv, \yv) \notag \\[1mm] %\label{eq:defjointentropy}  \\
&   =  - \int_{\R^M} \theta_{(\rxv, \ryv)}^{m}(\xv, \yv) \log \theta_{(\rxv, \ryv)}^{m}(\xv, \yv) \, \mathrm{d}\Hm{m}|_{\sE}(\xv, \yv) \label{eq:defjointentropy}
\ea
with $M=M_1+M_2$.
However, there are still some questions to answer:
%Rev1.7: reformulated questions to match the answer
\begin{itemize}
\item Assuming that $\rxv$, $\ryv$, and $(\rxv, \ryv)$ are $m_1$-, $m_2$-, and $m$-rectifiable, respectively, is there a relationship between the quantities $\eh{m_1}(\rxv)$, $\eh{m_2}(\ryv)$, and $\eh{m}(\rxv, \ryv)$ provided they exist?
\item Suppose we have an $m_1$-rectifiable random variable $\rxv$ and an $m_2$-rectifiable random variable $\ryv$ on the same probability space. 
Which additional assumptions ensure that $(\rxv, \ryv)$ is $(m_1+m_2)$-rectifiable?
\item Conversely, suppose we have an $m$-rectifiable random variable $(\rxv, \ryv)$. Which additional assumptions ensure that $\rxv$ and $\ryv$ are rectifiable?
\end{itemize}
In what follows, we will provide answers to these questions under appropriate  conditions on the involved random variables.

One important shortcoming of Hausdorff measures (in contrast to, e.g., the Lebesgue measure) is that the product of two Hausdorff measures is in general not again a Hausdorff measure. 
However, 
%% restricting the Hausdorff measures to specific rectifiable sets will guarantee that the product is again a Hausdorff measure.
our definition of the support of a rectifiable measure in Definition~\ref{def:support}  guarantees 
that the product of two Hausdorff measures restricted to the respective supports is again a Hausdorff measure.
%%  accordingly and, thus, obtain the following result.
%% \cha{In particular, we defined a support set of a rectifiable measure accordingly and, thus, obtain the following result.
\begin{lemma}\label{lem:prodcomp}
Let $\rxv$ be $m_1$-rectifiable with support $\sE_1$, and let $\ryv$ be $m_2$-rectifiable with support $\sE_2$.
Then $\sE_1\times \sE_2$ is $(m_1+m_2)$-rectifiable  and 
%\vspace{-1mm}
\be\label{eq:defprodcomp}
\Hm{m_1+m_2}|_{\sE_1\times \sE_2} = \Hm{m_1}|_{\sE_1}\times \Hm{m_2}|_{\sE_2}\,.
\ee
\end{lemma}
\begin{IEEEproof}
According to Definition~\ref{def:rectrandomvar}, we have $\sE_1=\bigcup_{k\in \N}f_k(\sA_k)$ and $\sE_2=\bigcup_{k\in \N}g_k(\sB_k)$ where, 
for $k\in \N$, $\sA_k$ and $\sB_k$ are bounded Borel sets and $f_k$ and $g_k$ are Lipschitz functions that are one-to-one on $\sA_k$ and $\sB_k$, respectively.
By \cite[Th.~15.1]{ke95}, the sets $f_k(\sA_k)$ and $g_k(\sB_k)$ are also Borel sets and, thus, \cite[Th.~3.2.23]{fed69} implies $\Hm{m_1+m_2}|_{f_k(\sA_k)\times g_k(\sB_k)} = \Hm{m_1}|_{f_k(\sA_k)}\times \Hm{m_2}|_{g_k(\sB_k)}$.
The result \eqref{eq:defprodcomp} then follows by
%% from 
the $\sigma$-additivity of Hausdorff measures.
\end{IEEEproof}

\subsection{Joint Entropy for Independent Random Variables}

We start our investigation of joint entropy with independent random variables.
%Rev on product-compatible supports.
In this case, it turns out that the $m$-dimensional entropy is additive.
\begin{theorem}\label{th:prodindeprect}
Let $\rxv\colon \Omega \to \R^{M_1}$ and $\ryv\colon \Omega \to \R^{M_2}$ be independent random variables on a probability space $(\Omega, \mathfrak{S}, \mu)$.
Furthermore, let $\rxv$ be $m_1$-rectifiable with support $\sE_1$ and let $\ryv$ be $m_2$-rectifiable  with support $\sE_2$.
%Rev , where $\sE_1$ and $\sE_2$ are product-compatible.
Then the following properties hold:
\begin{enumerate}
\item The random variable $(\rxv, \ryv)\colon \Omega \to \R^{M_1+M_2}$ is $(m_1+m_2)$-rectifiable. \label{en:indepprodrect}
\item The $(m_1+m_2)$-dimensional Hausdorff density of $(\rxv, \ryv)$ satisfies
\be\label{eq:prodhddensity}
\theta_{(\rxv, \ryv)}^{m_1+m_2}(\xv, \yv)=\theta_{\rxv}^{m_1}(\xv)\,\theta_{\ryv}^{m_2}(\yv)
\ee
$\Hm{m_1+m_2}$-almost everywhere.\label{en:indepprodfact} 
\item The set  $\sE_1\times \sE_2$ is $(m_1+m_2)$-rectifiable and satisfies $\mu(\rxv, \ryv)^{-1}\ll \Hm{m_1+m_2}|_{\sE_1\times \sE_2}$. \label{en:indepprodsupport}
\item If $\eh{m_1}(\rxv)$ and $\eh{m_2}(\ryv)$ are finite, then the $(m_1+m_2)$-dimensional entropy of the random variable $(\rxv,\ryv)$ is given by \label{en:indepentsum}
\be \notag %\label{eq:jointentindep}
\eh{m_1+m_2}(\rxv, \ryv)=\eh{m_1}(\rxv)+\eh{m_2}(\ryv)\,.
\ee
\end{enumerate}
\end{theorem}
\begin{IEEEproof}
See Appendix~\ref{app:proofprodindeprect}.
\end{IEEEproof}

A corollary of Theorem~\ref{th:prodindeprect} is a result for finite sequences of independent random variables.
Such sequences will be important for our discussion of typical sets in Section~\ref{sec:aep}.
\begin{corollary}\label{cor:independentidentseq}
Let $\rxv_{1:n}\triangleq (\rxv_1, \dots, \rxv_n)$ be a finite sequence of independent random variables, where $\rxv_i \in \R^{M_i}$, $i\in \{1, \dots, n\}$  is $m_i$-rectifiable  with support $\sE_i$ and $m_i$-dimensional Hausdorff density $\theta_{\rxv_i}^{m_i}$.
%Rev1
%Furthermore, let $\sE_1\times \sE_2\times \dots \times \sE_{i-1}$ and ${\sE_{i}}$ be product-compatible for all $i\in \{2, \dots, n\}$.
Then $\rxv_{1:n}$  is an $m$-rectifiable random variable on $\R^{M}$, where $m=\sum_{i=1}^{n} m_i$ and $M=\sum_{i=1}^{n} M_i$,
 and the set $\sE\triangleq \sE_1\times \dots \times \sE_{n}$ is $m$-rectifiable and satisfies $\mu(\rxv_{1:n})^{-1}\ll \Hm{m}|_{\sE}$.
Moreover, the $m$-dimensional Hausdorff density of $\rxv_{1:n}$ is given by
\be \notag % \label{eq:indepproddens}
\theta_{\rxv_{1:n}}^{m}(\xv_{1:n})=\prod_{i=1}^{n}\theta_{\rxv_i}^{m_i}(\xv_i)\,.
\ee
Finally, if $\eh{m_i}(\rxv_i)$ is finite for $i\in \{1, \dots, n\}$, then 
\be\label{eq:indepprodentropy}
\eh{m}(\rxv_{1:n})=\sum_{i=1}^{n}\eh{m_i}(\rxv_i)\,.
\ee
\end{corollary}
\begin{IEEEproof}
The corollary follows by inductively applying Theorem~\ref{th:prodindeprect} to the two random variables $(\rxv_1, \dots,$  $\rxv_{i-1})$ and $\rxv_{i}$.
%We only have to show that $\sE^{n-1}\times \sE$ is $mn$-rectifiable. Then Theorem~\ref{th:prodindeprect} gives the desired result.
\end{IEEEproof}

%\begin{figure*}[!t]
%% ensure that we have normalsize text
%\normalsize
%% Store the current equation number.
%\newcounter{MYtempeqncnt}
%\setcounter{MYtempeqncnt}{\value{equation}}
%% Set the equation number to one less than the one
%% desired for the first equation here.
%% The value here will have to changed if equations
%% are added or removed prior to the place these
%% equations are referenced in the main text.
%\setcounter{equation}{40}
%\be\label{eq:marginalentropy}
%\eh{m_2}(\ryv)= -\int_{\sE} \theta_{(\rxv, \ryv)}^{m}(\xv, \yv)\log \bigg(\int_{\sE^{(\yv)}}\frac{\theta_{(\rxv, \ryv)}^{m}(\xvt, \yv)}{\Jm^{\sE}_{\proj_{\ryv}}(\xvt, \yv)}\, \mathrm{d}\Hm{m-m_2}(\xvt)\bigg)  \mathrm{d}\Hm{m}(\xv, \yv)
%\ee
%% Restore the current equation number.
%\setcounter{equation}{\value{MYtempeqncnt}}
%% IEEE uses as a separator
%\hrulefill
%% The spacer can be tweaked to stop underfull vboxes.
%\vspace*{4pt}
%\end{figure*}

\subsection{Dependent Random Variables}\label{sec:deprandvar}

The case of dependent random variables is more involved. 
The rectifiability of $\rxv$ and $\ryv$ does not necessarily imply the rectifiability of $(\rxv, \ryv)$ (which is  expected, since the marginal distributions carry only a small part of the information carried by the joint distribution).
In general, even for continuous random variables $\rxv$ and $\ryv$, we cannot calculate the joint differential entropy $h(\rxv, \ryv)$ from the mere knowledge of the differential entropies $h(\rxv)$ and $h(\ryv)$. 
However, it is always possible to bound the differential entropy according to \cite[eq.~(8.63)]{Cover91}
\be\label{eq:entunionbound}
h(\rxv, \ryv)\leq h(\rxv)+h(\ryv)\,.
\ee
In general, no bound resembling~\eqref{eq:entunionbound} holds for our entropy definition. 
The following simple setting provides a counterexample.
\begin{example}\label{ex:entbounduniform} 
We continue 
%% Example~\ref{ex:entunitcircle}  
our example of a random variable on the unit circle (see Section \ref{sec:entropyunitcircle}) for the special case of a uniform distribution of $\rz$ on $[0,2 \pi)$.
From \eqref{eq:finalentunitcircle}, we obtain
%% It is straightforward to see that 
\be\label{eq:calchxy}
\eh{1}(\rxv)  = h(\rz)=\log(2\pi)\,.
\ee
We can now analyze the components%
\footnote{To conform with the notation $(\rxv, \ryv)$ used  in our treatment of joint entropy, we change the component notation from $(\rx_1 \; \rx_2)^{\trans}$ to $(\rx \;\, \ry)^{\trans}$.}
 $\rx$ and $\ry$ of the random variable $\rxv = (\rx \;\ist \ry)^{\trans} = (\cos \rz \; \sin \rz)^{\trans}$.
One can easily see that $\rx$ is a continuous random variable and its probability density function is given by  $f_{\rx}(x)=1/(\pi\sqrt{1-x^2})$. 
By symmetry, the same holds for $\ry$, i.e., $f_{\ry}(y)=1/(\pi\sqrt{1-y^2})$.
Basic calculus then yields for the differential entropy of $\rx$ and $\ry$
\be\label{eq:entexampy}
h(\rx)= h(\ry) = \log\bigg(\frac{\pi}{2}\bigg)\,.
\ee
Since $\rx$ and $\ry$ are continuous random variables, it follows from  Theorem~\ref{th:specialcaseentropy} that $\eh{1}(\rx)=h(\rx)$ and  $\eh{1}(\ry)=h(\ry)$.
Thus,
\be \notag 
\eh{1}(\rx) + \eh{1}(\ry) = 2\log\bigg(\frac{\pi}{2}\bigg)
 < \log (2\pi)\,.
\ee
Comparing with \eqref{eq:calchxy}, we see that
$\eh{1}(\rx, \ry)> \eh{1}(\rx) + \eh{1}(\ry)$. 
\hfill \IEEEQED
\end{example}

The reason for this seemingly unintuitive behavior of our entropy are the geometric properties of the projection
%Similar to the problem of products of Hausdorff measures, the projection of a rectifiable set is not as straightforward as it might seem. 
%In the following, we will consider
$\proj_{\ryv}\colon \R^{M_1+M_2}\to \R^{M_2}$, 
$\proj_{\ryv}(\xv,\yv)=\yv$,
i.e., the projection of $\R^{M_1+M_2}$ to the last $M_2$ components.
Although $\proj_{\ryv}$ is linear and has a Jacobian determinant $\Jm_{\proj_{\ryv}}$ of $1$ everywhere on $\R^{M_1+M_2}$, things get more involved once we consider $\proj_{\ryv}$ as a mapping between rectifiable sets and want to calculate the Jacobian determinant $\Jm^{\sE}_{\proj_{\ryv}}$ of the tangential differential of $\proj_{\ryv}$ which maps an $m$-rectifiable set $\sE\subseteq \R^{M_1+M_2}$ to an $m_2$-rectifiable set $\sE_2\subseteq \R^{M_2}$ \cite[Sec.~3.2.16]{fed69}.
In this setting, $\Jm^{\sE}_{\proj_{\ryv}}$ is not necessarily constant and may also become zero. 
%Thus, we will often assume that $\Jm^{\sE}_{\proj_{\ryv}}$ is nonzero.
%This issue and the fact that~\eqref{eq:entunionbound} does not hold for our entropy definition is illustrated by the following simple example.
Thus, the marginalization of an $m$-dimensional Hausdorff density is not as easy as the marginalization of a probability density function.
The following theorem shows how to marginalize Hausdorff densities and describes the implications for $m$-dimensional entropy.

%On the other hand, we can start with a joint distribution of $(\rxv, \ryv)$ that is $m$-rectifiable with support $\sE\subseteq \R^{M_1+M_2}$.
%As we will see in Theorem~\ref{th:marginals} below, mild assumptions guarantee  that $\rxv$ and $\ryv$ are again rectifiable and their dimensions equal the dimensions of the projections of $\sE$ onto $\R^{M_1}$ and $\R^{M_2}$, respectively.

\begin{theorem}\label{th:marginals}
Let $(\rxv, \ryv)\in \R^{M_1+M_2}$ be an $m$-rectifiable random variable ($m\leq M_1+M_2$) with $m$-dimensional Hausdorff density $\theta_{(\rxv, \ryv)}^{m}$ and support $\sE$. 
%Denote by $\proj_{\ryv}\colon \R^{M_1+M_2}\to \R^{M_2}$ the projection of $\R^{M_1+M_2}$ to the last $M_2$ components.
Furthermore, let $\sEt_2\triangleq\proj_{\ryv}(\sE)\subseteq \R^{M_2}$ be $m_2$-rectifiable ($m_2\leq m$, $m_2\leq M_2$),  $\Hm{m_2}(\sEt_2)<\infty$, 
and $\Jm^{\sE}_{\proj_{\ryv}}> 0$ $\Hm{m}|_{\sE}$-almost everywhere.
%and $\sE=\bigcup_{i\in \sI}\sF_i$ with mutually disjoint sets $\sF_i$ satisfying $\Hm{m}(\sF_i)<\infty$, where $\sI$ is a finite or countably infinite index set.
Then the following properties hold:
\begin{enumerate}
\item \label{en:margisrect}The random variable $\ryv$ is $m_2$-rectifiable.
\item \label{en:margsupport}There exists a support $\sE_2\subseteq \sEt_2$ of $\ryv$.
\item \label{en:marghddens}The $m_2$-dimensional Hausdorff density of $\ryv$ is given by
\be\label{eq:hddensmarginal}
\theta_{\ryv}^{m_2}(\yv)=\int_{\sE^{(\yv)}}\frac{\theta_{(\rxv, \ryv)}^{m}(\xv, \yv)}{\Jm^{\sE}_{\proj_{\ryv}}(\xv, \yv)}\, \mathrm{d}\Hm{m-m_2}(\xv)
\ee
$\Hm{m_2}$-almost everywhere, where $\sE^{(\yv)}\triangleq \{\xv\in \R^{M_1}: (\xv, \yv)\in \sE\}$. 
\item \label{en:margentropy}An expression of the $m_2$-dimensional entropy of $\ryv$  is given by
\be
\eh{m_2}(\ryv)
 = -\int_{\sE} \theta_{(\rxv, \ryv)}^{m}(\xv, \yv)\log \theta_{\ryv}^{m_2}(\yv) \, \mathrm{d}\Hm{m}(\xv, \yv) \label{eq:marginalentropy}
\ee
%\ba
%\hspace{-8mm}\eh{m_2}(\ryv)
%& = -\int_{\sE} \theta_{(\rxv, \ryv)}^{m}(\xv, \yv)\log \bigg(\int_{\sE^{(\yv)}}\frac{\theta_{(\rxv, \ryv)}^{m}(\xvt, \yv)}{\Jm^{\sE}_{\proj_{\ryv}}(\xvt, \yv)} \notag \\
%& \rule{23mm}{0mm} \times  \mathrm{d}\Hm{m-m_2}(\xvt)\bigg)  \mathrm{d}\Hm{m}(\xv, \yv) \label{eq:marginalentropy}
%\ea
provided the integral on the right-hand side exists and is finite.
\end{enumerate}
Under the assumptions that $\sEt_1\triangleq \proj_{\rxv}(\sE)$ is $m_1$-rectifiable ($m_1\leq m$, $m_1\leq M_1$), $\Hm{m_1}(\sEt_1)<\infty$, 
and $\Jm^{\sE}_{\proj_{\rxv}}> 0$ $\Hm{m}|_{\sE}$-almost everywhere, analogous results hold for $\rxv$.
\end{theorem}
\begin{IEEEproof}
See Appendix~\ref{app:proofmarginals}.
\end{IEEEproof}

%\begin{remark}
%Some assumptions we made in Theorem~\ref{th:marginals} might not be necessary. In particular, we need the assumption $\Hm{m_2}(\sE_2)<\infty$
%only to be able to apply a modified version of the coarea formula (see Theorem~\ref{th:mycaf} in Appendix~\ref{app:proofmarginals}). 
%%\cite[Th.~3.2.22]{fed69}.
%%Unfortunately, no more general coarea formula (e.g., for sets of infinite Hausdorff measure) seems to be available for mappings between rectifiable sets.
%\end{remark}
%

%\begin{figure*}[!t]
%% ensure that we have normalsize text
%\normalsize
%% Store the current equation number.
%\newcounter{MYtempeqncnt}
%\setcounter{MYtempeqncnt}{\value{equation}}
%% Set the equation number to one less than the one
%% desired for the first equation here.
%% The value here will have to changed if equations
%% are added or removed prior to the place these
%% equations are referenced in the main text.
%\setcounter{equation}{43}
%\be\label{eq:marginalentropycompy}
 %\eh{m_2}(\ryv) = -\int_{\sE} \theta_{(\rxv, \ryv)}^{m_1+m_2}(\xv, \yv)\log \bigg(\int_{\sE_1} \theta_{(\rxv, \ryv)}^{m_1+m_2}(\xvt, \yv) \, \mathrm{d}\Hm{m_1}(\xvt)\bigg)  \mathrm{d}\Hm{m_1+m_2}(\xv, \yv) 
%\ee% Restore the current equation number.
%\setcounter{equation}{\value{MYtempeqncnt}}
%% IEEE uses as a separator
%\hrulefill
%% The spacer can be tweaked to stop underfull vboxes.
%\vspace*{4pt}
%\end{figure*}

We will illustrate the main findings of Theorem~\ref{th:marginals} in the setting of Example~\ref{ex:entbounduniform}.
\begin{example}\label{ex:projuniform} 
As in  Example~\ref{ex:entbounduniform}, we consider $(\rx, \ry)\in \R^2$ uniformly distributed on the unit circle $\sS_1$.
By~\eqref{eq:hddunitcircle}, $\theta^1_{(\rx, \ry)}(x,y)=1/(2\pi)$ $\Hm{1}$-almost everywhere on $\sS_1$.
In Example~\ref{ex:entbounduniform}, we already obtained $\eh{1}(\ry)= \log (\pi/2)$ (there, we used the fact that $\ry$ is a continuous random variable and that,  by Theorem~\ref{th:specialcaseentropy},  $\eh{1}(\ry)=h(\ry)$).
Let us now calculate $\eh{1}(\ry)$ using Theorem~\ref{th:marginals}.
Note first that $\proj_{\ry}(\sS_1)=[-1,1]$, which is $1$-rectifiable and satisfies $\Hm{1}([-1,1])=2<\infty$.
Next, we calculate the Jacobian determinant $\Jm^{\sS_1}_{\proj_{\ryv}}(x, y)$.
Consider an arbitrary point on the unit circle, which can always be expressed as $\big({\pm}\sqrt{1-y^2}, \pm y\big)$ with $y\in [0,1]$. 
At that point, the projection $\proj_{\ry}$ restricted to the tangent space of $\sS_1$ can be shown to amount to a multiplication by the factor $\sqrt{1-y^2}$. 
Thus, $\Jm^{\sS_1}_{\proj_{\ry}}\big({\pm}\sqrt{1-y^2}, \pm y\big)=\sqrt{1-y^2}$.
Hence, we obtain from \eqref{eq:marginalentropy} 
\ba
&\eh{1}(\ry) \notag \\
& \rule{3mm}{0mm}  = -\int_{\sS_1} \theta_{(\rx, \ry)}^{1}(x, y) \notag \\
& \rule{11mm}{0mm} \times \log \bigg(\int_{\sS_1^{(y)}}\frac{\theta_{(\rx, \ry)}^{1}(\tilde{x}, y)}{\Jm^{\sS_1}_{\proj_{\ry}}(\tilde{x}, y)}\, \mathrm{d}\Hm{1-1}(\tilde{x})\bigg) \, \mathrm{d}\Hm{1}(x, y) \notag \\
&  \rule{3mm}{0mm} = - \int_{\sS_1} \frac{1}{2\pi} \log \bigg(\int_{\sS_1^{(y)}}\frac{\frac{1}{2\pi}}{\sqrt{1-y^2}}\, \mathrm{d}\Hm{0}(\tilde{x})\bigg) \mathrm{d}\Hm{1}(x, y)  \notag \\
&  \rule{3mm}{0mm}\stackrel{\hidewidth (a)\hidewidth}= -  \frac{1}{2\pi}\int_{\sS_1} \log \bigg(\sum_{\tilde{x}\in\sS_1^{(y)}}\frac{\frac{1}{2\pi}}{\sqrt{1-y^2}}\bigg) \, \mathrm{d}\Hm{1}(x, y)  \notag \\
&  \rule{3mm}{0mm}\stackrel{\hidewidth (b)\hidewidth}= - \frac{1}{2\pi}\int_{\sS_1}  \log \bigg(2\frac{\frac{1}{2\pi}}{\sqrt{1-y^2}}\bigg) \, \mathrm{d}\Hm{1}(x, y)  \notag \\[1mm]
%&  \rule{3mm}{0mm}= - \frac{1}{2\pi} \int_{\sS_1} \log \bigg(\frac{1}{\pi\sqrt{1-y^2}}\bigg) \, \mathrm{d}\Hm{1}(x, y)  \notag \\
&  \rule{3mm}{0mm}= - \frac{1}{2\pi} \int_0^{2\pi} \log \bigg(\frac{1}{\pi \abs{\cos(\phi)}}\bigg) \, \mathrm{d}\phi  \notag \\[1mm]
&  \rule{3mm}{0mm}= \log\bigg(\frac{\pi}{2}\bigg)  \label{eq:marginalexample}
\ea
where $(a)$ holds because $\Hm{0}$ is the counting measure and $(b)$ holds because $\sS_1^{(y)}=\{x\in \R: (x,y)\in \sS_1\}=\big\{\sqrt{1-y^2}, -\sqrt{1-y^2}\big\}$ contains two points for all $y\in (-1,1)$.
Note that our above result for $\eh{1}(\ry)$ coincides with the result previously obtained in Example~\ref{ex:entbounduniform}.
\hfill \IEEEQED
%Note that we already obtained the same value for $\eh{1}(\ry)$ in  where we used the fact that $\ry$ is a continuous random variable and that,  by Theorem~\ref{th:specialcaseentropy},  $\eh{1}(\ry)=h(\ry)$.
\end{example}
 
\subsection{Product-Compatible Random Variables}
There are special settings in which $m$-dimensional entropy more closely matches the behavior we know from (differential) entropy. 
In these cases, the three random variables $\rxv$, $\ryv$, and $(\rxv, \ryv)$ are rectifiable with ``matching'' dimensions, and we will see that an inequality similar to~\eqref{eq:entunionbound} holds.

\begin{definition}\label{def:productcomprvs}
Let $\rxv$ be an $m_1$-rectifiable random variable on $\R^{M_1}$ with support $\sE_1$, and  let $\ryv$ be an $m_2$-rectifiable random variable on $\R^{M_2}$ with support $\sE_2$.
The random variables $\rxv$ and $\ryv$ are called \textit{product-compatible} if 
%$\sE_1$ and $\sE_2$ are product-compatible and
 $(\rxv, \ryv)$ is an $(m_1+m_2)$-rectifiable random variable on $\R^{M_1+M_2}$. %with support $\sE\subseteq \sE_1\times \sE_2$. 
%and $(m_1+m_2)$-dimensional Hausdorff density $\theta_{(\rxv, \ryv)}^{m_1+m_2}$.
%and then .
\end{definition}
It is easy to see that for product-compatible random variables  $\rxv$ and $\ryv$, $\mu(\rxv, \ryv)^{-1}\ll \Hm{m_1+m_2}|_{\sE_1\times \sE_2}$.
Thus,  by Property~\ref{en:corexistsupport} in Corollary~\ref{cor:propsrectrandvar}, there exists a support $\sE\subseteq \sE_1\times \sE_2$.

The most important part of Definition~\ref{def:productcomprvs} is that the dimensions of $\rxv$ and $\ryv$ add up to the joint dimension of $(\rxv, \ryv)$. 
Note that this was not the case in Example~\ref{ex:projuniform}, where $\rx$ and $\ry$ ``shared'' the dimension $m=1$ of $(\rx, \ry)$.
A simple example of product-compatible random variables is the case of an $m_1$-rectifiable random variable $\rxv$ and an independent $m_2$-rectifiable random variable $\ryv$.
% with product-compatible supports $\sE_1$ and $\sE_2$.
Indeed, by Theorem~\ref{th:prodindeprect}, $(\rxv, \ryv)$ is $(m_1+m_2)$-rectifiable.
% and $\mu(\rxv, \ryv)^{-1}\ll \Hm{m_1+m_2}|_{\sE_1\times \sE_2}$.
%Thus, by Property~\ref{en:corexistsupport} in Corollary~\ref{cor:propsrectrandvar}, there exists a support $\sE\subseteq \sE_1\times \sE_2$.

Another example of product-compatible random variables can be deduced from Theorem~\ref{th:marginals}.
Let $(\rxv, \ryv)$ be $(m_1+m_2)$-rectifiable. Assume that $\sEt_2\triangleq\proj_{\ryv}(\sE)\subseteq \R^{M_2}$ is $m_2$-rectifiable,  $\Hm{m_2}(\sEt_2)<\infty$, 
and $\Jm^{\sE}_{\proj_{\ryv}}> 0$ $\Hm{m}|_{\sE}$-almost everywhere. 
Furthermore, assume that $\sEt_1\triangleq \proj_{\rxv}(\sE)\subseteq \R^{M_1}$ is $m_1$-rectifiable, $\Hm{m_1}(\sEt_1)<\infty$, 
and $\Jm^{\sE}_{\proj_{\rxv}}> 0$ $\Hm{m}|_{\sE}$-almost everywhere.
By Theorem~\ref{th:marginals}, $\rxv$ is $m_1$-rectifiable and $\ryv$ is $m_2$-rectifiable.
Thus, 
%if in addition $\sEt_1$ and $\sEt_2$ are product-compatible, then 
$\rxv$ and $\ryv$ are product-compatible.

%As we will see  in Section~\ref{sec:mutualinformation}, the mutual information between rectifiable random variables can only be finite in the product-compatible setting. 
The setting of product-compatible random variables will be especially important for our discussion of mutual information in Section~\ref{sec:mutualinformation}.
However, already for joint entropy, we obtain some useful results.

\begin{theorem}\label{th:marginalscomp}
Let $\rxv$ be an $m_1$-rectifiable random variable on $\R^{M_1}$ with support $\sE_1$,
 and  let $\ryv$ be an $m_2$-rectifiable random variable on $\R^{M_2}$ with support $\sE_2$.
Furthermore, let $\rxv$ and $\ryv$ be product-compatible.
Denote by $\theta_{(\rxv, \ryv)}^{m_1+m_2}$ the $(m_1+m_2)$-dimensional Hausdorff density of $(\rxv, \ryv)$ and by $\sE\subseteq \sE_1\times \sE_2$ a support of $(\rxv, \ryv)$.
Then the following properties hold:
\begin{enumerate}
%\item \label{en:marghddenscompx}The $m_1$-dimensional Hausdorff density of $\rxv$ is given by
%\be\label{eq:hddensmarginalcompx}
%\theta_{\rxv}^{m_1}(\xv)=\int_{\sE_2} \theta_{(\rxv, \ryv)}^{m_1+m_2}(\xv, \yv)\, \mathrm{d}\Hm{m_2}(\yv)
%\ee
%$\Hm{m_1}$-almost everywhere. 
\item \label{en:marghddenscompy}The $m_2$-dimensional Hausdorff density of $\ryv$ is given by
\be \notag %\label{eq:hddensmarginalcompy}
\theta_{\ryv}^{m_2}(\yv)=\int_{\sE_1} \theta_{(\rxv, \ryv)}^{m_1+m_2}(\xv, \yv)\, \mathrm{d}\Hm{m_1}(\xv)
\ee
$\Hm{m_2}$-almost everywhere. 
%\item \label{en:margentropycompx}An expression of the $m_1$-dimensional entropy of $\rxv$ is
%\be\label{eq:marginalentropycompx}
%\eh{m_1}(\rxv)= -\int_{\sE} \theta_{(\rxv, \ryv)}^{m_1+m_2}(\xv, \yv)  \log \bigg(\int_{\sE_2} \theta_{(\rxv, \ryv)}^{m_1+m_2}(\xv, \yvt) \, \mathrm{d}\Hm{m_2}(\yvt)\bigg) \, \mathrm{d}\Hm{m_1+m_2}(\xv, \yv) 
%\ee
%provided the integral on the right-hand side exists.
\item \label{en:margentropycompy}An expression of the $m_2$-dimensional entropy of $\ryv$
is given by
\ba
 \eh{m_2}(\ryv) & = -\int_{\sE} \theta_{(\rxv, \ryv)}^{m_1+m_2}(\xv, \yv) \notag \\
& \rule{15mm}{0mm} \times\log \theta_{\ryv}^{m_2}(\yv) \, \mathrm{d}\Hm{m_1+m_2}(\xv, \yv) \notag % \label{eq:marginalentropycompy}
\ea
%\ba\label{eq:marginalentropycompy}
%\hspace{-5mm}\eh{m_2}(\ryv) & = -\int_{\sE} \theta_{(\rxv, \ryv)}^{m_1+m_2}(\xv, \yv) \notag \\ 
%& \rule{10mm}{0mm}\times \log \bigg(\int_{\sE_1} \theta_{(\rxv, \ryv)}^{m_1+m_2}(\xvt, \yv) \, \mathrm{d}\Hm{m_1}(\xvt)\bigg) \notag \\ 
%& \rule{42mm}{0mm}\times  \mathrm{d}\Hm{m_1+m_2}(\xv, \yv) 
%\ea
provided the integral on the right-hand side exists and is finite.
%\item \label{en:indepboundcomp}The inequality
%\be\label{eq:indepboundcomppre}
%\eh{m_1+m_2}(\rxv, \ryv)\leq \eh{m_1}(\rxv) + \eh{m_2}(\ryv)
%\ee
%holds, provided $\eh{m_1+m_2}(\rxv, \ryv)$ exists and is finite.
\end{enumerate}
Due to symmetry, analogous properties hold for $\theta_{\rxv}^{m_1}$ and $\eh{m_1}(\rxv)$.
\end{theorem}
\begin{IEEEproof}
The proof follows along the lines of the proof of Theorem \ref{th:marginals} in Appendix \ref{app:proofmarginals}.
However, due to the product-compatibility of $\rxv$ and $\ryv$, one can use Fubini's theorem in place of \eqref{eq:mycaf}.
\end{IEEEproof}
For product-compatible random variables, also the inequality $\eh{m_1+m_2}(\rxv, \ryv)\leq \eh{m_1}(\rxv) + \eh{m_2}(\ryv)$ holds. 
However, the proof of this inequality will be much easier once we considered the mutual information between rectifiable random variables.
Thus, we postpone a formal presentation of the inequality  to Corollary~\ref{cor:indepboundandcondrecent} in Section~\ref{sec:mutualinformation}.

%%%%%%%%%%%%%%%%%%%%%%%%%%%%%%%%%%%%%%%%%%%%%%%%%%%%%%%%%%%
\section{Conditional Entropy}\label{sec:conditionalentropy}
%%%%%%%%%%%%%%%%%%%%%%%%%%%%%%%%%%%%%%%%%%%%%%%%%%%%%%%%%%%

%Rev1.8: Avoid conditional random variables
In contrast to joint entropy, conditional entropy is a nontrivial extension of entropy.
We would like to define the entropy for a random variable $\rxv$ on $\R^{M_1}$ under the condition that a dependent random variable $\ryv$ on $\R^{M_2}$ is known.
%This ``conditional'' random variable is usually denoted  as $(\rxv\condi\ryv=\yv)$. 
For discrete and---under appropriate assumptions---for continuous random variables, the distribution of $(\rxv\condi\ryv=\yv)$ is well defined and so is the associated entropy $H(\rxv\condi\ryv=\yv)$ or differential entropy $h(\rxv\condi\ryv=\yv)$.
Averaging over all $\yv$ then results in the well-known definitions of conditional entropy $H(\rxv\condi \ryv)$, involving only the probability mass functions $p_{(\rxv, \ryv)}$ and $p_{\ryv}$, or of  conditional differential  entropy $h(\rxv\condi\ryv)$, involving only the probability density functions $f_{(\rxv, \ryv)}$ and $f_{\ryv}$.
Indeed, if $\rxv$ and $\ryv$ are discrete random variables, we have 
\ba
 H(\rxv\condi \ryv) & = \sum_{j\in \N}p_{\ryv}(\yv_j)\,H(\rxv\condi\ryv=\yv_j)  \notag \\%\label{eq:disccondentropyalt}\\
& = -\sum_{i, j\in \N}p_{(\rxv, \ryv)}(\xv_i, \yv_j) \log \bigg(\frac{p_{(\rxv, \ryv)}(\xv_i,\yv_j)}{p_{\ryv}(\yv_j)}\bigg)  \notag \\
& = -\E_{(\rxv, \ryv)}\bigg[\log \bigg(\frac{p_{(\rxv, \ryv)}(\rxv, \ryv)}{p_{\ryv}(\ryv)}\bigg)\bigg]\label{eq:disccondentropy}
\ea
and, if $\rxv$ and $\ryv$ are continuous random variables, we have 
\ba
h(\rxv\condi \ryv)  
&  = \int_{\R^{M_2}}f_{\ryv}(\yv)\,h(\rxv\condi\ryv=\yv) \, \mathrm{d}\yv \notag \\%\label{eq:contcondentropyalt} \\
& = -\int_{\R^{M_1+M_2}}f_{(\rxv, \ryv)}(\xv, \yv) \log \bigg(\frac{f_{(\rxv, \ryv)}(\xv, \yv)}{f_{\ryv}(\yv)}\bigg)  \mathrm{d}(\xv, \yv) \notag \\
& = -\E_{(\rxv, \ryv)}\bigg[\log \bigg(\frac{f_{(\rxv, \ryv)}(\rxv, \ryv)}{f_{\ryv}(\ryv)}\bigg)\bigg]\,.\label{eq:contcondentropy}
\ea
A straightforward generalization to rectifiable measures would be to mimic the right-hand sides of \eqref{eq:disccondentropy} and \eqref{eq:contcondentropy} using Hausdorff densities. 
However, it will turn out that this naive approach is only partly correct:
due to the geometric subtleties of the projection discussed in Section~\ref{sec:deprandvar}, we may have to include a correction term that reflects the geometry of the conditioning process.

\subsection{Conditional Probability}
%Rev1.8: Avoid conditional random variables
For general random variables $\rxv$ and $\ryv$, 
we recall the concept of conditional probabilities, which can be summarized as follows (a detailed account can be found in \cite[Ch.~5]{gr10}):
For a pair of random variables $(\rxv, \ryv)$ on $\R^{M_1+M_2}$, there exists a regular conditional probability $\Pr\{\rxv\in \sA\condi \ryv=\yv\}$, i.e., for each measurable set $\sA\subseteq \R^{M_1}$, the function  $\yv \mapsto \Pr\{\rxv\in \sA\condi \ryv=\yv\}$ is measurable and $\Pr\{\rxv\in \cdot \condi \ryv=\yv\}$ defines a probability measure for each $\yv\in \R^{M_2}$.
Furthermore, the regular conditional probability $\Pr\{\rxv\in \sA\condi \ryv=\yv\}$ satisfies
\be
 \Pr\{(\rxv, \ryv)\in \sA_1\! \times\! \sA_2\}
%\notag \\ &  \rule{20mm}{0mm}
= 
\int_{\sA_2}\Pr\{\rxv\in \sA_1\condi \ryv=\yv\} \, \mathrm{d}\mu\ryv^{-1}(\yv)\,.\label{eq:condprop}
\ee
The  regular conditional probability $\Pr\{\rxv\in \sA\condi \ryv=\yv\}$ involved in \eqref{eq:condprop} is not unique.
Nevertheless, we can still use \eqref{eq:condprop} in a definition of conditional entropy because any version of the regular conditional probability satisfies \eqref{eq:condprop}.
For the remainder of this section, we consider a fixed version of the regular conditional probability $\Pr\{\rxv\in \sA\condi \ryv=\yv\}$.

\subsection{Definition of Conditional Entropy}
In order to define a conditional entropy $\eh{m-m_2}(\rxv\condi\ryv)$, we first show that $\Pr\{\rxv\in \cdot\condi \ryv=\yv\}$ is  a rectifiable measure. 
The next theorem establishes sufficient conditions such that $\Pr\{\rxv\in \cdot\condi \ryv=\yv\}$ is rectifiable for almost every $\yv$.
As before, we denote by $\proj_{\ryv}\colon \R^{M_1+M_2}\to \R^{M_2}$ the projection of $\R^{M_1+M_2}$ to the last $M_2$ components, i.e., $\proj_{\ryv}(\xv, \yv)=\yv$.
\begin{theorem}\label{th:condproprect}
Let $(\rxv, \ryv)$ be an $m$-rectifiable random variable on $\R^{M_1+M_2}$  with $m$-dimensional Hausdorff density $\theta_{(\rxv, \ryv)}^{m}$ and support $\sE$. 
%Denote by $\proj_{\ryv}\colon \R^{M_1+M_2}\to \R^{M_2}$ the projection of $\R^{M_1+M_2}$ to the last $M_2$ components.
Furthermore, let $\sEt_2\triangleq\proj_{\ryv}(\sE)\subseteq \R^{M_2}$ be $m_2$-rectifiable ($m_2\leq m$, $m_2\leq M_2$, $m-m_2\leq M_1$), $\Hm{m_2}(\sEt_2)<\infty$, and $\Jm^{\sE}_{\proj_{\ryv}}> 0$ $\Hm{m}|_{\sE}$-almost everywhere.
% and $\sE=\bigcup_{i\in \sI}\sF_i$ with mutually disjoint sets $\sF_i$ satisfying  $\Hm{m}(\sF_i)<\infty$, where $\sI$ is a finite or countably infinite index set.
Then the following properties hold:

\begin{enumerate}
\item \label{en:condisrect}The measure $\Pr\{\rxv\in \cdot\condi \ryv=\yv\}$ is $(m-m_2)$-rectifiable for $\Hm{m_2}|_{\sE_2}$-almost every $\yv\in \R^{M_2}$, where $\sE_2\subseteq \sEt_2$ is a support%
\footnote{By Theorem~\ref{th:marginals}, the random variable $\ryv$ is $m_2$-rectifiable with Hausdorff density $\theta_{\ryv}^{m_2}$ (given by \eqref{eq:hddensmarginal}) and some support $\sE_2\subseteq \sEt_2$.} 
of $\ryv$.
\item \label{en:conddensity}The $(m-m_2)$-dimensional Hausdorff density 
of the measure $\Pr\{\rxv\in \cdot\condi \ryv=\yv\}$ is given by
\be  \label{eq:conddensity}
\theta_{\Pr\{\rxv\in \cdot\condi \ryv=\yv\}}^{m-m_2}(\xv)=
\frac{\theta_{(\rxv, \ryv)}^{m}(\xv, \yv)}{\Jm^{\sE}_{\proj_{\ryv}}(\xv, \yv)\, \theta_{\ryv}^{m_2}(\yv)}
\ee
$\Hm{m-m_2}|_{\sE^{(\yv)}}$-almost everywhere,
for $\Hm{m_2}|_{\sE_2}$-al\-most every $\yv\in \R^{M_2}$. 
Here, as before,  $\sE^{(\yv)}\triangleq \{\xv\in \R^{M_1}: (\xv, \yv)\in \sE\}$.
%\item \label{en:condentropy}For $\Hm{m_2}|_{\sE_2}$-almost every $\yv\in \R^{M_2}$, the $(m-m_2)$-dimensional entropy of $(\rxv\condi\ryv=\yv)$ is given by
%\be\label{eq:entgivenyeqy}
%\eh{m-m_2}(\rxv\condi \ryv=\yv)= -\int_{\sE^{(\yv)}}\frac{\theta_{(\rxv, \ryv)}^{m}(\xv, \yv)}{\Jm^{\sE}_{\proj_{\ryv}}(\xv, \yv)\, \theta_{\ryv}^{m_2}(\yv)} \log\bigg(\frac{\theta_{(\rxv, \ryv)}^{m}(\xv, \yv)}{\Jm^{\sE}_{\proj_{\ryv}}(\xv, \yv)\, \theta_{\ryv}^{m_2}(\yv)}\bigg)\, \mathrm{d}\Hm{m-m_2}(\xv)\,.
%\ee
\end{enumerate}
\end{theorem}
\begin{IEEEproof}
See Appendix~\ref{app:proofcondproprect}.
\end{IEEEproof}

As for joint entropy, the case of product-compatible random variables (see Definition~\ref{def:productcomprvs}) is of special interest and results in a more intuitive characterization of the Hausdorff density of $\Pr\{\rxv\in \cdot\condi \ryv=\yv\}$.
\begin{theorem}\label{th:condproprectcomp}
Let $\rxv$ be an $m_1$-rectifiable random variable on $\R^{M_1}$ with support $\sE_1$, and  let $\ryv$ be an $m_2$-rectifiable random variable on $\R^{M_2}$ with support $\sE_2$.
Furthermore, let $\rxv$ and $\ryv$ be product-compatible.
Then the following properties hold:
\begin{enumerate}
\item \label{en:condisrectcomp}The measure $\Pr\{\rxv\in \cdot\condi \ryv=\yv\}$ is $m_1$-rectifiable for $\Hm{m_2}|_{\sE_2}$-almost every $\yv\in \R^{M_2}$.
\item \label{en:conddensitycomp}The $m_1$-dimensional Hausdorff density 
of $\Pr\{\rxv\in \cdot\condi \ryv=\yv\}$ is given by
\be \label{eq:conddensitycomp}
\theta_{\Pr\{\rxv\in \cdot\condi \ryv=\yv\}}^{m_1}(\xv)=
\frac{\theta_{(\rxv, \ryv)}^{m_1+m_2}(\xv, \yv)}{\theta_{\ryv}^{m_2}(\yv)}
\ee
$\Hm{m_1}|_{\sE_1}$-almost everywhere, for $\Hm{m_2}|_{\sE_2}$-almost every $\yv\in \R^{M_2}$.
%\item \label{en:condentropycomp}For $\Hm{m_2}|_{\sE_2}$-almost every $\yv\in \R^{M_2}$, the $m_1$-dimensional entropy of $(\rxv\condi\ryv=\yv)$ is given by
%\be\label{eq:entgivenyeqycomp}
%\eh{m_1}(\rxv\condi \ryv=\yv)= -\int_{\sE_1}\frac{\theta_{(\rxv, \ryv)}^{m_1+m_2}(\xv, \yv)}{\theta_{\ryv}^{m_2}(\yv)} \log\bigg(\frac{\theta_{(\rxv, \ryv)}^{m_1+m_2}(\xv, \yv)}{\theta_{\ryv}^{m_2}(\yv)}\bigg)\, \mathrm{d}\Hm{m_1}(\xv)\,.
%\ee
\end{enumerate}
\end{theorem}
%Rev1.1:Omit proof, just list differences to other Theorem
\begin{IEEEproof}
The proof follows along the lines of the proof of Theorem \ref{th:condproprect} in Appendix \ref{app:proofcondproprect}.
However, due to the product-compatibility of $\rxv$ and $\ryv$, one can use Fubini's theorem in place of \eqref{eq:mycaf}.
%See Appendix~\ref{app:proofcondproprectcomp}.
\end{IEEEproof}

Note that Theorems~\ref{th:condproprect} and \ref{th:condproprectcomp} hold for any version of the regular conditional probability $\Pr\{\rxv\in \sA\condi \ryv=\yv\}$.
However, for different versions, the statement ``for $\Hm{m_2}|_{\sE_2}$-almost every $\yv\in \R^{M_2}$'' may refer to different sets of $\Hm{m_2}|_{\sE_2}$-measure zero; e.g., \eqref{eq:conddensity} may hold for different $\yv\in \R^{M_2}$. 
Thus, results that are independent of the version of  the regular conditional probability can
only be obtained if we can avoid these ``almost everywhere''-statements.
To this end, we will define conditional entropy as an expectation over $\ryv$. % similar to \eqref{eq:disccondentropy} and \eqref{eq:contcondentropy}.
%To this end, we will calculate the expectation of $\eh{m-m_2}(\rxv\condi \ryv=\yv)$.
%The resulting expression will no longer depend on the specific version of the regular conditional probability.
%Anticipating this independence (cf.\ Theorem~\ref{th:condentropyasexp}) and motivated by \eqref{eq:disccondentropyalt} and \eqref{eq:contcondentropyalt},
%we next define conditional entropy for rectifiable random variables.

\begin{definition}\label{def:condentropy}
Let $(\rxv, \ryv)$ be an $m$-rectifiable random variable on $\R^{M_1+M_2}$ such that $\ryv$ is $m_2$-rectifiable with $m_2$-dimensional Hausdorff density $\theta_{\ryv}^{m_2}$ and support $\sE_2$.
The \textit{conditional entropy of $\rxv$ given $\ryv$} is defined as%
\footnote{The inner integral in \eqref{eq:defcondentropy} can be intuitively interpreted as an entropy $\eh{m-m_2}(\rxv\condi \ryv=\yv)$. 
However, such an entropy is not well defined in general and depends on the choice of the conditional probability.}
\ba
& \eh{m-m_2}(\rxv\condi\ryv) \notag \\ 
& \rule{3mm}{0mm} \triangleq\, -
\int_{\sE_2} \theta_{\ryv}^{m_2}(\yv)\, \int_{\sE^{(\yv)}}\theta_{\Pr\{\rxv\in \cdot\condi \ryv=\yv\}}^{m-m_2}(\xv) \notag \\ 
& \rule{9mm}{0mm}\times \log\theta_{\Pr\{\rxv\in \cdot\condi \ryv=\yv\}}^{m-m_2}(\xv)\, \mathrm{d}\Hm{m-m_2}(\xv) \, \mathrm{d}\Hm{m_2}(\yv)\label{eq:defcondentropy}
\ea
provided the right-hand side of \eqref{eq:defcondentropy} exists and coincides for all versions of the regular conditional probability $\Pr\{\rxv\in \sA\condi \ryv=\yv\}$.
\end{definition}
\begin{remark}\label{rem:indcond}
For independent random variables $\rxv$ and $\ryv$, inserting \eqref{eq:prodhddensity} into \eqref{eq:conddensitycomp} implies that $\theta_{\Pr\{\rxv\in \cdot\condi \ryv=\yv\}}^{m_1}(\xv)= \theta_{\rxv}^{m_1}(\xv)$.
Thus, \eqref{eq:defcondentropy} reduces to $\eh{m_1}(\rxv\condi\ryv)=\eh{m_1}(\rxv)$.
\end{remark}

The following theorem gives a characterization of conditional entropy and sufficient conditions for \eqref{eq:defcondentropy} to be well-defined in the sense that the right-hand side  of \eqref{eq:defcondentropy} coincides for all versions of the regular conditional probability $\Pr\{\rxv\in \sA\condi \ryv=\yv\}$.
\begin{theorem}\label{th:condentropyasexp}
Let $(\rxv, \ryv)$ be an $m$-rectifiable random variable on $\R^{M_1+M_2}$ with $m$-dimensional Hausdorff density $\theta_{(\rxv, \ryv)}^{m}$ and support $\sE$. 
Furthermore, let $\sE_2\triangleq\proj_{\ryv}(\sE)$ be $m_2$-rectifiable, $\Hm{m_2}(\sE_2)<\infty$, and $\Jm^{\sE}_{\proj_{\ryv}}> 0$ $\Hm{m}|_{\sE}$-almost everywhere.
Then
\ba
\eh{m-m_2}(\rxv\condi\ryv)
&= -\E_{(\rxv, \ryv)}\bigg[\log\bigg(\frac{\theta_{(\rxv, \ryv)}^{m}(\rxv, \ryv)}{\theta_{\ryv}^{m_2}(\ryv)}\bigg)\bigg] 
\notag \\[1mm] & \rule{23mm}{0mm}
+ \E_{(\rxv, \ryv)}\big[\log \Jm^{\sE}_{\proj_{\ryv}}(\rxv, \ryv) \big]\label{eq:condentropyasexp}
\ea
provided the right-hand side of \eqref{eq:condentropyasexp} exists and is finite.
\end{theorem}
\begin{IEEEproof}
See Appendix~\ref{app:proofcondentropyasexp}.
\end{IEEEproof}
Note the difference between \eqref{eq:condentropyasexp} and  the expressions \eqref{eq:disccondentropy} and \eqref{eq:contcondentropy} of $H(\rxv \condi \ryv)$ and $h(\rxv \condi \ryv)$, respectively: 
in the case of rectifiable random variables, we generally have to include the geometric correction term $\E_{(\rxv, \ryv)}\big[\log \Jm^{\sE}_{\proj_{\ryv}}(\rxv, \ryv) \big]$.
However, we will show next that, in the special case of product-compatible rectifiable random variables, this correction term does not appear.
\begin{theorem}\label{th:condentropyasexpcomp}
Let the $m_1$-rectifiable random variable $\rxv$ on $\R^{M_1}$ and  the $m_2$-rectifiable random variable $\ryv$ on $\R^{M_2}$ be product-compatible.
%Denote by $\theta_{(\rxv, \ryv)}^{m_1+m_2}$ the $m_1+m_2$-dimensional Hausdorff density of $(\rxv, \ryv)$ and by $\theta_{\ryv}^{m_2}$ the $m_2$-dimensional Hausdorff density of $\ryv$.
Then
\be\label{eq:condentropyasexpcomp}
\eh{m_1}(\rxv\condi\ryv)
= -\E_{(\rxv, \ryv)}\bigg[\log\bigg(\frac{\theta_{(\rxv, \ryv)}^{m_1+m_2}(\rxv, \ryv)}{\theta_{\ryv}^{m_2}(\ryv)}\bigg)\bigg]
\ee
provided the right-hand side of \eqref{eq:condentropyasexpcomp} exists and is finite.
\end{theorem}
%Rev1.1:Omit proof, just list differences to other Theorem
\begin{IEEEproof}
The proof follows along the lines of the proof of Theorem \ref{th:condentropyasexp} in Appendix \ref{app:proofcondentropyasexp}.
However, due to the product-compatibility of $\rxv$ and $\ryv$, one can use Fubini's theorem in place of \eqref{eq:mycaf}.
\end{IEEEproof}

\subsection{Chain Rule for Rectifiable Random Variables}\label{sec:chainrule}
As in the case of entropy and differential entropy, we can give a chain rule for $m$-dimensional entropy.
%However,  the geometric correction term appears again.
\begin{theorem}\label{th:chainrule}
Let $(\rxv, \ryv)$ be an $m$-rectifiable random variable on $\R^{M_1+M_2}$ with $m$-dimensional Hausdorff density $\theta_{(\rxv, \ryv)}^{m}$ and support $\sE$. 
Furthermore, let $\sE_2\triangleq\proj_{\ryv}(\sE)$ be $m_2$-rectifiable, $\Hm{m_2}(\sE_2)<\infty$, and $\Jm^{\sE}_{\proj_{\ryv}}> 0$ $\Hm{m}|_{\sE}$-almost everywhere.
Then 
\be\label{eq:chainrule}
\eh{m}(\rxv, \ryv)= \eh{m_2}(\ryv)+ \eh{m-m_2}(\rxv\condi\ryv) - \E_{(\rxv, \ryv)}\big[\log \Jm^{\sE}_{\proj_{\ryv}}(\rxv, \ryv) \big]
\ee
provided the corresponding integrals exist and are finite.
\end{theorem}
\begin{IEEEproof}
By the definition of $\eh{m}(\rxv, \ryv)$ in \eqref{eq:defjointentropyexp} and the definition of $\eh{m_2}(\ryv)$ in~\eqref{eq:defmdimentropy}, we have
\ba\label{eq:jointmargcorriscond}
& \eh{m}(\rxv, \ryv) - \eh{m_2}(\ryv) + \E_{(\rxv, \ryv)}\big[\log \Jm^{\sE}_{\proj_{\ryv}}(\rxv, \ryv) \big] \notag \\
& \rule{6mm}{0mm} = - \E_{(\rxv, \ryv)}\big[ \log \theta_{(\rxv, \ryv)}^{m}(\rxv, \ryv)\big] + \E_{\ryv}\big[\log \theta_{\ryv}^{m_2}(\ryv) \big] \notag \\
& \rule{50mm}{0mm} + \E_{(\rxv, \ryv)}\big[\log \Jm^{\sE}_{\proj_{\ryv}}(\rxv, \ryv) \big]  \notag \\
& \rule{6mm}{0mm} = - \E_{(\rxv, \ryv)}\bigg[\log\bigg(\frac{\theta_{(\rxv, \ryv)}^{m}(\rxv, \ryv)}{\theta_{\ryv}^{m_2}(\ryv)}\bigg)\bigg] 
%\notag \\ & \rule{40mm}{0mm} 
+ \E_{(\rxv, \ryv)}\big[\log \Jm^{\sE}_{\proj_{\ryv}}(\rxv, \ryv) \big]\,.  
\ea
Because we assumed in the theorem that the integrals corresponding to the terms on the left-hand side of  \eqref{eq:jointmargcorriscond} are finite, the right-hand side of \eqref{eq:jointmargcorriscond} is also finite. 
By~\eqref{eq:condentropyasexp}, the right-hand side of \eqref{eq:jointmargcorriscond} equals $\eh{m-m_2}(\rxv\condi\ryv)$. 
Thus, \eqref{eq:chainrule} holds.
\end{IEEEproof}
Next, we continue Examples~\ref{ex:entbounduniform} and \ref{ex:projuniform}  from Section~\ref{sec:deprandvar}.
We will see that the geometric correction term  in the chain rule, $\E_{(\rxv, \ryv)}\big[\log \Jm^{\sE}_{\proj_{\ryv}}(\rxv, \ryv) \big]$, is indeed necessary.
\begin{example}\label{ex:chainruleuniform}
%Rev1.8: avoid conditional random variables
As in Examples~\ref{ex:entbounduniform} and \ref{ex:projuniform}, we consider $(\rx, \ry)\in \R^2$ uniformly distributed on the unit circle $\sS_1$, i.e., $\theta^1_{(\rx, \ry)}(x,y)=1/(2\pi)$ $\Hm{1}$-almost everywhere on $\sS_1$.
According to \eqref{eq:marginalexample},
\be\label{eq:calchy}
\eh{1}(\ry) = \log\bigg(\frac{\pi}{2}\bigg)
\ee
and according to \eqref{eq:calchxy},
\be \label{eq:calchxy2}
\eh{1}(\rx, \ry) 
 = \log(2\pi)\,.
\ee
To calculate the conditional entropy $\eh{0}(\rx\condi \ry)$ (note that $m-m_2=1-1=0$), we consider the regular conditional probability $\Pr\{\rx\in \sA\condi \ry=y\}$.
It is easy to see that one possible version of $\Pr\{\rx\in \sA\condi \ry=y\}$ is the following: for $y\in (-1,1)$, 
%% the probability 
$\Pr\{\rx=x\condi \ry=y\}=1/2$ for  $x=\pm \sqrt{1-y^2}$ and $\Pr\{\rx\in \sA\condi \ry=y\}=0$ if $\pm \sqrt{1-y^2}\notin \sA$. 
The probabilities for $\abs{y}\geq 1$ are irrelevant because $\Pr\{\ry\notin (-1,1)\}=0$.
Hence, by \eqref{eq:defcondentropy}, we obtain
\ba
& \eh{0}(\rx\condi \ry) \notag \\ 
& \rule{4mm}{0mm}= -
\int_{(-1,1)} \theta_{\ry}^{1}(y)  \int_{\{\pm \sqrt{1-y^2}\}}\frac{1}{2} \log\frac{1}{2}\, \mathrm{d}\Hm{0}(x) \, \mathrm{d}\Hm{1}(y)  \notag \\[1mm] 
& \rule{4mm}{0mm}= -
\int_{(-1,1)} \theta_{\ry}^{1}(y)    \log\frac{1}{2}  \, \mathrm{d}\Hm{1}(y)  \notag \\ 
&  \rule{4mm}{0mm}=\log 2\,.\label{eq:calchxgiveny}
\ea
This differs from $\eh{1}(\rx, \ry)- \eh{1}(\ry) = \log(2\pi) - \log(\pi/2)$, and therefore the conjecture that there holds a chain rule without a correction term is wrong. 
To calculate the correction term, which according to \eqref{eq:chainrule} is given by $\E_{(\rx, \ry)}\big[\log \Jm^{\sS_1}_{\proj_{\ry}}(\rx, \ry) \big]$, we recall from Example~\ref{ex:projuniform} that $\Jm^{\sS_1}_{\proj_{\ry}}\big(\pm\sqrt{1-y^2}, \pm y\big)=\sqrt{1-y^2}$ or, more conveniently, 
$\Jm^{\sS_1}_{\proj_{\ry}}(\cos \phi, \sin \phi)=\abs{\cos\phi}$.
Thus, we obtain
\ba\label{eq:calcexpjacobian}
\E_{(\rx, \ry)}\big[\log \Jm^{\sS_1}_{\proj_{\ry}}(\rx, \ry) \big]
& = \int_{\sS_1} \frac{1}{2\pi} \log  \Jm^{\sS_1}_{\proj_{\ry}}(x,   y)  \, \mathrm{d}\Hm{1}(x, y) \notag \\
& =  \int_{0}^{2\pi} \frac{1}{2\pi} \log  \abs{\cos\phi}  \, \mathrm{d}\phi  \notag \\
& = -\log 2 \,.
\ea
 We finally verify that \eqref{eq:calcexpjacobian} is consistent with the chain rule \eqref{eq:chainrule}.
 Starting from \eqref{eq:calchxy2}, we obtain
\ba
\eh{1}(\rx, \ry) & = \log(2\pi) \notag \\
& = \log\bigg(\frac{\pi}{2}\bigg) +  \log 2  - (-\log 2 )\notag \\
& 
= \eh{1}(\ry) + \eh{0}(\rx\condi \ry) - \E_{(\rx, \ry)}\big[\log \Jm^{\sS_1}_{\proj_{\ry}}(\rx, \ry) \big] \notag 
\ea
where the final expansion is obtained by using  \eqref{eq:calchy},  \eqref{eq:calchxgiveny}, and  \eqref{eq:calcexpjacobian}.
\hfill \IEEEQED
\end{example}
Example~\ref{ex:chainruleuniform} also provides a counterexample to the rule ``conditioning does not increase entropy,''
which holds for the entropy of discrete random variables and the differential entropy of continuous random variables. 
Indeed, comparing~\eqref{eq:entexampy} and \eqref{eq:calchxgiveny}, we see that for the components of a uniform distribution on the unit circle, we have $\eh{1}(\rx)<\eh{0}(\rx\condi \ry)$.
However, as we will see in Corollary~\ref{cor:indepboundandcondrecent} in Section~\ref{sec:mutualinformation}, this is only due to a ``reduction of dimensions'':
if $\rxv$ and $\ryv$ are product-compatible, which implies that $\eh{m_1}(\rxv)$ and $\eh{m-m_2}(\rxv\condi \ryv)$ are of the same dimension $m_1=m-m_2$, conditioning will indeed not increase entropy, i.e., $\eh{m_1}(\rxv\condi\ryv)\leq \eh{m_1}(\rxv)$.
Also the chain rule \eqref{eq:chainrule} reduces to its traditional form, as stated next.
\begin{theorem}\label{th:chainrulecomp}
Let the $m_1$-rectifiable random variable $\rxv$ on $\R^{M_1}$ and  the $m_2$-rectifiable random variable $\ryv$ on $\R^{M_2}$ be product-compatible.
Then 
\be\label{eq:chainrulecomp}
\eh{m_1+m_2}(\rxv, \ryv)= \eh{m_2}(\ryv)+ \eh{m_1}(\rxv\condi\ryv)
\ee
provided the entropies $\eh{m_1+m_2}(\rxv, \ryv)$ and $\eh{m_2}(\ryv)$ exist and are finite.
\end{theorem}
\begin{IEEEproof}
By the definition of $\eh{m_1+m_2}(\rxv, \ryv)$ in \eqref{eq:defjointentropyexp} and the definition of $\eh{m_2}(\ryv)$ in~\eqref{eq:defmdimentropy}, we have
\ba\label{eq:jointmargcorriscondcomp}
 & \eh{m_1+m_2}(\rxv, \ryv) - \eh{m_2}(\ryv)  \notag \\
& \rule{8mm}{0mm} = - \E_{(\rxv, \ryv)}\big[ \log \theta_{(\rxv, \ryv)}^{m_1+m_2}(\rxv, \ryv)\big] + \E_{\ryv}\big[\log \theta_{\ryv}^{m_2}(\ryv) \big]   \notag \\
& \rule{8mm}{0mm} = - \E_{(\rxv, \ryv)}\bigg[\log\bigg(\frac{\theta_{(\rxv, \ryv)}^{m_1+m_2}(\rxv, \ryv)}{\theta_{\ryv}^{m_2}(\ryv)}\bigg)\bigg]\,.  
\ea
By~\eqref{eq:condentropyasexpcomp}, the right-hand side of \eqref{eq:jointmargcorriscondcomp} equals  $\eh{m_1}(\rxv\condi\ryv)$. 
Thus, \eqref{eq:chainrulecomp} holds.
\end{IEEEproof}

Using an induction argument,
we can extend the chain rule~\eqref{eq:chainrulecomp} to a sequence of random variables. 
\begin{corollary}
\label{cor:chainrule}
Let $\rxv_{1:n}\triangleq (\rxv_1, \dots, \rxv_n)$ be a sequence of  random variables where each $\rxv_i\in \R^{M_i}$ is $m_i$-rectifiable.
Assume that $\rxv_{1:i-1}$ and $\rxv_{i}$ are product-compatible for $i\in \{2, \dots, n\}$.
Then
\be  \label{eq:chainruleextendedcomp}
\eh{m}(\rxv_{1:n})= \eh{m_1}(\rxv_1)+ \sum_{i=2}^n \eh{m_i}(\rxv_{i}\condi\rxv_{1:i-1}) 
\ee
with $m= \sum_{i=1}^n m_i$, provided the corresponding integrals exist and are finite.
\end{corollary}
We note that, consistently with Remark~\ref{rem:indcond}, \eqref{eq:indepprodentropy} is a special case of \eqref{eq:chainruleextendedcomp}.
%Rev1.1:Omit proof
%\begin{IEEEproof}
%We prove \eqref{eq:chainruleextendedcomp} by induction.
%For $n=2$, \eqref{eq:chainruleextendedcomp} reduces to \eqref{eq:chainrulecomp}.
%Thus, we only have to show the inductive step.
%Assume that \eqref{eq:chainruleextendedcomp} holds for $n-1$ random variables, i.e., 
%\be \label{eq:chainrulefornmin1}
%\eh{\widetilde{m}}(\rxvt_{1:n-1})= \eh{\widetilde{m}_1}(\rxvt_1)+ \sum_{i=2}^{n-1} \eh{\widetilde{m}_i}(\rxvt_{i}\condi\rxvt_{1:i-1})
%\ee
 %for $\rxvt_{1:n-1}$ such that each $\rxvt_i\in \R^{\widetilde{M}_i}$ is $\widetilde{m}_i$-rectifiable, $\widetilde{m}=\sum_{i=1}^{n-1} \widetilde{m}_i$, and $\rxvt_{1:i-1}$ and $\rxvt_{i}$ are product-compatible for $i\in \{2, \dots, n-1\}$.
%Choosing $\rxvt_1\triangleq (\rxv_1, \rxv_2)$ and $\rxvt_i\triangleq \rxv_{i+1}$ for $i\in \{2, \dots, n-1\}$, 
%we have $\rxvt_{1:n-1}=\rxv_{1:n}$, $\widetilde{m}_1=m_1+m_2$, and $\widetilde{m}_i=m_{i+1}$ for $i\in \{2, \dots, n-1\}$.
%Thus, $\widetilde{m}=\sum_{i=1}^n m_i=m$ and 
%\eqref{eq:chainrulefornmin1} implies
%\be\label{eq:chainrulenmin1comp}
%\eh{m}(\rxv_{1:n})= \eh{m_1+m_2}(\rxv_1, \rxv_2)+ \sum_{i=3}^n \eh{m_i}(\rxv_{i}\condi\rxv_{1:i-1}) \,.
%\ee
%By~\eqref{eq:chainrulecomp}, we also have
%\be\label{eq:chainrulex1x2comp}
%\eh{m_1+m_2}(\rxv_1, \rxv_2)= \eh{m_1}(\rxv_1)+ \eh{m_2}(\rxv_2\condi\rxv_1)\,.
%\ee
%Combining \eqref{eq:chainrulenmin1comp} and \eqref{eq:chainrulex1x2comp}, we obtain \eqref{eq:chainruleextendedcomp}.
%\end{IEEEproof}

%%%%%%%%%%%%%%%%%%%%%%%%%%%%%%%%
\section{Mutual Information} \label{sec:mutualinformation}   
%%%%%%%%%%%%%%%%%%%%%%%%%%%%%%%%
The basic definition of mutual information is for discrete random variables $\rxv$ and $\ryv$ with probability mass functions $p_{\rxv}(\xv_i)$ and $p_{\ryv}(\yv_j)$ and joint probability mass function $p_{(\rxv, \ryv)}(\xv_i, \yv_j)$.
The mutual information between $\rxv$ and $\ryv$ is given by \cite[eq.~(2.28)]{Cover91}
\be\label{eq:mutinfdisc}
I(\rxv; \ryv) \triangleq \sum_{i,j}p_{(\rxv, \ryv)}(\xv_i, \yv_j) \log \bigg(\frac{p_{(\rxv, \ryv)}(\xv_i, \yv_j)}{p_{\rxv}(\xv_i)p_{\ryv}(\yv_j)}\bigg)\,.
\ee
However, mutual information is also defined between arbitrary random variables $\rxv$ and $\ryv$ on a common probability space.
This definition is based on \eqref{eq:mutinfdisc} and quantizations $[\rxv]_{\mathfrak{Q}}$ and $[\ryv]_{\mathfrak{R}}$ \cite[eq.~(8.54)]{Cover91}.
We recall from Section~\ref{sec:renyiandepsent} that for a measurable, finite partition $\mathfrak{Q}=\{\sA_1, \dots, \sA_N\}$ of $\R^{M_1}$ (i.e., $\R^{M_1}=\bigcup_{i=1}^N\sA_i$ with $\sA_i\in \mathfrak{Q}$ mutually disjoint and measurable), 
the quantization $[\rxv]_{\mathfrak{Q}}\in \{1, \dots, N\}$ is defined as the discrete random variable with probability mass function  $p_{[\rxv]_{\mathfrak{Q}}}(i)=\Pr\{[\rxv]_{\mathfrak{Q}}=i\}=\Pr\{\rxv\in \sA_i\}$ for $i\in \{1, \dots, N\}$.
\begin{definition}[\mbox{\cite[eq.~(8.54)]{Cover91}}]
Let $\rxv\colon \Omega \to \R^{M_1}$ and $\ryv\colon \Omega \to \R^{M_2}$ be random variables on a common probability space $(\Omega, \mathfrak{S}, \mu)$.
The mutual information between $\rxv$ and $\ryv$ is defined as
\be \notag 
I(\rxv;\ryv)\triangleq \sup_{\mathfrak{Q}, \mathfrak{R}} I([\rxv]_{\mathfrak{Q}}; [\ryv]_{\mathfrak{R}})
\ee
where the supremum is taken over all measurable, finite partitions $\mathfrak{Q}$ of $\R^{M_1}$ and $\mathfrak{R}$ of $\R^{M_2}$.
\end{definition}
The Gelfand-Yaglom-Perez theorem \cite[Lem.~5.2.3]{Gray1990Entropy} provides an expression of  mutual information in terms of Radon-Nikodym derivatives:
for random variables $\rxv\colon \Omega \to \R^{M_1}$ and $\ryv\colon \Omega \to \R^{M_2}$ on a common probability space $(\Omega, \mathfrak{S}, \mu)$,
\ba
I(\rxv; \ryv) & =
\int_{\R^{M_1+M_2}}\log \Big(\frac{\mathrm{d}\mu(\rxv, \ryv)^{-1}}{\mathrm{d}\big(\mu\rxv^{-1}\times \mu\ryv^{-1}\big)}(\xv,\yv) \Big) \notag \\
& \rule{40mm}{0mm}\times  \mathrm{d}\mu(\rxv, \ryv)^{-1}(\xv,\yv) 
\label{eq:gyptheorem1}
\ea
if $\mu(\rxv, \ryv)^{-1}\ll  \mu\rxv^{-1}\times \mu\ryv^{-1}$, and 
\be
I(\rxv; \ryv) =
\infty
\label{eq:gyptheorem2}
\ee
if $\mu(\rxv, \ryv)^{-1}\centernot\ll  \mu\rxv^{-1}\times \mu\ryv^{-1}$.

For the special cases of discrete and continuous random variables, there exist  expressions of mutual information in terms of entropy and differential entropy, respectively.
We will extend these expressions  to the case of rectifiable random variables.
The resulting generalization will involve the entropies $\eh{m_1}(\rxv)$, $\eh{m_2}(\ryv)$, and $\eh{m}(\rxv,\ryv)$.
\begin{theorem}\label{th:mutinf}
Let $\rxv$ be an $m_1$-rectifiable random variable with support $\sE_1\subseteq \R^{M_1}$, 
let $\ryv$ be an $m_2$-rectifiable random variable  with support $\sE_2\subseteq \R^{M_2}$, 
 and let $(\rxv, \ryv)$ be $m$-rectifiable with support $\sE\subseteq \sE_1\times \sE_2$.
%and assume that $\sE_1$ and $\sE_2$ are product-compatible.
The mutual information $I(\rxv; \ryv)$ satisfies:
\begin{enumerate}
\item \label{en_mutinfcasecomp}If $\rxv$ and $\ryv$ are product-compatible (i.e., $m=m_1+m_2$), then
\ba\label{eq:mutinfasdens}
& I(\rxv; \ryv)=
 \int_{\sE}\theta_{(\rxv, \ryv)}^m(\xv, \yv)  \notag \\
& \rule{15mm}{0mm}\times \log \bigg(\frac{\theta_{(\rxv, \ryv)}^m(\xv, \yv)}{\theta_{\rxv}^{m_1}(\xv)\theta_{\ryv}^{m_2}(\yv)}\bigg) \, \mathrm{d}\Hm{m}(\xv, \yv) \,.
\ea
Furthermore,
\be
I(\rxv; \ryv) 
 =  \eh{m_1}(\rxv)+\eh{m_2}(\ryv) - \eh{m}(\rxv,\ryv) \label{eq:mutinfasjoint}
\ee
and
\be
I(\rxv; \ryv) 
 = \eh{m_1}(\rxv)-\eh{m_1}(\rxv\condi\ryv) %\notag \\
 =  \eh{m_2}(\ryv)-\eh{m_2}(\ryv\condi\rxv)  \label{eq:mutinfascondent}
\ee
provided  the entropies $\eh{m_1}(\rxv)$, $\eh{m_2}(\ryv)$, and $\eh{m}(\rxv,\ryv)$ exist and are finite.
\item  \label{en_mutinfcasenoncomp}If $m<m_1+m_2$, then $I(\rxv; \ryv) =\infty$.
\end{enumerate}
\end{theorem}
\begin{IEEEproof}
See Appendix~\ref{app:proofmutinf}.
\end{IEEEproof}
In Theorem~\ref{th:mutinf}, the case $m<m_1+m_2$ can be interpreted as $\rxv$ and $\ryv$ ``sharing'' at least one dimension.
In a communication scenario, this would imply that it is possible to reconstruct an at least one-dimensional component of $\rxv$  from $\ryv$ (and, also, to reconstruct an at least one-dimensional component of $\ryv$ from $\rxv$). 
Thus, an infinite amount of information could be transmitted over a channel $\rxv\longrightarrow \ryv$ (or $\ryv\longrightarrow \rxv$).
This is consistent with our result that $I(\rxv; \ryv)=\infty$.
%As an example, we again consider the components of a uniform distribution on the unit circle in $\R^2$.
%According to Example~\ref{ex:chainruleuniform}, if we observe $\ry=y$, we know that $\rx=\pm \sqrt{1-y^2}$

A corollary of Theorem~\ref{th:mutinf} states that for product-compati\-ble random variables, we can upper-bound the joint entropy by the sum of the individual entropies and prove that conditioning does not increase entropy.
\begin{corollary}\label{cor:indepboundandcondrecent}
Let the $m_1$-rectifiable random variable $\rxv$ on $\R^{M_1}$ and  the $m_2$-rectifiable random variable $\ryv$ on $\R^{M_2}$ be product-compatible.
 Then
\be\label{eq:indepboundcomp}
\eh{m_1+m_2}(\rxv, \ryv)\leq \eh{m_1}(\rxv) + \eh{m_2}(\ryv)
\ee
and
\be\label{eq:condredentcomp}
\eh{m_1}(\rxv\condi\ryv)\leq \eh{m_1}(\rxv)
\ee
 provided  the entropies $\eh{m_1}\rmv(\rxv)$, $\eh{m_2}\rmv(\ryv)$, and $\eh{m_1+m_2}\rmv(\rxv, \ryv)$ exist and are finite.
\end{corollary}
\begin{IEEEproof}
The inequality \eqref{eq:indepboundcomp} follows from \eqref{eq:mutinfasjoint} and the nonnegativity of mutual information.
Similarly, \eqref{eq:condredentcomp} follows from \eqref{eq:mutinfascondent} and the nonnegativity of mutual information.
\end{IEEEproof}

%%%%%%%%%%%%%%%%%%%%%%%%%%%%%%%%%%%%%%%%%%%%%%%%%%%%%%%%%%%
\section{Asymptotic Equipartition Property}\label{sec:aep}
%%%%%%%%%%%%%%%%%%%%%%%%%%%%%%%%%%%%%%%%%%%%%%%%%%%%%%%%%%%

Similar to classical entropy and differential entropy, the $m$-dimensional entropy $\eh{m}(\rxv)$ satisfies an asymptotic equipartition property (AEP). 
Let us consider a sequence $\rxv_{1:n}\triangleq (\rxv_1,   \dots, \rxv_n)$ of i.i.d.\ random variables $\rxv_i$.
Our main findings are similar to the discrete and continuous cases: 
based on $\eh{m}(\rxv)$, we define sets $\sA_{\varepsilon}^{(n)}$ of typical sequences $\xv_{1:n}$ and show that,  for sufficiently large $n$, a random sequence $\rxv_{1:n}$ belongs to  $\sA_{\varepsilon}^{(n)}$ with probability arbitrarily close to one.
Furthermore, we obtain upper and lower bounds on the size of $\sA_{\varepsilon}^{(n)}$ given by $e^{n(\eh{m}(\rxv)+ \varepsilon)}$ and $(1-\delta)e^{n(\eh{m}(\rxv)-\varepsilon)}$, respectively.
In the  case of classical entropy and differential entropy, these properties are useful in the proof of various coding theorems because they allow us to consider only typical sequences.

Our analysis  follows the steps in \cite[Sec.~8.2]{Cover91}.
However, whereas in the discrete case the size of a set of sequences $\xv_{1:n}$ is measured by its cardinality  and in the continuous case by its Lebesgue measure, in the present case of $m$-rectifiable random variables $\rxv_i$, we resort to the Hausdorff measure.
\begin{lemma}\label{th:mylln}
Let $\rxv_{1:n}=(\rxv_1,\dots, \rxv_n)$ be a sequence of i.i.d.\ $m$-rectifiable random variables $\rxv_i$ on $\R^M$, where each $\rxv_i$ has $m$-dimensional Hausdorff density $\theta_{\rxv}^m$ and $m$-dimensional entropy $\eh{m}(\rxv)$. 
The random variable $-(1/n) \sum_{i=1}^n\log \theta_{\rxv}^m(\rxv_i) $ converges to $\eh{m}(\rxv)$ in probability, i.e., for any $\varepsilon>0$  
\be\notag %\label{eq:mywlln}
\lim_{n\to \infty} \Pr\bigg\{\bigg\lvert - \frac{1}{n} \sum_{i=1}^n\log \theta_{\rxv}^m(\rxv_i) -\eh{m}(\rxv) \bigg\rvert>\varepsilon\bigg\}=0\,.
\ee
\end{lemma}
\begin{IEEEproof}
By \eqref{eq:defmdimentropy}, we have
$\eh{m}(\rxv) = - \E_{\rxv}\big[\log  \theta_{\rxv}^m(\rxv) \big]$,
and by the weak law of large numbers, the sample mean $-(1/n) \sum_{i=1}^n\log \theta_{\rxv}^m(\rxv_i) $ converges in probability to the expectation $- \E_{\rxv}\big[\log  \theta_{\rxv}^m(\rxv) \big]$.
\end{IEEEproof}
We can  define typical sets in the usual way \cite[Sec.~8.2]{Cover91}.
\begin{definition}
Let $\rxv$ be an $m$-rectifiable random variable on $\R^M$ with support $\sE$ and $m$-dimensional Hausdorff density $\theta_{\rxv}^m$. 
For $\varepsilon>0$ and $n\in \N$, the $\varepsilon$-typical set $\sA_{\varepsilon}^{(n)}\subseteq \R^{nM}$ is defined as
\be
\sA_{\varepsilon}^{(n)} 
  \triangleq   \bigg\{\xv_{1:n} \in \sE^n   :     \bigg\lvert -\frac{1}{n} \sum_{i=1}^n\log \theta_{\rxv}^m(\xv_i) - \eh{m}(\rxv)\bigg\rvert\leq\varepsilon\bigg\}.
	\notag %\label{eq:deftypicalset}
\ee
\end{definition}

%Rev1
%Note that $\sA_{\varepsilon}^{(n)}\subseteq \sE^n$.
%The assumption $\xv_{1:n}\in \sE^n$ simplifies working with $\sA_{\varepsilon}^{(n)}$. 
%This is not a strong restriction because, by Property~\ref{en:cordensnull} in Corollary~\ref{cor:propsrectrandvar},  
%$\theta_{\rxv}^m(\xv)=0$ $\Hm{m}$-almost everywhere on $\sE^c$.

The AEP for sequences of $m$-rectifiable random variables is expressed by the following central result.
%Rev1.1:Omit proof just mention similarities to continuous case
\begin{theorem}\label{th:aep}
Let $\rxv_{1:n}=(\rxv_1,\dots, \rxv_n)$ be a sequence of i.i.d.\ $m$-rectifiable random variables $\rxv_i$ on $\R^M$, where each $\rxv_i$ has $m$-dimensional Hausdorff density $\theta_{\rxv}^m$, support $\sE$, and $m$-dimensional entropy $\eh{m}(\rxv)$. 
%Furthermore, let $\sE^{n}$ and $\sE$ be product-compatible  for all $n\in \N$.
Then the typical set $\sA_{\varepsilon}^{(n)}$ satisfies the following properties.
\begin{enumerate}
\item  \label{en:lln}For $\delta>0$ and $n$ sufficiently large,
\be \notag 
\Pr\{\rxv_{1:n}\in \sA_{\varepsilon}^{(n)}\}>1-\delta\,.
\ee
\item For all $n\in \N$, \label{en:aepupperbound}
\be\notag %\label{eq:aepupperbound}
\Hm{nm}(\sA_{\varepsilon}^{(n)})\leq e^{n(\eh{m}(\rxv)+\varepsilon)}\,.
\ee
\item For $\delta>0$ and $n$ sufficiently large, \label{en:aeplowerbound}
\be\notag %\label{eq:aeplowerbound}
\Hm{nm}(\sA_{\varepsilon}^{(n)})> (1-\delta)e^{n(\eh{m}(\rxv)-\varepsilon)}\,.
\ee
\end{enumerate}
\end{theorem}
\begin{IEEEproof}
The proof is similar to that
%% follows along the same lines as 
in the continuous case \cite[Th.~8.2.2]{Cover91}, however with the Lebesgue measure replaced by the Hausdorff measure.
%See Appendix~\ref{app:proofaep}.
\end{IEEEproof}

%
%%%%%%%%%%%%%%%%%%%%%%%%%%%%%%%%%
\section{Entropy Bounds on Expected Codeword Length} \label{sec:sourcecoding}
%%%%%%%%%%%%%%%%%%%%%%%%%%%%%%%%%
%
%Rev1.9: change to  instantaneous source code and not one-to-one
A well-known result for discrete random variables is a connection between the minimal expected codeword length of an instantaneous 
source code and the entropy of the random variable \cite[Th.~5.4.1]{Cover91}. 
More specifically, let $\rxv$ be a discrete random variable on $\R^M$ with  possible realizations $\{\xv_i:i\in \sI\}$.
In variable-length source coding, a one-to-one function $f\colon \{\xv_i:i\in \sI\}\to \{0,1\}^*$,  
where $\{0,1\}^*$ denotes the set of all finite-length binary sequences, is used to represent each  realization $\xv_i$ by a  finite-length binary sequence $\sv_i=f(\xv_i)$.
This code is instantaneous (or prefix free) if no $f(\xv_i)$ coincides with the first bits of another $f(\xv_j)$.
The expected binary codeword length is defined as
\be \notag 
L_{f}(\rxv)\triangleq \E_{\rxv}[\ell(f(\rxv))]
\ee
where $\ell(\sv)$ denotes the length of a binary sequence $\sv\in \{0,1\}^*$.
The minimal expected binary codeword length $L^*(\rxv)$ is defined as the minimum of $L_{f}(\rxv)$ over the set of all possible  instantaneous codes $f$.
By \cite[Th.~5.4.1]{Cover91}, $L^*(\rxv)$ satisfies%
\footnote{The factor $\ld e$ appears because we defined entropy using the natural logarithm.}
\be\label{eq:sourcebounddiscrete}
H(\rxv) \ld e\leq L^*(\rxv) <H(\rxv) \ld e +1\,.
\ee

\subsection{Expected Codeword Length of an Integer-Dimen\-sional Random Variable}
\label{sec:sourcecoding_length}

For a nondiscrete $m$-rectifiable random variable $\rxv$ (i.e., $m\geq 1$), a one-to-one code of finite expected codeword length does not exist. 
However, quantizations of $\rxv$ can be encoded using finite-length binary sequences.
We will present results for the minimal expected codeword length of constrained quantizations of $\rxv$.
\begin{definition}\label{def:mdpartition}
Let $\sE\subseteq \R^M$ be an $m$-rectifiable set.
Furthermore, let $\mathfrak{Q}=\{\sA_1, \dots, \sA_N\}$ be a finite $\Hm{m}$-measurable partition  of $\sE$, i.e., all sets $\sA_i$ are mutually disjoint and $\Hm{m}$-measurable, and $\bigcup_{i=1}^N \sA_i=\sE$.
The partition $\mathfrak{Q}$ is said to be an \textit{$(m,\delta)$-partition of $\sE$} if
$\Hm{m}(\sA_i)\leq \delta$ for all $i\in \{1, \dots, N\}$.
The set of all $(m,\delta)$-partitions of $\sE$ is denoted $\mathfrak{P}_{m,\delta}^{(\sE)}$.
\end{definition}
Note that the definition of an $(m,\delta)$-partition of an $m$-rectifiable set $\sE$ does not involve a distortion function. 
On the one hand, this is convenient because we do not have to argue about a good distortion measure. 
On the other hand, the points in a set $\sA_i$ of a partition $\mathfrak{Q}\in\mathfrak{P}_{m,\delta}^{(\sE)}$ are not necessarily ``close'' to each other;
in fact, $\sA_i$ is not even necessarily connected. 
Thus, although the partitions in $\mathfrak{P}_{m,\delta}^{(\sE)}$ consist of measure-theoretically small sets, these sets might be considered large in terms of specific distortion measures.

In what follows, we will consider the quantized random variable $[\rxv]_{\mathfrak{Q}}$ for $\mathfrak{Q}\in\mathfrak{P}_{m,\delta}^{(\sE)}$.
We recall that  $[\rxv]_{\mathfrak{Q}}$ is the discrete random variable such that $\Pr\{[\rxv]_{\mathfrak{Q}}=i\}=\Pr\{\rxv\in \sA_i\}$ for $i\in \{1, \dots, N\}$.
Due to the interpretation of $\eh{m}(\rxv)$  as a generalized entropy (cf.\ Remark~\ref{rem:genent}), we can use \cite[eq.\ (1.8)]{cs73} to obtain the following result.
%
%?EARLIER?
%\begin{definition}[\mbox{\cite[eq.~(1.5)]{cs73}}]\label{def:genent}
%Let $\lambda$ be a $\sigma$-finite measure on $\R^M$. For a probability measure $\mu$ on $\R^M$, the generalized entropy of $\mu$ with respect to $\lambda$ is defined as
%\be
%H_{\lambda}(\mu)\triangleq 
%\begin{cases}
%- \displaystyle\int_{\R^{M}}\log \bigg(\frac{\mathrm{d}\mu}{\mathrm{d}\lambda}(\xv)\bigg) \, \mathrm{d}\mu(\xv) & \text{if }\mu \ll  \lambda \\
%\infty &  \text{else.}
%\end{cases}
%\ee
%\end{definition}
%We see that our entropy definition (Definition~\ref{def:mdimentropy}) coincides with Definition \ref{def:genent} for the choice $\lambda=\Hm{m}|_{\sE}$.
%Thus, we can use a basic result  to obtain the following lemma.}
%We first present an expression of the $m$-dimensional entropy of an $m$-rectifiable random variable $\rxv$ as the infimum of the entropy of quantizations  $[\rxv]_{\mathfrak{Q}}$.
%This expression 
%\cha{is based on an interpretation of $\eh{m}(\rxv)$ as a generalized entropy as described in \cite{cs73} and} will be used in the proof of Theorem~\ref{th:sourcelowerupperbound} further below.
\begin{lemma}\label{lem:ourentasinf}
Let $\rxv$ be an $m$-rectifiable random variable, i.e., $\mu\rxv^{-1}\ll \Hm{m}|_{\sE}$ for an $m$-rectifiable set $\sE\subseteq \R^M$,  with $m\geq 1$ and   $\Hm{m}(\sE)<\infty$. 
Let $\mathfrak{P}_{m,\infty}^{(\sE)}$ denote the set of all finite, $\Hm{m}$-measurable partitions of $\sE$.
Then
\ba
& \eh{m}(\rxv) \notag \\
& \rule{3mm}{0mm} = \inf_{\mathfrak{Q}\in \mathfrak{P}_{m,\infty}^{(\sE)}} \!\Bigg(-\sum_{\sA\in \mathfrak{Q}} \mu\rxv^{-1}(\sA) \log \bigg(\frac{\mu\rxv^{-1}(\sA)}{\Hm{m}|_{\sE}(\sA)}\bigg) \Bigg)\label{eq:ourentasinffirst} \\
& \rule{3mm}{0mm} = \inf_{\mathfrak{Q}\in \mathfrak{P}_{m,\infty}^{(\sE)}}\! \bigg(H([\rxv]_{\mathfrak{Q}})+ \sum_{\sA\in \mathfrak{Q}} \mu\rxv^{-1}(\sA) \log \Hm{m}|_{\sE}(\sA) \bigg)\label{eq:ourentasinf}
\,.
\ea
\end{lemma}
\begin{IEEEproof}
See Appendix~\ref{app:ourentasinfproofs}.
\end{IEEEproof}
The terms in \eqref{eq:ourentasinf} give an interesting interpretation of $m$-dimensional entropy. 
Looking for a quantization that minimizes the first term, $H([\rxv]_{\mathfrak{Q}})$, corresponds to minimizing the amount of data required to represent this quantization.
Of course, the minimum is simply obtained for the partition $\mathfrak{Q}=\{\sE\}$, which gives $H([\rxv]_{\mathfrak{Q}})=0$.
But in \eqref{eq:ourentasinf}, we also have an additional  term that penalizes a bad ``resolution'' of the quantization:
if the quantized random variable $[\rxv]_{\mathfrak{Q}}$ is 
with high probability---corresponding to $\mu\rxv^{-1}(\sA)$ being large---in a large quantization set $\sA$, then this is penalized by the term $ \mu\rxv^{-1}(\sA) \log \Hm{m}|_{\sE}(\sA)$.
Thus,  \eqref{eq:ourentasinf} shows that $m$-dimensional entropy can be interpreted in terms of a tradeoff between fine resolution and efficient representation.

We now turn to a generalization of \eqref{eq:sourcebounddiscrete} to rectifiable random variables.
%We first present a lower bound on the expected codeword length of any quantization $[\rxv]_{\mathfrak{Q}}$ of an $m$-rectifiable random variable with $\mathfrak{Q}\in \mathfrak{P}_{m,\delta}^{(\sE)}$.
\begin{theorem}\label{th:sourcelowerupperbound}
Let $\rxv$ be an $m$-rectifiable random variable, i.e., $\mu\rxv^{-1}\ll \Hm{m}|_{\sE}$ for an $m$-rectifiable set $\sE\subseteq \R^M$, with $m\geq 1$ and  $\Hm{m}(\sE)<\infty$. 
For any $\mathfrak{Q}\in \mathfrak{P}_{m,\delta}^{(\sE)}$, the minimal expected binary codeword length of the quantized random variable $[\rxv]_{\mathfrak{Q}}$ satisfies
\be\label{eq:sourcelowerbound}
L^*([\rxv]_{\mathfrak{Q}}) \geq \eh{m}(\rxv)\ld e -\ld \delta \,.
\ee
Furthermore, for each $\varepsilon > 0$, there exists $\delta_{\varepsilon}>0$ such that the following holds: for each $\delta\in (0,\delta_{\varepsilon})$, there exists a partition $\mathfrak{Q}_{\delta}\in \mathfrak{P}_{m,\delta}^{(\sE)}$ such that
\be\label{eq:sourceupperboundspecial}
 L^*([\rxv]_{\mathfrak{Q}_{\delta}})< \eh{m}(\rxv)\ld e -\ld \delta + 1 +\varepsilon\,.
\ee
%For each $\delta\in (0,\delta_{\varepsilon})$ there exists a partition $\mathfrak{Q}_{\delta}\in \mathfrak{P}_{m,\delta}^{(\sE)}$ such that
%\be\label{eq:sourceupperbound}
 %L^*([\rxv]_{\mathfrak{Q}_{\delta}})< \eh{m}(\rxv)\ld e -\ld \delta + 3 + \varepsilon\,.
%\ee
\end{theorem}
\begin{IEEEproof}
See Appendix~\ref{app:sourcecodingproofs}.
We note that the proof is based on \eqref{eq:sourcebounddiscrete} and the expression of $\eh{m}(\rxv)$ given in \eqref{eq:ourentasinf}.
\end{IEEEproof}
The lower bound \eqref{eq:sourcelowerbound} shows  the following: 
if we want a quantization $\mathfrak{Q}$ of $\rxv$ with   good resolution (in the sense that $\Hm{m}(\sA)\leq \delta$ for all $\sA\in \mathfrak{Q}$), then we have to use \emph{at least} $\eh{m}(\rxv)\ld e -\ld \delta$ bits to represent this quantized random variable using an instantaneous code. %Rev1.9: added instantaneous code
However, by the upper bound~\eqref{eq:sourceupperboundspecial}, we know that for a sufficiently fine resolution (i.e., $\delta<\delta_{\varepsilon}$), that resolution $\delta$ can be achieved by using  \emph{at most} $1+\varepsilon$ additional bits (in addition to the lower bound $\eh{m}(\rxv)\ld e -\ld \delta$).

\subsection{Expected Codeword Length of Sequences of Integer-Dimen\-sional Random Variables}

We will now apply Theorem~\ref{th:sourcelowerupperbound}  to sequences of i.i.d. random variables.
To this end, we consider quantizations of an entire sequence, $[\rxv_{1:n}]_{\mathfrak{Q}}=[(\rxv_1,\dots, \rxv_n)]_{\mathfrak{Q}}$ with%
\footnote{We choose partitions $\mathfrak{Q}$ of resolution $\delta^n$, i.e., the sets $\sA\in \mathfrak{Q}$ satisfy $\Hm{nm}(\sA)\leq \delta^n$.
This choice is made for consistency with the case of  partitions $\mathfrak{Q}$ of $\sE^n$ that are constructed as products of sets $\sA_i$ in $\mathfrak{Q}_1\in \mathfrak{P}_{m,\delta}^{(\sE)}$.
More specifically, for $\sA=\sA_1\times \dots \times \sA_n$ with $\sA_i \in \mathfrak{Q}_1$, we have $\Hm{m}(\sA_i)\leq \delta$ and $\Hm{nm}(\sA)\leq \delta^n$ and the sets $\sA$ cover $\sE^n$, i.e., $\mathfrak{Q}\triangleq \{\sA=\sA_1\times \dots \times \sA_n: \sA_i\in \mathfrak{Q}_1\}\in \mathfrak{P}_{nm,\delta^n}^{(\sE^n)}$.} 
$\mathfrak{Q}\in \mathfrak{P}_{nm,\delta^n}^{(\sE^n)}$.
We denote by 
\be\label{eq:mebclpss}
L^*_n([\rxv_{1:n}]_{\mathfrak{Q}})\triangleq \frac{L^*([\rxv_{1:n}]_{\mathfrak{Q}})}{n}
\ee
 the minimal expected binary codeword length per source symbol.
\begin{corollary}\label{cor:codeseq}
Let $\rxv_{1:n}=(\rxv_1,\dots, \rxv_n)$ be a sequence of i.i.d.\ $m$-rectifiable random variables ($m\geq 1$) on $\R^M$ with  $m$-dimensional entropy $\eh{m}(\rxv)$ and support $\sE$ satisfying $\Hm{m}(\sE)<\infty$.
%Furthermore, assume that $\sE^i$ and $\sE$ are product-compatible for $i\in \{1, \dots, n-1\}$.
Then, for each $\varepsilon > 0$, there exists $\delta_{\varepsilon}>0$ such that the following holds: 
for each $\delta\in (0,\delta_{\varepsilon})$, there exists a partition $\mathfrak{Q}\in \mathfrak{P}_{nm,\delta^n}^{(\sE^n)}$ such that the minimal expected binary codeword length per source symbol satisfies
\be\label{eq:sourceboundjoint}
\eh{m}(\rxv)\ld e -\ld \delta 
  \leq \! L^*_n([\rxv_{1:n}]_{\mathfrak{Q}}) \!
 \leq \eh{m}(\rxv)\ld e -\ld \delta + \frac{1+\varepsilon}{n_{}}.
\ee
%\ba\label{eq:sourceboundjoint}
%\eh{m}(\rxv)\ld e -\ld \delta 
%& \leq L^*_n([\rxv_{1:n}]_{\mathfrak{Q}}) \notag \\
%& \leq \eh{m}(\rxv)\ld e -\ld \delta + \frac{1+\varepsilon}{n}\,.
%\ea
%Here, $L^*_n([\rxv_{1:n}]_{\mathfrak{Q}})\triangleq L^*([\rxv_{1:n}]_{\mathfrak{Q}})/n$ denotes the minimal expected binary codeword length per symbol.
\end{corollary}
\begin{IEEEproof}
By Corollary~\ref{cor:independentidentseq}, the random variable $\rxv_{1:n}$ is $nm$-rectifiable with $\mu(\rxv_{1:n})^{-1}\ll \Hm{m}|_{\sE^n}$ and $nm$-di\-men\-sion\-al entropy $\eh{nm}(\rxv_{1:n})=n\eh{m}(\rxv)$.
Thus, by Theorem~\ref{th:sourcelowerupperbound}, there exists %for $\varepsilon > 0$ 
$\hat{\delta}_{\varepsilon}>0$ such that the following holds:
\begin{enumerate}
\renewcommand{\theenumi}{($*$)}
\renewcommand{\labelenumi}{($*$)\hspace{-5mm}}
\item \label{en:propdelteps}\hspace{5mm}For all $\hat{\delta}\in (0,\hat{\delta}_{\varepsilon})$, there exists a partition $\mathfrak{Q}\in \mathfrak{P}_{nm,\hat{\delta}}^{(\sE^n)}$ such that
\ba  %\label{eq:applysourceboundtriv}
n\eh{m}(\rxv)\ld e -\ld \hat{\delta} 
& \leq L^*([\rxv_{1:n}]_{\mathfrak{Q}}) \notag \\
& < n\eh{m}(\rxv)\ld e -\ld \hat{\delta} + 1 + \varepsilon\,.\notag
\ea
\end{enumerate}
%By \eqref{eq:indepprodentropy}, we have $\eh{nm}(\rxv_{1:n})=n \eh{m}(\rxv)$.
Define $\delta_{\varepsilon}\triangleq \hat{\delta}_{\varepsilon}^{1/n}$ and let $\delta \in (0, \delta_{\varepsilon})$.
We have that $\delta \in (0, \delta_{\varepsilon})$ is equivalent to  $\delta^n \in (0, \hat{\delta}_{\varepsilon})$.
Thus, by \ref{en:propdelteps} for the specific case $\hat{\delta}=\delta^n$, there exists  a partition $\mathfrak{Q}\in \mathfrak{P}_{nm,\delta^n}^{(\sE^n)}$ such that
\ba 
n \eh{m}(\rxv)\ld e - \ld \delta^n 
& \leq L^*([\rxv_{1:n}]_{\mathfrak{Q}}) \notag \\
& < n \eh{m}(\rxv)\ld e - \ld \delta^n + 1 + \varepsilon\,.\notag 
\ea
Dividing by $n$ and using \eqref{eq:mebclpss} gives \eqref{eq:sourceboundjoint}.
\end{IEEEproof}
Corollary~\ref{cor:codeseq} shows that the upper bound on the expected codeword length per source symbol becomes closer to the lower bound $\eh{m}(\rxv)\ld e -\ld \delta$ if we are allowed to quantize and code entire sequences. 
However, note that using the quantization $\mathfrak{Q}\in \mathfrak{P}_{nm,\delta^n}^{(\sE^n)}$ of the joint random variable $\rxv_{1:n}$, it is not guaranteed that we can reconstruct each  $\rxv_i$ to within a set $\sA_i$ satisfying $\Hm{m}(\sA_i)\leq \delta$. 
All we know is that each $\sA\in \mathfrak{Q}$ satisfies $\Hm{nm}(\sA)\leq \delta^n$, i.e., the overall resolution of the sequence is good, but   the resolution of each individual source symbol is not necessarily good too.

%%%%%%%%%%%%%%%%%%%%%%%%%%%%%%%
\section{Shannon Lower Bound for Integer-Dimensional Sources} \label{sec:ratedistortion}   
%%%%%%%%%%%%%%%%%%%%%%%%%%%%%%%
%In this section we present lower bounds on the rate distortion function based on the lower bounds given in \cite{csiszar74} in terms of the new entropy and maybe entropy maximizing measures.
As a second application of the proposed entropy definition, we present a lower bound on the rate-distortion (RD) function of integer-dimensional sources.
%The problem considered in rate-distortion (RD) theory is to represent a given random variable $\rxv$ %by a discrete random variable $\ryv$ 
%using as few values as possible while keeping the expected distortion below some threshold~\cite[Ch.~4]{gr90}.
The RD function for a source $\rxv$ and a distortion function $d(\cdot,\cdot)$ is defined as~\cite[eq.~(4.1.3)]{gr90}
\be \notag %\label{eq:defrd}
R(D) \triangleq \inf_{\E_{(\rxv,\ryv)}[d(\rxv,\ryv)] \leq D} I(\rxv;\ryv)
\ee
for $D\geq 0$,
where the  constrained infimum is taken over all  joint probability distributions of $(\rxv,\ryv)$ with the given probability distribution of $\rxv$ as the first marginal. 
We will consider throughout this section a source random variable $\rxv$ on $\R^M$ and
%Let $d(\cdot,\cdot)$ denote 
a translation invariant distortion function $d(\cdot,\cdot)$ on $\R^M \times \R^M$, i.e., $d(\xv, \yv)=d(\xv-\yv, \0v)$ for all $\xv, \yv\in \R^M$.
Furthermore, we assume that $d(\cdot,\cdot)$
satisfies
%\ba \notag 
$\inf_{\yv\in \R^M}d(\xv,\yv)=0$
%\ea
for each $\xv\in \R^M$.
%By the source coding theorem \cite[Sec.~9.6]{Gallager68}, the RD function characterizes the 
We also assume that there exist $D\geq 0$ such that $R(D)$ is finite, and we denote by $D_0$ the infimum of these $D$.
Finally, we assume that there exists a finite set $\sB\subseteq \R^M$ such that 
$
\E_{\rxv}\big[\min_{\yv\in \sB}d(\rxv,\yv)\big]<\infty$.
This assumption guarantees that there exists a finite quantization of $\rxv$ with bounded expected distortion.
Under these standard assumptions, we have the following 
characterization of the RD function \cite[Th.~2.3]{csiszar74}:
%\begin{theorem}[\mbox{\cite[Th.~2.3]{csiszar74}}]\label{th:csiszarrd}
For each $D>D_0$,
\be\label{eq:rdcharcs}
R(D)= \max_{s\geq 0}\max_{\alpha_s(\cdot)}\big(\!-sD + \E_{\rxv}[\log \alpha_s(\rxv)]\big)
\ee
where the second maximization is with respect to all functions%
%Rev1.10: changed footnote to reference to (1.23)
\footnote{Although in \cite[Th.~2.3]{csiszar74} $\alpha_s(\xv)\geq 1$ is assumed, \eqref{eq:rdcharcs} also holds for $\alpha_s(\xv)>0$ because of \cite[eq.\ (1.23)]{csiszar74}.} 
$\alpha_s\colon \R^M\to (0,\infty)$ satisfying
\be\label{eq:condonalpha}
\E_{\rxv}\big[\alpha_s(\rxv)e^{-s d(\rxv,\yv)}\big]\leq 1
\ee
for each $\yv\in \R^M$.
%\end{theorem}

\subsection{Shannon Lower Bound}
\label{sec:ratedistortion_bound}

The most common form of the traditional Shannon lower bound \cite[Sec.~4.3]{gr90} for a \emph{discrete} source $\rxv$ is the following inequality
\be\label{eq:tradshlb}
R(D)\geq H(\rxv)-\max H(\rwv)
\ee
where the maximum is taken over all discrete random variables $\rwv$ whose expected distortion relative to $\0v$ is equal to $D$, i.e., $\E_{\rwv}\big[d(\rwv,\0v)\big]=D$.
An important aspect of the bound~\eqref{eq:tradshlb} is that the contribution of the source $\rxv$ and the contribution of the distortion function $d(\cdot, \cdot)$  and distortion $D$ become  separated. 
For a fixed distortion function and a given distortion, we can calculate $\max H(\rwv)$ and then use the bound~\eqref{eq:tradshlb} for  different sources $\rxv$ simply by calculating their entropy $H(\rxv)$.

For a \emph{continuous} random variable $\rxv$ on $\R^M$, a bound similar to~\eqref{eq:tradshlb} can also be derived under certain assumptions.
However, it is more convenient to state the continuous Shannon lower bound in the following parametric form (i.e., involving a parameter $s\geq 0$) \cite[Sec.~4.6]{gr90}
\be\label{eq:contshlb}
R(D)\geq h(\rxv)-sD - \log \widetilde{\gamma}(s)
\ee
where
\be\label{eq:gammascont}
\widetilde{\gamma}(s)\triangleq \int_{\R^M} e^{-s d(\xv,\0v)}\, \mathrm{d}\Leb^M(\xv)
\ee
and \eqref{eq:contshlb} holds for all $s\geq 0$.
The right-hand side of  \eqref{eq:contshlb} can be maximized with respect to $s$, and it turns out that \cite[Lem.~4.6.2]{gr90} %under certain conditions 
\be \notag 
\min_{s\geq 0} \big(sD+\log \widetilde{\gamma}(s)\big)= \max h(\rwv)
\ee
where the maximum is taken over all continuous random variables $\rwv$ such that $\E_{\rwv}\big[d(\rwv,\0v)\big]=D$.
This results again in the simple formula (cf.~\eqref{eq:tradshlb})
\be \notag 
R(D)\geq h(\rxv)- \max h(\rwv)\,.
\ee
Because the parametric bound \eqref{eq:contshlb} is more convenient in most cases and already allows us to separate the source from the distortion, we will concentrate on a generalization of \eqref{eq:contshlb} to rectifiable random variables.
%
%To establish a lower bound on the RD function of rectifiable random variables, 
To this end, we will use the characterization of the RD function in \eqref{eq:rdcharcs} with a specific choice of the function $\alpha_s$. 
%The proof follows along the same lines as the proofs of the traditional Shannon lower bounds for discrete \cite[Ch.~4.3]{gr90} and continuous \cite[Ch.~4.6]{gr90} random variables.

\begin{theorem}\label{th:shannonlb}
The RD function of an $m$-rectifiable random variable $\rxv$ on $\R^M$ with support $\sE$ is lower bounded by
\be\label{eq:slboundrd} 
R(D)\geq R_{\text{SLB}}(D,s)\triangleq\eh{m}(\rxv)- sD -\log \gamma(s) 
\ee
for each $s\geq 0$, where 
\be\label{eq:defyvt}
\gamma(s)\triangleq \sup_{\yv\in \R^M} \int_{\sE} e^{-s d(\xv,\yv)}\, \mathrm{d}\Hm{m}(\xv), \qquad  s\geq 0 \,.
\ee
\end{theorem}
\begin{IEEEproof}
%Denote by $\yvt(s)$ any $\yv$ taking on the maximum in  \eqref{eq:defyvt}.
We start by noting that \eqref{eq:defyvt} implies
\be\label{eq:defyvtimpy}
\int_{\sE} e^{-s d(\xv,\yv)}\, \mathrm{d}\Hm{m}(\xv) \leq  \gamma(s)
\ee
for all $\yv\in \R^M$.
Let $s\geq 0$ be fixed.
By \eqref{eq:rdcharcs}, % Theorem~\ref{th:csiszarrd}, 
\be\label{eq:rdboundspecs}
R(D)\geq -sD + \E_{\rxv}[\log \alpha_s(\rxv)]
\ee
for every function $\alpha_s$  satisfying \eqref{eq:condonalpha}.
We have (cf.~\eqref{eq:corexpectation})
\ba %\label{eq:condgamma}
& \E_{\rxv}\bigg[\frac{1}{\theta_{\rxv}^m(\rxv)\gamma(s)}e^{-s d(\rxv,\yv)}\bigg]\notag \\
& \rule{15mm}{0mm} = \int_{\sE}\frac{1}{\theta_{\rxv}^m(\xv)\gamma(s)} e^{-s d(\xv,\yv)} \theta_{\rxv}^m(\xv)\, \mathrm{d}\Hm{m}(\xv) \notag \\[1mm]
&  \rule{15mm}{0mm}= \frac{1}{\gamma(s)} \int_{\sE} e^{-s d(\xv,\yv)}\, \mathrm{d}\Hm{m}(\xv)  \notag  \\[1mm]
&  \rule{15mm}{0mm}\stackrel{\hidewidth  \eqref{eq:defyvtimpy} \hidewidth }\leq \;\frac{\gamma(s)}{\gamma(s)} \notag  \\
&  \rule{15mm}{0mm}= \, 1 \notag %
\ea
 for all $\yv\in \R^M$.
Therefore, the choice  $\alpha_s(\xv)\triangleq \frac{1}{\theta_{\rxv}^m(\xv)\gamma(s)}$ satisfies  \eqref{eq:condonalpha}. 
Inserting $\alpha_s(\xv)=\frac{1}{\theta_{\rxv}^m(\xv)\gamma(s)}$ into \eqref{eq:rdboundspecs}, we obtain
\ba %\label{eq:rdasmax}
R(D) & \geq  -sD + \E_{\rxv}\bigg[\log \frac{1}{\theta_{\rxv}^m(\rxv)\gamma(s)}\bigg] \notag \\[1mm]
& = -\E_{\rxv}[\log \theta_{\rxv}^m(\rxv) ]-sD -\E_{\rxv}[\log \gamma(s) ]  \notag \\[1mm]
%& = \eh{m}(\rxv)-sD +\E_{\rxv}[\log(\gamma(s))]  \notag \\
& = \eh{m}(\rxv)-sD -\log \gamma(s) \,. %\notag \\[-10mm]
 \notag  
\ea
\end{IEEEproof}
For a continuous random variable $\rxv$ with positive probability density function almost everywhere (i.e., $M$-rectifiable with support $\R^M$), the definitions of $\widetilde{\gamma}(s)$ in \eqref{eq:gammascont} and $\gamma(s)$ in \eqref{eq:defyvt} coincide. 
Indeed, because $d(\xv, \yv)=d(\xv-\yv, \0v)$ and a translation of the integrand by $\yv$ does not change the value of the integral over $\R^M$, the right-hand side of \eqref{eq:gammascont} can be written as (recall that $\Hm{M}=\Leb^M$)
\be \label{eq:gammaeqgammat} 
\int_{\R^M} e^{-s d(\xv,\0v)}\, \mathrm{d}\Leb^M(\xv)
= \int_{\R^M} e^{-s d(\xv,\yv)}\, \mathrm{d}\Hm{M}(\xv)
\ee
for any $\yv\in \R^M$. 
Because the left-hand side of \eqref{eq:gammaeqgammat} does not depend on $\yv$, taking the supremum over $\yv\in \R^M$ in \eqref{eq:gammaeqgammat} results in 
\be \notag %\label{eq:gammaeqgammat} 
\int_{\R^M} \!e^{-s d(\xv,\0v)}  \mathrm{d}\Leb^M(\xv)
= \sup_{\yv\in \R^M}\int_{\R^M} \!e^{-s d(\xv,\yv)}  \mathrm{d}\Hm{M}(\xv)
\ee
which is \eqref{eq:defyvt}.
Thus, for a continuous random variable $\rxv$ with positive probability density function almost everywhere, the Shannon lower bounds~\eqref{eq:contshlb} and \eqref{eq:slboundrd} coincide.
However, for a continuous random variable $\rxv$ whose support $\sE$ is a proper subset of $\R^M$ we have $\gamma(s)\leq \widetilde{\gamma}(s)$, and thus the Shannon lower bound \eqref{eq:slboundrd} is tighter (i.e., larger) than \eqref{eq:contshlb}.
This is due to the fact that \eqref{eq:slboundrd} incorporates the additional information that the random variable is restricted to $\sE$.

\subsection{Maximizing the Shannon Lower Bound}\label{sec:maxslb}

\begin{figure}[t]
\centering
\begin{tikzpicture}
\begin{axis}[
width=8cm,
xmin=0,xmax=6,
ymin=0,ymax=5.7,
%title=Convergence Plot,
xlabel={$D$},
ylabel={$R$},
ylabel style={rotate=-90},
tick label style={font=\small},
grid=major, 
legend entries={{$\sF_1$}, {$\sF_2$}, $\bar{\sF}$},
legend style={
legend pos=north east, font=\small
}
]
\addplot+[red!50, no markers, line width=1pt] table {upenv1.dat}; %chigen
\addplot+[blue!30, dotted, no markers, line width=1pt] table {upenv2.dat}; %chigen
\addplot+[black, no markers,dashed,  line width=1.5pt] table {upenv3.dat}; %chigen
\addplot+[black, no markers,dashed,  line width=1.5pt] table {upenv4.dat}; %chigen
\addplot+[black, no markers,dashed,  line width=1.5pt] table {upenv5.dat}; %chigen
%\addplot+[red, name path=A1, no markers] table {datachiratiorestricted100.dat}; %chigen
%\addplot+[black, name path=B1, no markers] table {datachiratiorestricted10.dat}; %chigen
%\addplot+[red, name path=A2, no markers] table {datachiratiorestrictedup100.dat}; %chigen
%\addplot+[black, name path=B2, no markers] table {datachiratiorestrictedup10.dat}; %chigen
%\addplot[pattern=north west lines, pattern color=red!50] fill between[of=A1 and A2];
%\addplot[pattern=north east lines] fill between[of=B1 and B2];
\end{axis}
\end{tikzpicture}
\vspace{-3mm}
\caption{Illustration of the sets $\sF_1$, $\sF_2$, and $\bar{\sF}$ (assuming $D_0=1$).}
\label{fig:upenvelope}
\end{figure}
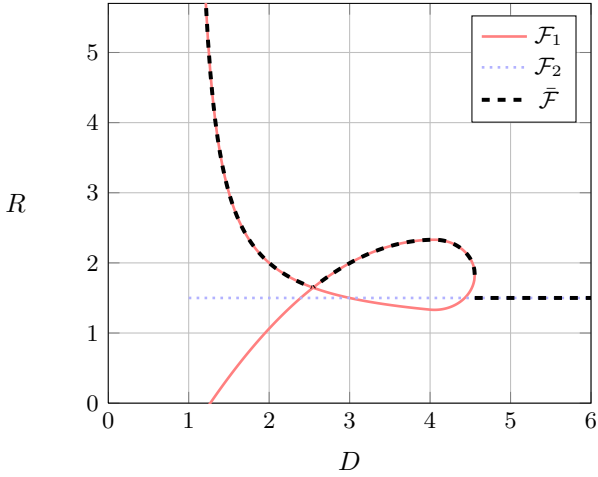

The optimal choice of $s$ in \eqref{eq:slboundrd} depends on  $D$ and is hard to find in general.
At least, the following lemma states that the optimal (i.e., largest) lower bound  in \eqref{eq:slboundrd},
\be\notag %\label{eq:largestslb}
R^*_{\text{SLB}}(D)\triangleq \, \sup_{s\geq 0} R_{\text{SLB}}(D,s)
\ee
  is achieved for a finite $s$.
We recall that $D_0$ is the infimum of all $D\geq 0$ such that $R(D)$ is finite.
\begin{lemma}\label{lem:rslbtoinfty}
Let $\rxv$ be an $m$-rectifiable random variable with support $\sE$ and finite $m$-dimensional entropy $\eh{m}(\rxv)$.
Then for $D>D_0$ the lower bound $R_{\text{SLB}}(D,s)$ in \eqref{eq:slboundrd} satisfies 
\be\notag %\label{eq:rslbtoinfty}
\lim_{s\to \infty}R_{\text{SLB}}(D,s)=-\infty\,.
\ee
\end{lemma}
\begin{IEEEproof}
See Appendix~\ref{app:rslbtoinfty}.
\end{IEEEproof}
If $R_{\text{SLB}}(D,s)$ is a continuous function of $s$, Lemma~\ref{lem:rslbtoinfty} implies that for a fixed $D>D_0$,  the global maximum of $R_{\text{SLB}}(D,s)$ with respect to $s$ exists and is either a local maximum or the boundary point $s=0$, i.e., $R^*_{\text{SLB}}(D)=R_{\text{SLB}}(D,s)$ for some finite $s\geq 0$.
Moreover, if $\gamma(s)$ in  \eqref{eq:defyvt} is differentiable, we can characterize  the local maxima of $R_{\text{SLB}}(D,s)$ as follows.
\begin{theorem}\label{th:gammadiff}
Let $\rxv$ be an $m$-rectifiable random variable with support $\sE$, and let 
$\gamma(s)$ be differentiable.
Then for $D>D_0$, the lower bound $R_{\text{SLB}}(D,s)$ in \eqref{eq:slboundrd}  is maximized either for $s=0$ or for some $s>0$ satisfying $\widetilde{D}(s)= D$, where
\be\notag%\label{eq:defoptd}
\widetilde{D}(s)\triangleq -\frac{\gamma'(s)}{\gamma(s)}\,.
\ee
That is, the largest lower bound is given by
\be\label{eq:charrstar}
R^*_{\text{SLB}}(D)=\max\bigg\{R_{\text{SLB}}(D,0), \!\!\sup_{s>0: \widetilde{D}(s)=D}\!\!R_{\text{SLB}}(D,s)\bigg\}\,.
\ee
%\item $R_{\text{SLB}}(D,s)$ is monotonically increasing in $s$.
%\end{enumerate}
\end{theorem}
\begin{IEEEproof}
We recall from \eqref{eq:slboundrd} that $R_{\text{SLB}}(D,s)=\eh{m}(\rxv)- sD -\log \gamma(s)$.
Thus, because $\gamma(s)$ is differentiable,  a necessary condition for a local maximum of $R_{\text{SLB}}(D,s)$ with respect to $s$ is obtained by
setting to zero the derivative of
 $R_{\text{SLB}}(D,s)$ with respect to $s$.
Solving the resulting equation for $D$ yields $\widetilde{D}(s)= D$.
%\be\label{eq:defoptd}
%D= -\frac{\gamma'(s)}{\gamma(s)}\,.
%\ee
Thus, for a given $D>D_0$,
 $R_{\text{SLB}}(D,s)$ can only have a local maximum at $s \in (0,\infty)$ satisfying $\widetilde{D}(s)= D$.
By Lemma~\ref{lem:rslbtoinfty}, the global maximum either is a local maximum or is achieved for $s=0$, which concludes the proof.
\end{IEEEproof}
If $\gamma(s)$ is differentiable, Theorem~\ref{th:gammadiff} provides a  ``parame\-tri\-za\-tion''  of the graph of the largest bound $R^*_{\text{SLB}}(D)$, i.e., we can characterize the set
\be\label{eq:defsg}
\sG\triangleq \big\{\big(D, R^*_{\text{SLB}}(D)\big)\in \R^2: D>D_0\big\}\,.
\ee
% of all pairs $(D, R^*_{\text{SLB}}(D))$ for $D>D_0$.
As a basis for this characterization, we define the sets 
\ba%\label{eq:graphoptshlb1}
\sF_1 & \triangleq
\big\{\big(\widetilde{D}(s), R_{\text{SLB}}(\widetilde{D}(s),s)\big): s>0\big\}
\notag \\[1mm]
\sF_2 & \triangleq \big\{\big(D, \eh{m}(\rxv)-  \log \Hm{m}(\sE)\big):D>D_0\big\}
\label{eq:graphoptshlb2}
\ea
which are illustrated in Fig.~\ref{fig:upenvelope}.
Note that $\sF_1$ is not necessarily the graph of a function, whereas $\sF_2$ constitutes a horizontal line in the $(D,R)$ plane.
\begin{corollary}\label{cor:graphoptshlb}
Let $\rxv$ be an $m$-rectifiable random variable with support $\sE$, and let  $\gamma(s)$ be differentiable.
Define $\sF\triangleq \sF_1\cup \sF_2$.
Then $\sG=\bar{\sF}$, where $\bar{\sF}$ is the upper envelope of $\sF$ given by
\be\label{eq:upenveqg}
\bar{\sF}\triangleq \bigg\{(D, R)\in \sF: R=\max_{(D,R')\in \sF}R'\bigg\}\,.
\ee
\end{corollary}
\begin{IEEEproof}
All elements $(D, R)\in  \sF$ can be written as $(D, R)=\big(D, R_{\text{SLB}}(D,s)\big)$ for some $s\geq 0$. 
Indeed, for $(D, R)\in  \sF_1$ this is obvious, and for $(D, R)\in  \sF_2$ we have 
\be
R  = \eh{m}(\rxv)- \log \Hm{m}(\sE) 
 \stackrel{(a)}=  \eh{m}(\rxv)- \log \gamma(0)
 \stackrel{\eqref{eq:slboundrd}} = 
R_{\text{SLB}}(D,0) \label{eq:drgraphforszero}
\ee
where $(a)$ holds because $\gamma(0)\stackrel{\eqref{eq:defyvt}}=\int_{\sE} 1\, \mathrm{d}\Hm{m}(\xv)=\Hm{m}(\sE)$. 
Hence, for all $(D, R)\in  \sF$, we obtain 
\be\label{eq:rprimeleq}
R\leq \sup_{s\geq 0} R_{\text{SLB}}(D,s)= R^*_{\text{SLB}}(D)\,.
\ee
Because $\bar{\sF}\subseteq \sF$, \eqref{eq:rprimeleq} also holds for $(D, R)\in \bar{\sF}$.

Consider now the pair  $(D, R)\in \bar{\sF}$ for a fixed $D>D_0$.
By~\eqref{eq:upenveqg}, for a pair $(D,R')\in \sF$ we obtain $R\geq R'$.
In particular, for $s>0$ satisfying $\widetilde{D}(s)=D$, the pair $\big(D,R_{\text{SLB}}(D,s)\big)$ belongs to $\sF_1\subseteq \sF$, and thus
\be\label{eq:rprime1}
R\geq R_{\text{SLB}}(D,s)\,.
\ee
Similarly, $\big(D, \eh{m}(\rxv)- \log \Hm{m}(\sE)\big)\in \sF_2\subseteq \sF$, and thus
\be\label{eq:rprime2}
R\geq \eh{m}(\rxv)- \log \Hm{m}(\sE)\stackrel{\eqref{eq:drgraphforszero}}=R_{\text{SLB}}(D,0)\,.
\ee
Combining \eqref{eq:rprime1} for all $s>0$ satisfying $\widetilde{D}(s)=D$ and \eqref{eq:rprime2}, we obtain
\be
R  \geq \max\bigg\{R_{\text{SLB}}(D,0), \!\!\sup_{ s>0: \widetilde{D}(s)=D}\!\!R_{\text{SLB}}(D,s)\bigg\} 
%\notag \\ & 
 \stackrel{\eqref{eq:charrstar}}= R^*_{\text{SLB}}(D)\,.\label{eq:rprimegeq}
\ee
Combining \eqref{eq:rprimeleq} and \eqref{eq:rprimegeq} for an arbitrary $(D,R)\in \bar{\sF}$ implies that $R= R^*_{\text{SLB}}(D)$.
By~\eqref{eq:defsg}, this yields $(D,R)\in \sG$ and thus $\bar{\sF}\subseteq \sG$.
Because both sets $\sG$ and $\bar{\sF}$ contain exactly one element $(D,R)$ for each $D>D_0$, we obtain $\bar{\sF}=\sG$.
\end{IEEEproof}
%Corollary~\ref{cor:graphoptshlb} implies that we can construct  $\sG$  by constructing $\sF_1$, i.e., the pairs $\big(\widetilde{D}(s) , R_{\text{SLB}}  (\widetilde{D}(s) ,s   ) \big)$ for all $s>0$, and $\sF_2$, i.e., the pairs $(D, \eh{m}(\rxv)- \log \Hm{m}(\sE))$ for all $D>D_0$ (note that ), and taking the upper envelope of the resulting set.
%We illustrate the sets $\sF_1$, $\sF_2$, and $\bar{\sF}$ in Fig.~\ref{fig:upenvelope}. 

In certain cases, it may not be possible to differentiate $\gamma(s)$, and thus the direct calculation of $\widetilde{D}(s)=-\gamma'(s)/\gamma(s)$ is not possible.
However, one can show that, under certain smoothness conditions, the supremum in \eqref{eq:defyvt} is in fact a maximum, i.e.,
\be\label{eq:gammaasmax}
\gamma(s)=\max_{\yv\in \R^M} \int_{\sE} e^{-s d(\xv,\yv)}\, \mathrm{d}\Hm{m}(\xv)
\ee
 and $\widetilde{D}(s)$ can be rewritten as
\be \notag 
\widetilde{D}(s)= \Dt(s) \triangleq \frac{1}{\gamma(s)}\int_{\sE}d(\xv,\yvt(s))e^{-sd(\xv,\yvt(s))} \, \mathrm{d}\Hm{m}(\xv)
\ee
where $\yvt(s)$ is the maximizing value in the definition of $\gamma(s)$ (cf.~\eqref{eq:defyvt}):
\be\notag % \label{eq:realdefyt}
\yvt(s)\triangleq \argmax_{\yv\in \R^M} \int_{\sE} e^{-s d(\xv,\yv)}\, \mathrm{d}\Hm{m}(\xv)\,.
\ee
(Thus, $\gamma(s)= \int_{\sE} e^{-s d(\xv,\yvt(s))}\, \mathrm{d}\Hm{m}(\xv)$.)
The following corollary shows that even if we do not know whether $\gamma(s)$ is differentiable, we can construct a set $\sFt$ of lower bounds on the RD function.
To this end, we define $\sFt\triangleq \sFt_1\cup \sF_2$, where
\be\notag %\label{eq:graphoptshlb1}
\sFt_1\triangleq
\big\{\big(\Dt(s), R_{\text{SLB}}(\Dt(s),s)\big): s>0\big\}
\ee
and $\sF_2$
was defined in \eqref{eq:graphoptshlb2}. 
\begin{corollary}\label{cor:graphnonoptshlb}
Let $\rxv$ be an $m$-rectifiable random variable with support $\sE$.
Then $\sFt$ is a set of lower bounds on the RD function, i.e., for each $(D,R)\in \sFt$, we have  $R(D)\geq R$.
\end{corollary}
\begin{IEEEproof}
Let $(D, R)\in \sFt$. 
%By Theorem~\ref{th:gammadiff}, this implies $s=0$ or $D= \widetilde{D}(s)$.

\emph{Case $(D, R)\in \sFt_1$:}
In this case, we have $(D, R)=\big(\Dt(s), R_{\text{SLB}}(\Dt(s),s)\big)$ for some $s>0$.
Thus, $R=R_{\text{SLB}}(\Dt(s),s)=R_{\text{SLB}}(D,s)$ and, by \eqref{eq:slboundrd},
$R\leq R(D)$.

\emph{Case $(D, R)\in \sF_2$:}
In this case, as in \eqref{eq:drgraphforszero}, we have
$
R = 
R_{\text{SLB}}(D,0) 
$. 
By~\eqref{eq:slboundrd}, we have $R_{\text{SLB}}(D,0)\leq R(D)$, which  implies $R\leq R(D)$.

In either case $R\leq R(D)$, which concludes the  proof.
\end{IEEEproof}

By Corollary~\ref{cor:graphnonoptshlb}, we can  use the sets $\sFt_1$ and $\sF_2$ to construct lower bounds on  the RD function.%
\footnote{
If $\sFt_1=\sF_1$, we obtain by Corollary~\ref{cor:graphoptshlb} that these bounds will be the best Shannon lower bounds.
However, explicit smoothness conditions that guarantee  $\sFt_1=\sF_1$ are difficult to find.} 
More specifically, these bounds are obtained via the following program:
\begin{enumerate}
\renewcommand{\theenumi}{(P\arabic{enumi})}
\renewcommand{\labelenumi}{(P\arabic{enumi})\hspace{-3.5mm}}
\item \hspace{3mm}\label{en:sbard}Calculate $\Dt(s)$ for $s\in (0,\infty)$.
\item \hspace{3mm}\label{en:plotscurve}Plot the $s$-parametrized curve $\big(\Dt(s) , R_{\text{SLB}}  (\Dt(s) ,s   ) \big)$ for $s\in  (0,\infty)$.
\item \hspace{3mm}\label{en:plothor}Plot the horizontal line $\big(D, \eh{m}(\rxv)- \log \Hm{m}(\sE)\big)$ for $D\in (D_0,\infty)$.
\item \hspace{3mm}\label{en:upperenvelope}Take the upper envelope of these two curves.
\end{enumerate}
%\begin{remark}\label{rem:dstar}
%One can show that, under certain smoothness conditions, the supremum in \eqref{eq:defyvt} is in fact a maximum, and $\widetilde{D}(s)$ can be rewritten as
%\ba \notag 
%\widetilde{D}(s)= \Dt(s)\triangleq \frac{1}{\gamma(s)}\int_{\sE}d(\xv,\yvt(s))e^{-sd(\xv,\yvt(s))} \, \mathrm{d}\Hm{m}(\xv)
%\ea
%where $\yvt(s)$ is the maximizing value in the definition of $\gamma(s)$ (see \eqref{eq:defyvt}):
%\be \label{eq:realdefyt}
%\yvt(s)\triangleq \argmax_{\yv\in \R^M} \int_{\sE} e^{-s d(\xv,\yv)}\, \mathrm{d}\Hm{m}(\xv)\,.
%\ee
%(Thus, $\gamma(s)= \int_{\sE} e^{-s d(\xv,\yvt(s))}\, \mathrm{d}\Hm{m}(\xv)$.)
%Therefore, we can also apply the program \ref{en:sbard}--\ref{en:upperenvelope} with $\Dt(s)$ in place of $\widetilde{D}(s)$.
%In fact, even  if $\gamma(s)$ is not differentiable, we can use the program \ref{en:sbard}--\ref{en:upperenvelope} with $\Dt(s)$ to obtain a lower bound on the RD function (although, we do not know the optimality of the resulting bound). 
%Indeed, by Theorem~\ref{th:shannonlb}, the points calculated in steps \ref{en:plotscurve} and \ref{en:plothor} indicate lower bounds on the RD function.
%\end{remark}
In the subsequent Section~\ref{sec:slbunitcircle}, we will apply the program \ref{en:sbard}--\ref{en:upperenvelope} to a specific example.

\subsection{Shannon Lower Bound on the Unit Circle}\label{sec:slbunitcircle}
To demonstrate the practical relevance of Theorem~\ref{th:shannonlb}, we apply 
%the program \ref{en:sbard}--\ref{en:upperenvelope} (with $\Dt(s)$) 
it to the simple example given by
 $\sE=\sS_{1}$, i.e., the unit circle in $\R^2$, and squared error distortion, i.e., $d(\xv, \yv)= \lVert \xv-\yv \rVert^2$.
In order to calculate $\gamma(s)$, we first show that it can be expressed as in \eqref{eq:gammaasmax}, i.e.,  
%While it is straightforward to show Property~\ref{en:diffdistortion} in Theorem~\ref{th:shannonlb},
%Property~\ref{en:existenceyvt} requires considerably more work.
%We have to show that 
\be\notag
\gamma(s)=\max_{\yv\in \R^{2}}\int_{\sS_{1}} e^{-s \lVert\xv-\yv\rVert^2}\, \mathrm{d}\Hm{1}(\xv)
\ee
 for all $s\geq 0$.
Let $s\geq 0$ be arbitrary but fixed. 
Note that we can restrict to $\yv=(y_1\; 0)^{\trans}$, with $y_1\geq 0$, because the problem is invariant under rotations.
Thus, 
\be \notag 
\int_{\sS_{1}} e^{-s \lVert\xv-\yv\rVert^2}\, \mathrm{d}\Hm{1}(\xv)  = \int_{\sS_{1}} e^{-s ((x_1-y_1)^2+x_2^2)}\, \mathrm{d}\Hm{1}(\xv) 
\ee
and therefore we have to maximize the function
\be \notag 
f_s(y_1)\triangleq \int_{\sS_{1}} e^{-s ((x_1-y_1)^2+x_2^2)}\, \mathrm{d}\Hm{1}(\xv)
\ee
on $[0,\infty)$.
To this end, we consider the derivative $f_s'$ and  change the order of differentiation and integration 
(according to \cite[Cor.~5.9]{Bartle95}, this is justified because $\Hm{1}|_{\sS_1}$ is a finite measure and $0<e^{-s ((x_1-y_1)^2+x_2^2)}\leq 1$ for $(x_1\; x_2)^{\trans}\in \sS_1$).
This results in the expression
\be\label{eq:difffs}
f_s'(y_1)= \int_{\sS_{1}} 2 s(x_1-y_1)\, e^{-s ((x_1-y_1)^2+x_2^2)}\, \mathrm{d}\Hm{1}(\xv)\,.
\ee
Because $x_1\leq 1$ for $\xv\in \sS_1$, we have $f_s'(y_1)<0$ for $y_1> 1$, i.e., $f_s$ is monotonically decreasing on $(1,\infty)$.
Thus, the function $f_s$ can only attain its maximum in the compact interval $[0,1]$.
Because  $f_s$ is a continuous function,
we conclude that $\gamma(s)=\max_{\yv\in \R^{2}}\int_{\sS_{1}} e^{-s \lVert\xv-\yv\rVert^2}\, \mathrm{d}\Hm{1}(\xv)$ exists for each $s\geq 0$.
 
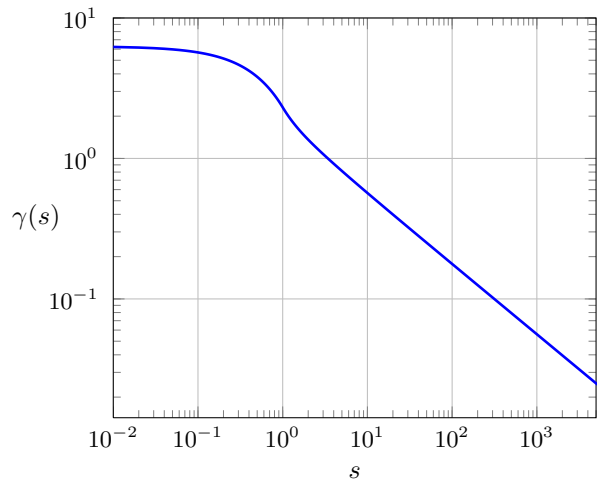
\begin{figure}[t]
\centering
\begin{tikzpicture}
\begin{loglogaxis}[
width=8cm,
xmin=0.01,
xmax=5000,
ymax=10,
%title=Convergence Plot,
xlabel={$s$},
ylabel={$\gamma(s)$},
ylabel style={text width=3em,
rotate=-90,align=right},
yticklabel style={text width=2.1em,align=right},
grid=major, 
tick label style={font=\small},
%legend entries={$y_1(s)$},
%legend style={
%legend pos=south east,
%}
]
\addplot+[blue, no markers, line width=1pt] table {gammas.dat}; %chigen
%\addplot+[red, no markers, line width=1pt] table {upperb.dat}; %chigen
%\addplot+[red, name path=A1, no markers] table {datachiratiorestricted100.dat}; %chigen
%\addplot+[black, name path=B1, no markers] table {datachiratiorestricted10.dat}; %chigen
%\addplot+[red, name path=A2, no markers] table {datachiratiorestrictedup100.dat}; %chigen
%\addplot+[black, name path=B2, no markers] table {datachiratiorestrictedup10.dat}; %chigen
%\addplot[pattern=north west lines, pattern color=red!50] fill between[of=A1 and A2];
%\addplot[pattern=north east lines] fill between[of=B1 and B2];
\end{loglogaxis}
\end{tikzpicture}
\vspace{-3mm}
\caption{Graph of $\gamma(s)$  for  $s\in [0.01, 5000]$.}
\label{fig:optys}
\end{figure}

To characterize $\gamma(s)$ in more detail, we consider the equation $f_s'(y_1)=0$ to find local maxima.
By \eqref{eq:difffs} and because $x_1^2+x_2^2=1$ for $\xv\in \sS_1$, $f_s'(y_1)=0$ is equivalent to
\be\label{eq:difffszero1}
 2  s  e^{-s (1+y_1^2)}\int_{\sS_{1}}(x_1-y_1)\, e^{2s x_1 y_1}\, \mathrm{d}\Hm{1}(\xv)=0\,.
\ee
Furthermore, because $2  s  e^{-s (1+y_1^2)}>0$ and using the transformation $x_1=\cos \phi$, $x_2= \sin\phi$, we obtain that \eqref{eq:difffszero1} is equivalent to
\be\label{eq:solvefsprime}
\int_0^{2\pi}(\cos \phi-y_1)\, e^{2s  y_1\cos \phi}\, \mathrm{d}\phi=0\,.
\ee
Because we know that the function $f_s'$ can only have zeros on $[0,1]$, we can solve \eqref{eq:solvefsprime} numerically for any fixed $s\geq 0$ and compare the values  $f_s(y_1)$ at the different solutions $y_1$ and at the boundary points $y_1=0$ and $y_1=1$ to find $\gamma(s)$. 
In Fig.~\ref{fig:optys},  the values of $\gamma(s)$ are depicted for $s\in [0.01,5000]$.

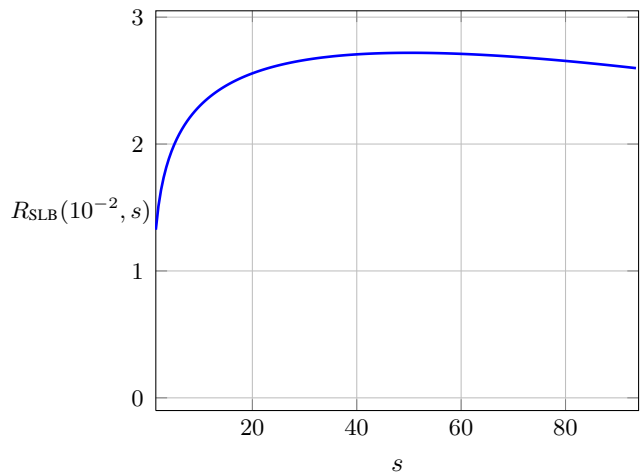
\begin{figure}[t]
\centering
\begin{tikzpicture}
\begin{axis}[
width=8cm,
xmin=1.5,
xmax=94,
ymin=-0.1,ymax=3.05,
%title=Convergence Plot,
xlabel={$s$},
ylabel={$R_{\text{SLB}}(10^{-2},s)$},
tick label style={font=\small},
ylabel style={text width=4em,rotate=-90,align=right,font=\small},
yticklabel style={text width=1.1em,align=right},
grid=major, 
%legend entries={$y_1(s)$},
%legend style={
%legend pos=south east,
%}
]
\addplot+[blue, no markers, line width=1pt] table {lowerbdpercent.dat}; %chigen
%\addplot+[red, no markers, line width=1pt] table {upperb.dat}; %chigen
%\addplot+[red, name path=A1, no markers] table {datachiratiorestricted100.dat}; %chigen
%\addplot+[black, name path=B1, no markers] table {datachiratiorestricted10.dat}; %chigen
%\addplot+[red, name path=A2, no markers] table {datachiratiorestrictedup100.dat}; %chigen
%\addplot+[black, name path=B2, no markers] table {datachiratiorestrictedup10.dat}; %chigen
%\addplot[pattern=north west lines, pattern color=red!50] fill between[of=A1 and A2];
%\addplot[pattern=north east lines] fill between[of=B1 and B2];
\end{axis}
\end{tikzpicture}
\vspace{-3mm}
\caption{Shannon lower bound $R_{\text{SLB}}(10^{-2},s)$ for   $s\in [1,94]$.}
\label{fig:exampledpercent}
\end{figure}

We now have all the ingredients to calculate the parametric lower bound $R_{\text{SLB}}(D,s)$ in \eqref{eq:slboundrd} for any given distortion $D$ and an arbitrary source $\rxv$ on $\sS_1$. 
In particular, let us consider  a uniform distribution of $\rxv$ on $\sS_1$, where $\eh{1}(\rxv)=\log (2\pi)$ (see~\eqref{eq:calchxy}).
In Fig.~\ref{fig:exampledpercent}, we show the lower bound $R_{\text{SLB}}(D,s)$ for $s\in [1,94]$ and distortion $D=10^{-2}$.
It can be seen that the maximal lower bound  $R_{\text{SLB}}(10^{-2},s)$  is obtained for $s\approx 50$.

\begin{figure}[t]
\centering
\begin{tikzpicture}
\begin{semilogxaxis}[
width=8cm,
xmax=2,
ymin=0,ymax=6,
%title=Convergence Plot,
xlabel={$D$},
ylabel={$R$},
ylabel style={text width=4em,rotate=-90,align=right},
yticklabel style={text width=1.1em,align=right},
tick label style={font=\small},
grid=major, 
legend entries={ upper bound on $R(D)$, lower bound on $R(D)$},
legend style={
legend pos=north east, font=\small
}
]
\addplot+[red, no markers, line width=1pt] table {upperb.dat}; %chigen
\addplot+[blue, no markers, dashed, line width=1pt] table {newshlb.dat}; %chigen
%\addplot+[red, name path=A1, no markers] table {datachiratiorestricted100.dat}; %chigen
%\addplot+[black, name path=B1, no markers] table {datachiratiorestricted10.dat}; %chigen
%\addplot+[red, name path=A2, no markers] table {datachiratiorestrictedup100.dat}; %chigen
%\addplot+[black, name path=B2, no markers] table {datachiratiorestrictedup10.dat}; %chigen
%\addplot[pattern=north west lines, pattern color=red!50] fill between[of=A1 and A2];
%\addplot[pattern=north east lines] fill between[of=B1 and B2];
\end{semilogxaxis}
\end{tikzpicture}
\vspace{-3mm}
\caption{Shannon lower bound constructed by \ref{en:sbard}--\ref{en:upperenvelope} and  upper bound \eqref{eq:rdboundatdbar}  for a source $\rxv$ on $\R^2$ uniformly distributed on the unit circle and squared error distortion.}
\label{fig:lubounduniform}
\end{figure}
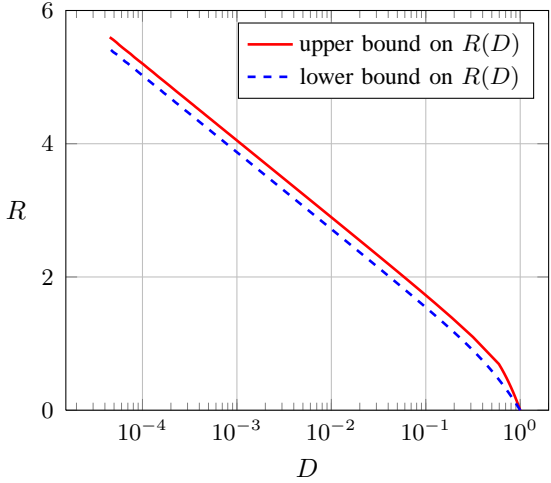

To plot Fig.~\ref{fig:exampledpercent}, we had to calculate $\gamma(s)$ for many different values of $s$.
We also used ``trial and error'' to find the region of $s$ where the maximal  lower bound $R_{\text{SLB}}(10^{-2},s)$ arises. 
To avoid this tedious optimization procedure, which would have to be carried out  for each value of $D$ under consideration, we can  use the program \ref{en:sbard}--\ref{en:upperenvelope} formulated in Section~\ref{sec:maxslb}.
In Fig.~\ref{fig:lubounduniform}, we show the lower bounds on $R(D)$ resulting from this program for $s\in [1,10^5]$, which corresponds to $D\in [5\cdot 10^{-5},1]$. 
%This calculation consists of the following steps:
%\begin{enumerate}
%\item Choose $s>0$;
%\item Calculate $\tilde{y}_1(s)$ by numerically solving \eqref{eq:solvefsprime};
%\item Calculate $\Dt(s)$ from \eqref{eq:assoptcircle};
%\item Calculate the maximal entropy $\sup_{\rwv} \eh{1}(\rwv)$ using \eqref{eq:optentropy};
%\item Calculate the entropy $\eh{1}(\rxv)$ of the source $\rxv$;
%\end{enumerate}
%For the case of a uniform distribution of $\rxv$ on $\sS_1$, this results in the lower bound depicted in Fig.~\ref{fig:lubounduniform}.
%
%
%
We also show in Fig.~\ref{fig:lubounduniform} an upper bound  on $R(D)$ using the following result.
\begin{theorem}\label{th:rdupperbound}
Let the random variable $\rxv$ on $\R^2$ be uniformly distributed on the unit circle, and consider squared error distortion, i.e., $d(\xv, \yv)=\lVert\xv-\yv\rVert^2$. 
For any $n\in \N$,
\be\label{eq:rdboundatdbar}
R(\bar{D}_n)\leq \log  n 
\ee
where
\be\label{eq:defdbarn}
\bar{D}_n= 1 - \bigg(\frac{n}{\pi} \sin \frac{\pi}{n}\bigg)^2\,.
\ee
\end{theorem}
\begin{IEEEproof}
See Appendix~\ref{app:rdupperbound}.
\end{IEEEproof}
The upper bound depicted in Fig.~\ref{fig:lubounduniform} was obtained by linearly interpolating the  upper bounds \eqref{eq:rdboundatdbar} corresponding to different values of $n$ (and, hence, of $\bar{D}_n$).
This is justified by the convexity of the RD function \cite[Lem.~10.4.1]{Cover91}.
Note that the lower and upper bounds shown in Fig.~\ref{fig:lubounduniform} are quite close, and thus they provide a rather accurate characterization of the RD function of $\rxv$.

%
%%%%%%%%%%%%%%%%%%%%%%%%%%%%%%%%%
%\section{Quantization of $m$-rectifiable Random Variables} \label{sec:quantization}   
%%%%%%%%%%%%%%%%%%%%%%%%%%%%%%%%%
%
%In this section we want to make the intuitive approach to our entropy definition rigorous.
%
%\ba
%H([\rxv]_{\mathfrak{P}}) 
%& = -\sum_{i\in \N}\Pr\{\rxv\in \sA_i\} \log \Pr\{\rxv\in \sA_i\} \\
%& = -\sum_{i\in \N} \int_{\sA_i} \theta_{\rxv}^m(\xv) \, \mathrm{d}\Hm{m}|_{\sE}(\xv)
%\log\Bigg(\int_{\sA_i} \theta_{\rxv}^m(\xvt) \, \mathrm{d}\Hm{m}|_{\sE}(\xvt)\Bigg) \\
%& = -\int \theta_{\rxv}^m(\xv) \log\Bigg(\int_{\sA_i(\xv)} \theta_{\rxv}^m(\xvt) \, \mathrm{d}\Hm{m}|_{\sE}(\xvt)\Bigg) \, \mathrm{d}\Hm{m}|_{\sE}(\xv) \\
%& = -\int \theta_{\rxv}^m(\xv) \log\Big(\theta_{\rxv}^m(\xvt)\, \Hm{m}(\sA_i(\xv) \cap \sE)\Big) \, \mathrm{d}\Hm{m}|_{\sE}(\xv)
%\ea
%
%
%We have a weak convergence of rectifiable measures
%\be
%\mu_{\xv, r} \stackrel{*}{\rightharpoonup} \theta_{\mu}^m(\xv)\Hm{m}|_{V(\xv)}
%\ee
%where $\mu_{\xv, r}(\sA)=\mu(\xv+r\sA)$, i.e., 
%for any function $\phi\in C_c(\R^M)$ we have
%\be
%\lim_{r\to 0} \int \phi \, \mathrm{d}\mu_{\xv, r} = \theta_{\mu}^m(\xv)\int \phi \, \mathrm{d}\Hm{m}|_{V(\xv)}
%\ee
%
%

%%%%%%%%%%%%%%%%%%%%%%%%%%%%%%%%
\section{Conclusion} \label{sec:conclusion}   
%%%%%%%%%%%%%%%%%%%%%%%%%%%%%%%%

We presented a generalization of entropy to singular random variables supported on integer-dimensional subsets of Euclidean space. 
More specifically, we considered random variables distributed according to a rectifiable measure. 
Similar to continuous random variables, these rectifiable random variables can be described by a density. 
However, in contrast to continuous random variables, the density is nonzero only on a lower-dimensional subset and has to be integrated with respect to a Hausdorff measure to calculate probabilities.
Our entropy definition is based on this Hausdorff density but otherwise resembles the usual definition of differential entropy. 
However, this formal similarity has to be interpreted with caution 
because Hausdorff measures and projections of   rectifiable sets do not always conform to intuition.
We thus emphasized mathematical rigor and carefully stated all the assumptions underlying our results. 

We showed that for the special cases of rectifiable random variables given by discrete and continuous random variables, our entropy definition reduces to  classical entropy and differential entropy, respectively.
Furthermore, we established a connection between our entropy and differential entropy for a rectifiable random variable that is obtained from a continuous random variable through a one-to-one transformation.
For joint and conditional entropy, our analysis showed that the geometry of the support sets of the random variables plays an important role.
This role is evidenced by the facts that  the chain rule may contain a geometric correction term and conditioning may increase entropy. 

Random variables that are neither discrete nor continuous are not only of theoretical interest. 
Continuity of a random variable cannot be assumed if there are deterministic dependencies reducing the intrinsic dimension of the random variable, which is especially likely to occur in high-dimensional problems. 
As two basic examples, we considered a random variable $\rxv\in \R^2$ supported on the unit circle, which is intrinsically only one-dimensional, and the class of positive semidefinite rank-one random matrices.
In both cases, the differential entropy is not defined and, in fact, classical information theory does not provide a rigorous definition of entropy for these random variables. 

As an application of our entropy definition to source coding, we provided
a characterization of the minimal codeword length for quantizations of integer-dimensional sources.
Furthermore, we presented a result in rate-distortion theory that generalizes the Shannon lower bound for discrete and continuous random variables to the larger class of rectifiable random variables. 
The usefulness of this bound was demonstrated by the example of a uniform source on the unit circle. 
The resulting bound appears to be the first rigorous lower bound on the rate-distortion function for that distribution.

Possible directions for future work include the extension of our entropy definition to distributions mixing different dimensions (e.g., discrete-continuous mixtures).
The extension to noninteger-dimensional singular distributions seems to be possible only in terms of upper and lower entropies, which could be defined based on the upper and lower Hausdorff densities%
\footnote{The  upper and lower Hausdorff densities exist for arbitrary distributions, whereas, by Preiss' Theorem \cite[Th.~5.6]{preiss87}, the existence of the Hausdorff density implies that the measure is rectifiable.} 
\cite[Def.~2.55]{Ambrosio2000Functions}.
%It would also be interesting to establish an interpretation of our entropy as a limit of ever finer quantizations.
%This would constitute a connection of our entropy definition to the limit of $\varepsilon$-entropy. 
Furthermore, our entropy can be extended to infinite-length sequences of rectifiable random variables,
which leads to the definition of an entropy rate generalizing the (differential) entropy rate of a sequence of discrete or continuous random variables.
Finally, applications of our entropy to source coding and channel coding problems involving integer-dimensional singular random variables are largely unexplored.

\appendices
\renewcommand*\thesubsectiondis{\thesection.\arabic{subsection}}
\renewcommand*\thesubsection{\thesection.\arabic{subsection}}

\section{Proof of Lemma~\ref{lem:support}} \label{app:proofsupport}
%%%%%%%%%%%%%%%%%%%%%%%%%%%%%%%
To prove the existence of a support $\sEt\subseteq \sE$, we have to construct a set $\sEt$ that satisfies (cf.\ Definition~\ref{def:support})
\begin{enumerate}
\renewcommand{\theenumi}{(\roman{enumi})}
\renewcommand{\labelenumi}{(\roman{enumi})}
	\item $\sEt=\bigcup_{k\in \IN}f_k(\sC_k)$ where, for $k\in \N$, $\sC_k\subseteq \R^m$ is a bounded Borel set and $f_k\colon \R^m\to \R^M$ is a Lipschitz function that is one-to-one on $\sC_k$;\label{prop:cupborel}
	\item $\sEt\subseteq \sE$; \label{prop:etsube}
	\item $\mu \ll  \Hm{m}|_{\sEt}$; \label{prop:abscont}
	\item $\frac{\mathrm{d}\mu}{\mathrm{d}\Hm{m}|_{\sEt}}>0$ $\Hm{m}|_{\sEt}$-almost everywhere.\label{prop:rdg0}
\end{enumerate}

To prove \ref{prop:cupborel}, we note that, by~\eqref{eq:rectascupborel}, the $m$-rectifiable set $\sE$ satisfies
$
\sE\subseteq\sE_0 \cup \bigcup_{k\in \IN}f_k(\sA_k)
$
with bounded Borel sets $\sA_k\subseteq \R^m$, Lipschitz functions  $f_k\colon \R^m\to \R^M$  that are one-to-one on $\sA_k$, and $\Hm{m}(\sE_0)=0$.
Because  $\mu\ll \Hm{m}|_{\sE}$, we obtain $\mu\ll \Hm{m}|_{\sE^*}$ where
$
\sE^*\triangleq \bigcup_{k\in \IN}f_k(\sA_k)$.
Thus, the Radon-Nikodym derivative $\frac{\mathrm{d}\mu}{\mathrm{d}\Hm{m}|_{\sE^*}}$ exists.
Note that $\frac{\mathrm{d}\mu}{\mathrm{d}\Hm{m}|_{\sE^*}}$ is in fact an equivalence class of measurable functions and only defined up to a set of $\Hm{m}|_{\sE^*}$-measure zero.
Because $\mu(\sE^c)=0$ and $\mu((\sE^*)^c)=0$, we can choose a function $g$ in the equivalence class of $\frac{\mathrm{d}\mu}{\mathrm{d}\Hm{m}|_{\sE^*}}$ satisfying $g(\xv)=0$ on $(\sE\cap \sE^*)^c$.
Since $g$ is a measurable function, the set $g^{-1}(\{0\})$ is $\Hm{m}$-measurable.
Furthermore, because $\sE^*$ is $m$-rectifiable, Property~\ref{en:subsetrect} in Lemma~\ref{lem:proprectset}  implies that the subset $g^{-1}(\{0\})\cap  \sE^*$ is again $m$-rectifiable. 
By~\cite[Lem.~15.5(4)]{ma95}, there exists a Borel set $\sB_0$ satisfying
\be\label{eq:defb0}
\sB_0\supseteq g^{-1}(\{0\})\cap  \sE^*
\ee
and $\Hm{m}(\sB_0\setminus (g^{-1}(\{0\})\cap  \sE^*))=0$.
The absolute continuity $\mu\ll \Hm{m}|_{\sE^*}\ll \Hm{m}$ then implies 
\be\label{eq:mub0capzero}
\mu(\sB_0\setminus (g^{-1}(\{0\})\cap \sE^*))=0\,.
\ee
We further have 
\ba
\mu(\sB_0) 
& \leq \mu(\sB_0\setminus (g^{-1}(\{0\})\cap \sE^*)) + \mu(g^{-1}(\{0\})\cap \sE^*)) \notag \\
& \leq  \mu(\sB_0\setminus (g^{-1}(\{0\})\cap \sE^*)) + \mu(g^{-1}(\{0\})) \notag \\
& \stackrel{\hidewidth (a) \hidewidth}=0 \label{eq:mub0null}
\ea
where  $(a)$ holds by~\eqref{eq:mub0capzero} and because $\mu(g^{-1}(\{0\}))=\int_{g^{-1}(\{0\})}g(\xv) \, \mathrm{d}\Hm{m}|_{\sE^*}(\xv)=0$.
Let us define
\be \label{eq:defset} 
\sEt
\triangleq \bigcup_{k\in \IN}f_k(\sA_k\setminus f_k^{-1}(\sB_0))
\ee
where $\sA_k\setminus f_k^{-1}(\sB_0)$ are bounded Borel sets (this is because $\sA_k$ are bounded Borel sets, $f_k$ are continuous functions, and $\sB_0$ is a Borel set).
As $f_k$ are Lipschitz functions that are one-to-one on $\sA_k$, and thus also on $\sA_k\setminus f_k^{-1}(\sB_0)$, this shows \ref{prop:cupborel}.

Next, we prove \ref{prop:etsube}.
We have $\yv\in f_k(\sA_k\setminus f_k^{-1}(\sB_0))$ if and only if there exists $\xv \in \sA_k\setminus f_k^{-1}(\sB_0)$ such that $f_k(\xv)=\yv$, which in turn holds if and only if there exists  $\xv' \in \sA_k$ such that $f_k(\xv')=\yv$ and $\yv\notin \sB_0$.
Hence, $f_k(\sA_k\setminus f_k^{-1}(\sB_0))= f_k(\sA_k)\setminus \sB_0$.
We can thus rewrite $\sEt$ in \eqref{eq:defset} as
\be \label{eq:defset2} 
\sEt
= \bigcup_{k\in \IN}f_k(\sA_k)\setminus \sB_0
= \sE^* \setminus \sB_0 
 \subseteq \sE^* \setminus (g^{-1}(\{0\}) \cap \sE^*)
\ee
where the final inclusion holds by~\eqref{eq:defb0}.
Because we chose $g(\xv)=0$ on  $(\sE\cap \sE^*)^c=\sE^c \cup (\sE^*)^c$, we obtain $\sE^c\subseteq g^{-1}(\{0\})$.
Inserting this  into \eqref{eq:defset2} yields
\be
\sEt
\subseteq  \sE^* \setminus (\sE^c \cap \sE^*) = \sE^* \cap (\sE \cup (\sE^*)^c) = \sE^* \cap \sE \subseteq \sE
\notag 
\ee
which is \ref{prop:etsube}.

To prove \ref{prop:abscont}, we start with an arbitrary $\Hm{m}$-measurable set $\sA\subseteq \R^M$ with $\Hm{m}|_{\sEt}(\sA)=0$.
We have
\ba
\Hm{m}|_{\sE^*}(\sA \setminus \sB_0)
& =\Hm{m}(\sE^* \cap (\sA \setminus \sB_0)) \notag \\
& =\Hm{m}(\sE^* \cap \sA \cap \sB_0^c) \notag \\
& =\Hm{m}((\sE^*\setminus \sB_0) \cap \sA) \notag \\
& \stackrel{\hidewidth (a)\hidewidth}=\Hm{m}(\sEt \cap \sA) \notag \\
& =\Hm{m}|_{\sEt}(\sA) \notag \\
& =0 \notag 
\ea
where $(a)$ holds because  $\sEt=\sE^*\setminus \sB_0$ by~\eqref{eq:defset2}.
Because $\mu\ll \Hm{m}|_{\sE^*}$, this implies $\mu(\sA \setminus \sB_0)=0$ and, since $\mu(\sB_0)=0$ by~\eqref{eq:mub0null}, we obtain $\mu(\sA)=0$.
Thus, $\Hm{m}|_{\sEt}(\sA)=0$ implies $\mu(\sA)=0$, which proves \ref{prop:abscont}.

To prove \ref{prop:rdg0}, we first show that $g$ is also in the equivalence class of the Radon-Nikodym derivative $\frac{\mathrm{d}\mu}{\mathrm{d}\Hm{m}|_{\sEt}}$.
Indeed, we have for an arbitrary measurable set $\sA\subseteq \R^M$
\ba
\mu(\sA)& = \int_{\sA}g(\xv)\, \mathrm{d}\Hm{m}|_{\sE^*}(\xv) \notag \\[1mm]
& = \int_{\sA\cap \sEt}g(\xv)\, \mathrm{d}\Hm{m}|_{\sE^*}(\xv) +
\int_{\sA\cap \sEt^c}g(\xv)\, \mathrm{d}\Hm{m}|_{\sE*}(\xv) \notag \\[1mm]
& \stackrel{\hidewidth (a) \hidewidth}= \int_{\sA}g(\xv)\, \mathrm{d}\Hm{m}|_{\sEt}(\xv) +
\int_{\sA\cap \sEt^c}g(\xv)\, \mathrm{d}\Hm{m}|_{\sE*}(\xv) \notag \\[1mm]
& = \int_{\sA}g(\xv)\, \mathrm{d}\Hm{m}|_{\sEt}(\xv) +
\mu(\sA\cap \sEt^c) \notag \\[1mm]
& \stackrel{\hidewidth (b) \hidewidth}= \int_{\sA}g(\xv)\, \mathrm{d}\Hm{m}|_{\sEt}(\xv) \notag
\ea
where $(a)$ holds because $\sEt\subseteq \sE^*$ (see~\eqref{eq:defset2}) and $(b)$ holds because $\mu(\sA\cap \sEt^c)=0$ (indeed $\Hm{m}|_{\sEt}(\sA\cap \sEt^c)=\Hm{m}(\sEt\cap\sA\cap \sEt^c)=0$ implies $\mu(\sA\cap \sEt^c)=0$ by \ref{prop:abscont}).
%Thus, $g$ is also in the equivalence class of the Radon-Nikodym derivative $\frac{\mathrm{d}\mu}{\mathrm{d}\Hm{m}|_{\sEt}}$.
By~\eqref{eq:defset2}, we have $\sEt\subseteq \sE^*$, which implies
\be\label{eq:etcapestarc}
\Hm{m}|_{\sEt}((\sE^*)^c)=\Hm{m}(\sEt\cap(\sE^*)^c)\leq \Hm{m}(\sE^*\cap(\sE^*)^c) =0\,.
\ee
By~\eqref{eq:defb0}, we have $\sB_0^c\subseteq \big(g^{-1}(\{0\})\big)^c\cup (\sE^*)^c$.
Hence, for $\xv\in \sB_0^c$ we have either $\xv\in \big(g^{-1}(\{0\})\big)^c$---which is equivalent to $g(\xv)>0$---or $\xv\in (\sE^*)^c$.
By~\eqref{eq:etcapestarc}, we therefore have for $\Hm{m}|_{\sEt}$-almost all $\xv\in  \sB_0^c$ that $g(\xv)>0$.
In particular, because, by~\eqref{eq:defset2}, $\sEt=\sE^* \setminus \sB_0\subseteq \sB_0^c$, we obtain  $g(\xv)>0$ for $\Hm{m}|_{\sEt}$-almost all $\xv\in  \sEt$.
This proves \ref{prop:rdg0}.

Finally, we show that the support is unique up to sets of $\Hm{m}$-measure zero.
Let $\sE_1$ and $\sE_2$ be two support sets of an $m$-rectifiable  measure $\mu$, and denote the Radon-Nikodym derivative  $\frac{\mathrm{d}\mu}{\mathrm{d}\Hm{m}|_{\sE_2}}$ by $g_2$.
Then 
\be\label{eq:e2g20}
\int_{\sE_2\setminus \sE_1}g_2(\xv) \, \mathrm{d}\Hm{m}|_{\sE_2}(\xv)=\mu(\sE_2\setminus \sE_1)=0
\ee
 where the latter equality holds because $\mu(\sE_1^c)=0$ (indeed, $\Hm{m}|_{\sE_1}(\sE_1^c)=0$ implies $\mu(\sE_1^c)=0$ due to $\mu\ll \Hm{m}|_{\sE_1}$).
Since by Definition~\ref{def:support} $g_2>0$ on $\sE_2$ $\Hm{m}|_{\sE_2}$-almost everywhere, \eqref{eq:e2g20} implies $\Hm{m}(\sE_2\setminus \sE_1)=0$. 
By an analogous argument, we obtain $\Hm{m}(\sE_1\setminus \sE_2)=0$.
This shows that $\sE_1$ and $\sE_2$ coincide up to a set of $\Hm{m}$-measure zero.

%%%%%%%%%%%%%%%%%%%%%%%%%%%%%%%
\section{Proof of Theorem~\ref{th:specialcaserect}} \label{app:proofspecialcaserect}
%%%%%%%%%%%%%%%%%%%%%%%%%%%%%%%

\emph{Proof of Part~\ref{en:specialcasedisc}:}
Let $\rxv$ be $0$-rectifiable with support $\sE$, i.e., $\mu\rxv^{-1}\ll \Hm{0}|_{\sE}$ for a $0$-rectifiable set $\sE$.
Recall that  a $0$-rectifiable set $\sE$ is by definition countable, i.e., $\sE=\{\xv_i: i\in \sI\}$ for a countable index set $\sI$.
By~\eqref{eq:probineisone}, 
$\Pr\{\rxv\in \sE\}=1$, which implies that $\rxv$ is a discrete random variable.
Finally, 
\ba
p_{\rxv}(\xv_i) \, &= \, \Pr\{\rxv=\xv_i\}\notag \\
& \stackrel{\hidewidth \eqref{eq:definvmeasure}\hidewidth } = \, \mu\rxv^{-1}(\{\xv_i\}) \notag \\
& = \, \int_{\{\xv_i\}} \frac{\mathrm{d}\mu\rxv^{-1}}{\mathrm{d}\Hm{0}|_{\sE}}(\xv)\, \mathrm{d}\Hm{0}|_{\sE}(\xv) \notag \\
& \stackrel{\hidewidth(a)\hidewidth}=  \, 
%\sum_{\xv=\xv_i} \frac{\mathrm{d}\mu\rxv^{-1}}{\mathrm{d}\Hm{0}|_{\sE}}(\xv)= 
\frac{\mathrm{d}\mu\rxv^{-1}}{\mathrm{d}\Hm{0}|_{\sE}}(\xv_i) \notag \\
& \stackrel{\hidewidth \eqref{eq:defrvhddensity}\hidewidth }= \, \theta_{\rxv}^0(\xv_i) \notag 
\ea
where  $(a)$ holds because $\Hm{0}$ is the counting measure.

Conversely, let $\rxv$ be a discrete random variable taking on the values $\xv_i$, $i\in \sI$. 
We set $\sE\triangleq\{\xv_i:i\in \sI\}$, which is countable and, thus, $0$-rectifiable.
Because $\sE$ includes all possible values of $\rxv$, we have $\Pr\{\rxv\in \sE^c\}=\mu\rxv^{-1}(\sE^c)=0$. 
For $\sA\subseteq \R^M$, the measure $\Hm{0}|_{\sE}(\sA)$ counts the number of points in $\sA$ that also belong to $\sE$.
Thus, for any set $\sA$ such that $\Hm{0}|_{\sE}(\sA)=0$, we obtain that $\sA\cap \sE=\emptyset$ and hence $\sA\subseteq \sE^c$.
This implies $\mu\rxv^{-1}(\sA)\leq\mu\rxv^{-1}(\sE^c)=0$.
Thus, we showed that $\mu\rxv^{-1}(\sA)=0$ for any set $\sA$ with $\Hm{0}|_{\sE}(\sA)=0$, i.e., $\mu\rxv^{-1}\ll \Hm{0}|_{\sE}$. 
Hence, $\rxv$ is $0$-rectifiable.

\emph{Proof of Part~\ref{en:specialcasecont}:}
Let $\rxv$ be  $M$-rectifiable on $\R^M$, i.e., $\mu\rxv^{-1}\ll \Hm{M}|_{\sE}$ for an $M$-rectifiable set $\sE$. Because $\Hm{M}$ is equal to the Lebesgue measure $\Leb^M$ \cite[Th.~2.10.35]{fed69}, we obtain $\mu\rxv^{-1}\ll \Leb^M|_{\sE} \ll  \Leb^M$.
Thus, by the Radon-Nikodym theorem, there exists the Radon-Nikodym derivative $f_{\rxv}=\frac{\mathrm{d}\mu\rxv^{-1}}{\mathrm{d}\Leb^M}$ satisfying $\Pr\{\rxv\in \sA\}=\int_{\sA}f_{\rxv}(\xv)\, \mathrm{d}\Leb^M(\xv)$ for any measurable $\sA\subseteq \R^M$, i.e., $\rxv$ is a continuous random variable. 
By~\eqref{eq:defrvhddensity}, $\theta_{\rxv}^M=f_{\rxv}$ $\Leb^M$-almost everywhere.

Conversely, let $\rxv$ be a continuous random variable on $\R^M$ with  probability density function $f_{\rxv}$. 
For a measurable set $\sA\subseteq \R^M$ satisfying $\Leb^M(\sA)=0$, we obtain $\mu\rxv^{-1}(\sA)=\Pr\{\rxv\in \sA\}=\int_{\sA}f_{\rxv}(\xv)\, \mathrm{d}\Leb^M(\xv)=0$. 
Thus,  we have $\mu\rxv^{-1}\ll \Leb^M$.
Because $\Leb^M=\Hm{M}=\Hm{M}|_{\R^M}$, this is equivalent to $\mu\rxv^{-1}\ll \Hm{M}|_{\R^M}$.
Furthermore, by Property~\ref{en:rmmrectifiable} in Lemma~\ref{lem:proprectset}, $\R^M$ is $M$-rectifiable.
It then follows from Definition~\ref{def:rectrandomvar} that $\rxv$ is an  $M$-rectifiable random variable.

%%%%%%%%%%%%%%%%%%%%%%%%%%%%%%%
\section{Proof of Theorem~\ref{th:gentransformation}} \label{app:proofgentransformation}
%%%%%%%%%%%%%%%%%%%%%%%%%%%%%%%

We first note that the set $\phi(\sE)$ is $m$-rectifiable because $\sE$ is $m$-rectifiable and because of Property~\ref{en:conslipschitz} in Lemma~\ref{lem:proprectset}.
To prove that $\ryv$ is $m$-rectifiable, we will show that $\mu\ryv^{-1} \ll \Hm{m}|_{\phi(\sE)}$.
%To this end, we note that $\Pr\{\phi(\rxv)\in \sA\}=\Pr\{\phi(\rxv)\in \sA\cap \sE\}=\Pr\{\rxv\in \phi^{-1}(\sA\cap \sE)\}$ for any $\Hm{m}$ measurable set $\sA$. 
For a measurable set $\sA\subseteq \R^M$, we have
\ba
\mu\ryv^{-1}(\sA) & =
\Pr\{\phi(\rxv)\in \sA\} \notag \\
& = \Pr\{\rxv\in \phi^{-1}(\sA)\} \notag  \\
& \stackrel{\hidewidth \eqref{eq:calcprobina}\hidewidth}= \int_{\phi^{-1}(\sA)} \theta_{\rxv}^m(\xv) \, \mathrm{d}\Hm{m}|_{\sE}(\xv) \notag \\
& = \int_{\phi^{-1}(\sA)\cap \sE} \frac{\theta_{\rxv}^m(\xv)}{\Jm^{\sE}_{\phi}(\xv)}\Jm^{\sE}_{\phi}(\xv) \, \mathrm{d}\Hm{m}(\xv) \notag \\
& \stackrel{\hidewidth (a)\hidewidth}= \int_{\sA\cap\phi(\sE)} \frac{\theta_{\rxv}^m(\phi^{-1}(\yv))}{\Jm^{\sE}_{\phi}(\phi^{-1}(\yv))}\, \mathrm{d}\Hm{m}(\yv) \notag \\
& = \int_{\sA} \frac{\theta_{\rxv}^m(\phi^{-1}(\yv))}{\Jm^{\sE}_{\phi}(\phi^{-1}(\yv))}\, \mathrm{d}\Hm{m}|_{\phi(\sE)}(\yv)\,.\label{eq:gencovdensityall}
\ea
Here, $(a)$ holds because of the generalized area formula \cite[Th.~2.91]{Ambrosio2000Functions}, and $\phi^{-1}\colon \phi(\sE)\to \sE$ is well defined because $\phi$ is one-to-one on $\sE$.
For a measurable set $\sA\subseteq \R^M$ satisfying $\Hm{m}|_{\phi(\sE)}(\sA)=0$, \eqref{eq:gencovdensityall} implies $\mu\ryv^{-1}(\sA)=0$, i.e., $\mu\ryv^{-1} \ll \Hm{m}|_{\phi(\sE)}$.
Thus, $\ryv$ is an $m$-rectifiable random variable.

By \eqref{eq:gencovdensityall},  $\frac{\theta_{\rxv}^m(\phi^{-1}(\yv))}{\Jm^{\sE}_{\phi}(\phi^{-1}(\yv))}$ equals the Radon-Niko\-dym derivative $\frac{\mathrm{d}\mu\ryv^{-1}}{\mathrm{d}\Hm{m}|_{\phi(\sE)}}(\yv)$, and
thus we obtain
%because, by~, $\frac{\mathrm{d}\mu\ryv^{-1}}{\mathrm{d}\Hm{m}|_{\phi(\sE)}}(\yv)=\theta_{\ryv}^m(\yv)$ $\Hm{m}|_{\phi(\sE)}$-almost everywhere, we obtain 
\be\label{eq:transformeddens}
\frac{\theta_{\rxv}^m(\phi^{-1}(\yv))}{\Jm^{\sE}_{\phi}(\phi^{-1}(\yv))}
=\frac{\mathrm{d}\mu\ryv^{-1}}{\mathrm{d}\Hm{m}|_{\phi(\sE)}}(\yv)
\stackrel{ \eqref{eq:defrvhddensity} }=\theta_{\ryv}^m(\yv)
\ee
for $\Hm{m}|_{\phi(\sE)}$-almost every $\yv\in \R^M$.
We conclude that
\ba
\eh{m}(\ryv) 
& \stackrel{\hidewidth \eqref{eq:entwithsupport2} \hidewidth}=
- \int_{\phi(\sE)}\theta_{\ryv}^m(\yv) \log \theta_{\ryv}^m(\yv) \, \mathrm{d}\Hm{m}(\yv) \notag  \\
& \stackrel{\hidewidth \eqref{eq:transformeddens} \hidewidth}=
- \int_{\phi(\sE)}\frac{\theta_{\rxv}^m(\phi^{-1}(\yv))}{\Jm^{\sE}_{\phi}(\phi^{-1}(\yv))} \notag \\*
& \rule{23mm}{0mm}\times \log \bigg(\frac{\theta_{\rxv}^m(\phi^{-1}(\yv))}{\Jm^{\sE}_{\phi}(\phi^{-1}(\yv))}\bigg) \, \mathrm{d}\Hm{m}(\yv)  \notag \\
& \stackrel{\hidewidth (a) \hidewidth}= - \int_{\sE} \frac{\theta_{\rxv}^m(\xv)}{\Jm^{\sE}_{\phi}(\xv)} \log \bigg(\frac{\theta_{\rxv}^m(\xv)}{\Jm^{\sE}_{\phi}(\xv)}\bigg)\Jm^{\sE}_{\phi}(\xv) \, \mathrm{d}\Hm{m}(\xv)  \notag \\
& = - \int_{\sE} \theta_{\rxv}^m(\xv) \log  \theta_{\rxv}^m(\xv)  \, \mathrm{d}\Hm{m}(\xv)\notag \\*
& \rule{23mm}{0mm}
+ \int_{\sE} \theta_{\rxv}^m(\xv) \log  \Jm^{\sE}_{\phi}(\xv)  \, \mathrm{d}\Hm{m}(\xv)  \notag \\
& \stackrel{\hidewidth \eqref{eq:corexpectation} \hidewidth}= \,\eh{m}(\rxv)+ \E_{\rxv}[\log \Jm^{\sE}_{\phi}(\rxv)] \notag 
\ea
where $(a)$ holds because of the generalized area formula
 \cite[Th.~2.91]{Ambrosio2000Functions}.
\section{Proof of Theorem~\ref{th:prodindeprect}} \label{app:proofprodindeprect}
%%%%%%%%%%%%%%%%%%%%%%%%%%%%%%%

\emph{Proof of Properties~\mbox{\ref{en:indepprodrect} and \ref{en:indepprodsupport}}:}
We first show that for any $\mu(\rxv, \ryv)^{-1}$-measurable set $\sA\subseteq \R^{M_1+M_2}$
\ba
\mu(\rxv, \ryv)^{-1}(\sA) &  =
\Pr\{(\rxv,\ryv)\in \sA\} \notag \\
& =\int_{\sA}\theta_{\rxv}^{m_1}(\xv)\theta_{\ryv}^{m_2}(\yv) \, \mathrm{d}\Hm{m_1+m_2}|_{\sE_1\times \sE_2}(\xv,\yv)\,.\label{eq:toshowproddens}
\ea
To this end, we first consider the rectangles $\sA_1\times \sA_2$ with $\sA_1\subseteq \R^{M_1}$ $\Hm{m_1}$-measurable and $\sA_2\subseteq \R^{M_2}$ $\Hm{m_2}$-measurable.
We have 
\ba
& \Pr\{(\rxv,\ryv)\in  \sA_1\!\times\! \sA_2\} \notag \\
& \stackrel{\hidewidth (a)\hidewidth}=\Pr\{\rxv \in \sA_1\}\, \Pr\{\ryv \in \sA_2\} \notag \\
& \stackrel{\hidewidth \eqref{eq:calcprobina}\hidewidth}= \int_{\sA_1}\theta_{\rxv}^{m_1}(\xv)\, \mathrm{d}\Hm{m_1}|_{\sE_1}(\xv)   \int_{\sA_2} \theta_{\ryv}^{m_2}(\yv)\, \mathrm{d}\Hm{m_2}|_{\sE_2}(\yv)  \notag \\
& \stackrel{\hidewidth (b)\hidewidth}= \int_{\sA_1\times \sA_2}\theta_{\rxv}^{m_1}(\xv)\theta_{\ryv}^{m_2}(\yv) \, \mathrm{d}\big(\Hm{m_1}|_{\sE_1}\times \Hm{m_2}|_{\sE_2}\big)(\xv,\yv)  \notag \\
&  \stackrel{\hidewidth (c)\hidewidth}= \int_{\sA_1\times \sA_2}\theta_{\rxv}^{m_1}(\xv)\theta_{\ryv}^{m_2}(\yv) \, \mathrm{d}\Hm{m_1+m_2}|_{\sE_1\times \sE_2}(\xv,\yv)  \label{eq:prodrectdens}
\ea
where $(a)$ holds because $\rxv$ and $\ryv$ are independent, 
$(b)$ holds by Fubini's theorem,
 and $(c)$ holds by Lem\-ma~\ref{lem:prodcomp}.
Because the rectangles generate the $\mu(\rxv, \ryv)^{-1}$-mea\-sur\-able sets, \eqref{eq:prodrectdens} implies \eqref{eq:toshowproddens}.
For a $\mu(\rxv, \ryv)^{-1}$-mea\-sur\-able set $\sA\subseteq \R^{M_1+M_2}$ satisfying $\Hm{m_1+m_2}|_{\sE_1\times \sE_2}(\sA)=0$,
 \eqref{eq:toshowproddens} implies $\mu(\rxv, \ryv)^{-1}(\sA)=0$, i.e., $\mu(\rxv, \ryv)^{-1}\ll \Hm{m_1+m_2}|_{\sE_1\times \sE_2}$ (note that this is Property~\ref{en:indepprodsupport}).
Furthermore, since $\rxv$ is $m_1$-rectifiable and $\ryv$ is $m_2$-rectifiable, it follows from 
Lemma~\ref{lem:prodcomp}  that $\sE_1\times \sE_2$ is $(m_1+m_2)$-rectifiable.
Hence, according to Definition~\ref{def:rectrandomvar}, $(\rxv, \ryv)$ is an $(m_1+m_2)$-rectifiable random variable, thus proving Property~\ref{en:indepprodrect}. 
%\end{IEEEproof}

\emph{Proof of Property~\mbox{\ref{en:indepprodfact}}:}
Again due to~\eqref{eq:toshowproddens}, 
\be\notag
\theta_{\rxv}^{m_1}\theta_{\ryv}^{m_2}=\frac{\mathrm{d}\mu(\rxv, \ryv)^{-1}}{\mathrm{d}\Hm{m_1+m_2}|_{\sE_1\times \sE_2}} \stackrel{\eqref{eq:defrvhddensity}}=\theta_{(\rxv, \ryv)}^{m_1+m_2}\,.
\ee

\emph{Proof of Property~\mbox{\ref{en:indepentsum}}:}
%By Property~\ref{en:indepprodsupport}, $\sE_1\times \sE_2$ is a support of $(\rxv, \ryv)$. 
We have
\ba
& \eh{m_1+m_2}(\rxv, \ryv) \notag \\
& \rule{3mm}{0mm} \stackrel{\hidewidth \eqref{eq:defjointentropy} \hidewidth}=\;
- \int_{\R^{M_1+M_2}} \theta_{(\rxv,\ryv)}^{m_1+m_2}(\xv,\yv)\log \theta_{(\rxv,\ryv)}^{m_1+m_2}(\xv,\yv) \, \notag \\*
& \rule{45mm}{0mm} \times \mathrm{d}\Hm{m_1+m_2}|_{\sE_1\times \sE_2}(\xv,\yv) \notag \\
& \rule{3mm}{0mm} \stackrel{\hidewidth \eqref{eq:prodhddensity}\hidewidth}=\;  - \int_{\R^{M_1+M_2}} \theta_{\rxv}^{m_1}(\xv)\theta_{\ryv}^{m_2}(\yv)\log \big(\theta_{\rxv}^{m_1}(\xv)\theta_{\ryv}^{m_2}(\yv)\big) \, \notag \\*
& \rule{45mm}{0mm} \times \mathrm{d}\Hm{m_1+m_2}|_{\sE_1\times \sE_2}(\xv,\yv)  \notag \\
& \rule{3mm}{0mm} \stackrel{\hidewidth (a)\hidewidth}=\;  - \int_{\R^{M_1+M_2}} \theta_{\rxv}^{m_1}(\xv)\theta_{\ryv}^{m_2}(\yv)\log \big(\theta_{\rxv}^{m_1}(\xv)\theta_{\ryv}^{m_2}(\yv)\big) \,  \notag \\*
& \rule{35mm}{0mm} \times\mathrm{d}\big(\Hm{m_1}|_{\sE_1}\times \Hm{m_2}|_{\sE_2}\big)(\xv,\yv) \notag \\
%& =\;  - \int \int \theta_{\rxv}^{m_1}(\xv)\theta_{\ryv}^{m_2}(\yv) \log \big(\theta_{\rxv}^{m_1}(\xv)\theta_{\ryv}^{m_2}(\yv)\big) \, \mathrm{d}\Hm{m_1}|_{\sE_1}(\xv)\, \mathrm{d}\Hm{m_2}|_{\sE_2}(\yv) \notag  \\
& \rule{3mm}{0mm} \stackrel{\hidewidth (b)\hidewidth}=\;  - \int_{\R^{M_2}} \int_{\R^{M_1}} \theta_{\rxv}^{m_1}(\xv)\theta_{\ryv}^{m_2}(\yv) \Big( \log \theta_{\rxv}^{m_1}(\xv)  \notag \\*
& \rule{23mm}{0mm}+ \log\theta_{\ryv}^{m_2}(\yv) \Big) \, 
 %\notag \\* & \rule{38mm}{0mm} \times
\mathrm{d}\Hm{m_1}|_{\sE_1}(\xv)\, \mathrm{d}\Hm{m_2}|_{\sE_2}(\yv) \notag  \\
&\rule{3mm}{0mm}  \stackrel{\hidewidth (c)\hidewidth}=\;  - \int_{\R^{M_1}} \theta_{\rxv}^{m_1}(\xv) \log \theta_{\rxv}^{m_1}(\xv)  \, \mathrm{d}\Hm{m_1}|_{\sE_1}(\xv) \notag \\*
& \rule{20mm}{0mm}  
- \int_{\R^{M_2}} \theta_{\ryv}^{m_2}(\yv) \log \theta_{\ryv}^{m_2}(\yv) \, \mathrm{d}\Hm{m_2}|_{\sE_2}(\yv) \notag \\
&\rule{3mm}{0mm}  \stackrel{\hidewidth \eqref{eq:defmdimentropy} \hidewidth}=\;  \; \eh{m_1}(\rxv) + \eh{m_2}(\ryv)\,.  \notag 
\ea
Here, %$(a)$ holds because of \eqref{eq:prodhddensity}, 
$(a)$ holds by Lemma~\ref{lem:prodcomp}, 
$(b)$ holds by Fubini's theorem,
and $(c)$ holds because, by~\eqref{eq:probineisone}, 
\be
 \int_{\R^{M_1}} \! \theta_{\rxv}^{m_1}(\xv)\, \mathrm{d}\Hm{m_1}|_{\sE_1}(\xv)  
   =\! \int_{\R^{M_2}} \! \theta_{\ryv}^{m_2}(\yv)\, \mathrm{d}\Hm{m_2}|_{\sE_2}(\yv) %\notag \\ & 
=1.\notag
\ee
%\end{IEEEproof}

%%%%%%%%%%%%%%%%%%%%%%%%%%%%%%%
\section{Proof of Theorem~\ref{th:marginals}} \label{app:proofmarginals}
%%%%%%%%%%%%%%%%%%%%%%%%%%%%%%%
We will use the generalized coarea formula \cite[Th.~3.2.22]{fed69} several times in our proofs.
Because the classical version of the generalized coarea formula only holds for sets of finite Hausdorff measure, 
 we first present an adaptation  that is suited to our setting.
\begin{theorem}\label{th:mycaf}
Let $\sE\subseteq \R^{M_1+M_2}$ be an $m$-rectifiable set. 
Furthermore, let $\sEt_2\triangleq\proj_{\ryv}(\sE)\subseteq \R^{M_2}$ be $m_2$-rectifiable ($m_2\leq M_2$, $m-m_2\leq M_1$),  $\Hm{m_2}(\sEt_2)<\infty$, 
and $\Jm^{\sE}_{\proj_{\ryv}}> 0$ $\Hm{m}|_{\sE}$-almost everywhere.
Finally, assume that $g\colon \sE\to \R$ is $\Hm{m}$-measurable and satisfies either of the following properties:
\begin{enumerate}
\renewcommand{\theenumi}{(\roman{enumi})}
\renewcommand{\labelenumi}{(\roman{enumi})}
\item \label{en:caffubini} $g(\xv, \yv)\geq 0$ $\Hm{m}$-almost everywhere;
\item \label{en:caflebesgue} $\int_{\sE}\abs{g(\xv, \yv)}\, \mathrm{d}\Hm{m}(\xv, \yv)<\infty$.
\end{enumerate}
Then for all $\Hm{m-m_2}$-measurable sets $\sA_1\subseteq \R^{M_1}$ and $\Hm{m_2}$-measurable sets $\sA_2\subseteq \R^{M_2}$,
\ba
& \int_{(\sA_1\times \sA_2)\cap \sE} g(\xv, \yv)\, \mathrm{d}\Hm{m}(\xv, \yv) \notag \\
& \rule{2mm}{0mm} = \int_{\sA_2\cap \sEt_2}  \int_{\sA_1\cap \sE^{(\yv)}}\frac{g(\xv, \yv)}{\Jm^{\sE}_{\proj_{\ryv}}(\xv, \yv)}\, \mathrm{d}\Hm{m-m_2}(\xv)\,\mathrm{d}\Hm{m_2}(\yv)\label{eq:mycaf}
\ea
where $\sE^{(\yv)}\triangleq \{\xv\in \R^{M_1}: (\xv, \yv)\in \sE\}$.
Furthermore, the set $\sA_1\cap \sE^{(\yv)}$ is $(m-m_2)$-rectifiable for $\Hm{m_2}$-almost every $\yv\in \R^{M_2}$.
\end{theorem}
\begin{IEEEproof}
By Property~\ref{en:hmesigmafinite} in Lemma~\ref{lem:proprectset},  $\Hm{m}|_{\sE}$ is $\sigma$-finite. 
Thus, we can partition $\sE$ as $\sE=\bigcup_{i\in \N}\sF_i$ with pairwise disjoint sets $\sF_i$ satisfying $\Hm{m}(\sF_i)<\infty$.
For $\sA_1\subseteq \R^{M_1}$ $\Hm{m_1}$-measurable and $\sA_2\subseteq \R^{M_2}$  $\Hm{m_2}$-measurable, we have
\ba\label{eq:gcaapplygca}
& \int_{(\sA_1\times \sA_2) \cap \sE}g(\xv, \yv)\, \mathrm{d} \Hm{m}(\xv, \yv) \notag \\
& \rule{5mm}{0mm}= \sum_{i\in \N}\int_{(\sA_1\times \sA_2) \cap \sF_i}g(\xv, \yv)\, \mathrm{d} \Hm{m}(\xv, \yv)   \notag \\
& \rule{5mm}{0mm}\stackrel{\hidewidth (a) \hidewidth}= \sum_{i\in \N}\int\limits_{\sA_2\cap \sEt_2} \int\limits_{\substack{(\sA_1\times \sA_2)\cap \\  \proj_{\ryv}^{-1}(\{\yv\})\cap\sF_i}}\!\!\!\!\frac{g(\xv, \yv')}{\Jm^{\sE}_{\proj_{\ryv}}(\xv, \yv')}\, \mathrm{d}\Hm{m-m_2}(\xv, \yv')  
\notag \\[-6mm] & \rule{66mm}{0mm} \times 
\, \mathrm{d}\Hm{m_2}(\yv)  
\ea
where  $(a)$ holds by the classical version of the general coarea formula~\cite[Th.~3.2.22]{fed69} 
(note that $\sEt_2$ and $\sF_i$ have finite Hausdorff measure) and because $\Jm^{\sE}_{\proj_{\ryv}}> 0$ $\Hm{m}|_{\sE}$-almost everywhere.
By either \ref{en:caffubini} or \ref{en:caflebesgue}, we can apply Fubini's theorem in~\eqref{eq:gcaapplygca} and change the order of integration and summation.
We thus obtain  
\ba%\label{eq:gcaconclude}
& \int_{(\sA_1\times \sA_2) \cap \sE}g(\xv, \yv)\, \mathrm{d} \Hm{m}(\xv, \yv)
\notag \\
& \rule{3mm}{0mm} = \int\limits_{\sA_2\cap \sEt_2}  \Bigg( \sum_{i\in \N}\int\limits_{\substack{(\sA_1\times \sA_2)\cap \\ \proj_{\ryv}^{-1}(\{\yv\})\cap\sF_i}}\!\!\!\!\frac{g(\xv, \yv')}{\Jm^{\sE}_{\proj_{\ryv}}(\xv, \yv')}\, \mathrm{d}\Hm{m-m_2}(\xv, \yv')\Bigg)  \notag \\[-3mm]
& \rule{68mm}{0mm} \times  \mathrm{d}\Hm{m_2}(\yv)  \notag 
\\[1mm]
&  \rule{3mm}{0mm} = \int\limits_{\sA_2\cap \sEt_2}\int\limits_{\substack{(\sA_1\times \sA_2)\cap \\ \proj_{\ryv}^{-1}(\{\yv\})\cap\sE}}\!\!\!\!\frac{g(\xv, \yv')}{\Jm^{\sE}_{\proj_{\ryv}}(\xv, \yv')}\, \mathrm{d}\Hm{m-m_2}(\xv, \yv') 
% \notag \\[-5mm] & \rule{66mm}{0mm} \times 
\, \mathrm{d}\Hm{m_2}(\yv)  \notag 
\\[1mm] 
&  \rule{3mm}{0mm} \stackrel{\hidewidth (a) \hidewidth}= \int\limits_{\sA_2\cap \sEt_2} \int\limits_{\substack{(\sA_1\times \sA_2)\cap \\ \proj_{\ryv}^{-1}(\{\yv\})\cap\sE}}\!\!\!\!\frac{g(\xv, \yv)}{\Jm^{\sE}_{\proj_{\ryv}}(\xv, \yv)}\, \mathrm{d}\Hm{m-m_2}(\xv, \yv') 
% \notag \\[-5mm] & \rule{66mm}{0mm} \times  
\,\mathrm{d}\Hm{m_2}(\yv)  \notag 
\\[1mm]
&  \rule{3mm}{0mm} \stackrel{\hidewidth (b) \hidewidth}= \int_{\sA_2\cap \sEt_2} \int_{\sA_1\cap \sE^{(\yv)}}\frac{g(\xv, \yv)}{\Jm^{\sE}_{\proj_{\ryv}}(\xv, \yv)}\, \mathrm{d}\Hm{m-m_2}(\xv) \, \mathrm{d}\Hm{m_2}(\yv)  \notag
\ea
where $(a)$  holds because $\yv'=\yv$ for all $(\xv, \yv')\in \proj_{\ryv}^{-1}(\{\yv\})$,
%\ba
%(\sA_1\times \sA_2)\cap\proj_{\ryv}^{-1}(\{\yv\})\cap\sE
%& = \{(\xv, \yv')\in (\sA_1\times \sA_2)\cap\sE:\proj_{\ryv}(\xv, \yv')=\yv\} \notag \\
%& =\{(\xv, \yv')\in (\sA_1\times \sA_2)\cap\sE:\yv'=\yv\} \notag 
%\ea
 and $(b)$  holds because the Hausdorff measure does not depend on the ambient space \cite[Remark~2.48]{Ambrosio2000Functions}, i.e., integration with respect to $\Hm{m-m_2}$ on the affine subspace $\proj_{\ryv}^{-1}(\{\yv\})\subseteq{\R^{M_1+M_2}}$ and on $\R^{M_1}$ is identical.
Thus, we have shown \eqref{eq:mycaf}.

We now prove the second part of Theorem~\ref{th:mycaf}.
By~\cite[Th.~3.2.22]{fed69}, the sets $\proj_{\ryv}^{-1}(\{\yv\})\cap \sF_i$ are $(m-m_2)$-rectifiable for $\Hm{m_2}$-almost every $\yv\in \R^{M_2}$. 
By Property~\ref{en:unionrectsetsrect} in Lemma~\ref{lem:proprectset}, the same holds for their union $\bigcup_{i\in \N}\proj_{\ryv}^{-1}(\{\yv\})\cap \sF_i=\proj_{\ryv}^{-1}(\{\yv\})\cap \sE$.
The Lipschitz mapping $\proj_{\rxv}\colon \R^{M_1+M_2}\to \R^{M_1}$, $\proj_{\rxv}(\xv, \yv)=\xv$ satisfies 
\ba
& \proj_{\rxv}\big(\proj_{\ryv}^{-1}(\{\yv\})\cap \sE\big) \notag \\ 
& \rule{5mm}{0mm}= \{\xv\in \R^{M_1}: \exists \yv'\in \R^{M_2} 
%\notag \\ & \rule{30mm}{0mm}
\text{ with } (\xv, \yv')\in \proj_{\ryv}^{-1}(\{\yv\})\cap \sE\} \notag \\ 
& \rule{5mm}{0mm}= \{\xv\in \R^{M_1}: (\xv, \yv)\in  \sE\} \notag \\ 
& \rule{5mm}{0mm}=\sE^{(\yv)}\,. \notag 
\ea
Thus, $\sE^{(\yv)}$ is obtained via a Lipschitz mapping from the set $\proj_{\ryv}^{-1}(\{\yv\})\cap \sE$, which is $(m-m_2)$-rectifiable
 for $\Hm{m_2}$-almost every $\yv\in \R^{M_2}$. 
Therefore,  by Property~\ref{en:conslipschitz} in Lemma~\ref{lem:proprectset}, $\sE^{(\yv)}$ is again $(m-m_2)$-rectifiable%
\footnote{Note that $\sE^{(\yv)}$ is $\Hm{m-m_2}$-measurable because $\proj_{\ryv}^{-1}(\{\yv\})\cap \sE$ is $\Hm{m-m_2}$-measurable and the Hausdorff measure does not depend on the ambient space \cite[Remark~2.48]{Ambrosio2000Functions}.}
 for $\Hm{m_2}$-almost every $\yv\in \R^{M_2}$.
Finally, by Property~\ref{en:subsetrect} in Lemma~\ref{lem:proprectset}, the same is true for $\sA_1\cap \sE^{(\yv)}$.
\end{IEEEproof}

We now proceed to the proof of Theorem~\ref{th:marginals}.

\emph{Proof of Property~\mbox{\ref{en:margisrect}}:}
We have for any $\Hm{m_2}$-mea\-sur\-able set $\sA\subseteq \R^{M_2}$
\ba\label{eq:calcmarginaldensity}
& \mu\ryv^{-1}(\sA)
\notag \\ & \rule{1mm}{0mm}
 = \Pr\{\ryv\in \sA\}  \notag \\
& \rule{1mm}{0mm}= \Pr\{(\rxv, \ryv)\in \R^{M_1}\!\times \!\sA\} \notag \\
& \rule{1mm}{0mm}\stackrel{\hidewidth \eqref{eq:calcprobina} \hidewidth}= \int_{(\R^{M_1}\times \sA) \cap \sE}\theta_{(\rxv, \ryv)}^{m}(\xv, \yv)\, \mathrm{d} \Hm{m}(\xv, \yv)  \notag \\
& \rule{1mm}{0mm}\stackrel{\hidewidth (a) \hidewidth}= \int_{\sA\cap \sEt_2}  \int_{\sE^{(\yv)}}\frac{\theta_{(\rxv, \ryv)}^{m}(\xv, \yv)}{\Jm^{\sE}_{\proj_{\ryv}}(\xv, \yv)}\, \mathrm{d}\Hm{m-m_2}(\xv)\, \mathrm{d}\Hm{m_2}(\yv) \notag \\
& \rule{1mm}{0mm}= \int_{\sA} \int_{\sE^{(\yv)}}\frac{\theta_{(\rxv, \ryv)}^{m}(\xv, \yv)}{\Jm^{\sE}_{\proj_{\ryv}}(\xv, \yv)}\, \mathrm{d}\Hm{m-m_2}(\xv) \,\mathrm{d}\Hm{m_2}|_{\sEt_2}(\yv) 
\ea
where in $(a)$ we used \eqref{eq:mycaf} for  $g(\xv, \yv)=\theta_{(\rxv, \ryv)}^{m}(\xv, \yv)\geq 0$.
For an $\Hm{m_2}$-measurable set $\sA$ satisfying $\Hm{m_2}|_{\sEt_2}(\sA)=0$,
\eqref{eq:calcmarginaldensity} implies $\mu\ryv^{-1}(\sA)=0$, i.e., $\mu\ryv^{-1}\ll \Hm{m_2}|_{\sEt_2}$.
Thus,  according to Definition~\ref{def:rectrandomvar}, $\ryv$ is $m_2$-rectifiable. 
%\end{IEEEproof}

\emph{Proof of Property~\mbox{\ref{en:margsupport}}:}
Because $\mu\ryv^{-1}\ll \Hm{m_2}|_{\sEt_2}$, it follows from Property~\ref{en:corexistsupport} in Corollary~\ref{cor:propsrectrandvar} that there exists a support $\sE_2\subseteq \sEt_2$ of the random variable $\ryv$.
%\end{IEEEproof}

\emph{Proof of Property~\mbox{\ref{en:marghddens}}:}
From \eqref{eq:calcmarginaldensity}, we see that 
\be\notag 
\int_{\sE^{(\yv)}}\frac{\theta_{(\rxv, \ryv)}^{m}(\xv, \yv)}{\Jm^{\sE}_{\proj_{\ryv}}(\xv, \yv)}\, \mathrm{d}\Hm{m-m_2}(\xv)
= \frac{\mathrm{d}\mu\ryv^{-1}}{\mathrm{d}\Hm{m_2}|_{\sEt_2}}(\yv)  =\theta_{\ryv}^{m_2}(\yv)
\ee
where the latter equation holds because $\mu\ryv^{-1}\ll \Hm{m_2}|_{\sEt_2}$.
%, $\frac{\mathrm{d}\mu\ryv^{-1}}{\mathrm{d}\Hm{m_2}|_{\sEt_2}}=\theta_{\ryv}^{m_2}(\yv)$.
This implies \eqref{eq:hddensmarginal}.
%\end{IEEEproof}

\emph{Proof of Property~\mbox{\ref{en:margentropy}}:}
Using \eqref{eq:hddensmarginal} in \eqref{eq:entwithsupport2} and proceeding similarly to  \eqref{eq:calcmarginaldensity}, we obtain
\ba
& \eh{m_2}(\ryv) \notag \\
%Rev1.1:Omit details
%& \stackrel{\hidewidth \eqref{eq:entwithsupport} \hidewidth}=\;  - \int_{\sEt_2} \theta_{\ryv}^{m_2}(\yv) \log \theta_{\ryv}^{m_2}(\yv) \, \mathrm{d}\Hm{m_2}(\yv) \notag \\
& =\;  - \int_{\sEt_2} \bigg(\int_{\sE^{(\yv)}}\frac{\theta_{(\rxv, \ryv)}^{m}(\xv, \yv)}{\Jm^{\sE}_{\proj_{\ryv}}(\xv, \yv)}\, \mathrm{d}\Hm{m-m_2}(\xv) \bigg)\notag \\
& \rule{8mm}{0mm} \times 
\log \bigg(\int_{\sE^{(\yv)}}\frac{\theta_{(\rxv, \ryv)}^{m}(\xvt, \yv)}{\Jm^{\sE}_{\proj_{\ryv}}(\xvt, \yv)}\, \mathrm{d}\Hm{m-m_2}(\xvt)\bigg) 
%\notag \\ & \rule{66mm}{0mm} \times 
\mathrm{d}\Hm{m_2}(\yv)  \notag \\
%& =\;  - \int_{\sEt_2} \bigg(\int_{\sE^{(\yv)}}\frac{\theta_{(\rxv, \ryv)}^{m}(\xv, \yv)}{\Jm^{\sE}_{\proj_{\ryv}}(\xv, \yv)} %\notag \\
%& \rule{45mm}{0mm} \times 
%\log \bigg(\int_{\sE^{(\yv)}}\frac{\theta_{(\rxv, \ryv)}^{m}(\xvt, \yv)}{\Jm^{\sE}_{\proj_{\ryv}}(\xvt, \yv)}\, \mathrm{d}\Hm{m-m_2}(\xvt)\bigg) \, \mathrm{d}\Hm{m-m_2}(\xv) \bigg) \mathrm{d}\Hm{m_2}(\yv)  \notag \\
%& \stackrel{\hidewidth (a) \hidewidth}
& =\;  - \int_{\sE} \theta_{(\rxv, \ryv)}^{m}(\xv, \yv)  \notag \\
& \rule{8mm}{0mm} \times\log \bigg(\int_{\sE^{(\yv)}}\frac{\theta_{(\rxv, \ryv)}^{m}(\xvt, \yv)}{\Jm^{\sE}_{\proj_{\ryv}}(\xvt, \yv)}\, \mathrm{d}\Hm{m-m_2}(\xvt)\bigg) 
% \notag \\ & \rule{61mm}{0mm} \times 
 \mathrm{d}\Hm{m}(\xv, \yv) \,. \notag 
\ea
%where in $(a)$ we used \eqref{eq:mycaf} with $\sA_1=\R^{M_1}$, $\sA_2=\R^{M_2}$, and 
%\be \notag
%g(\xv, \yv)= \theta_{(\rxv, \ryv)}^{m}(\xv, \yv)  \log \bigg(\int_{\sE^{(\yv)}}\frac{\theta_{(\rxv, \ryv)}^{m}(\xvt, \yv)}{\Jm^{\sE}_{\proj_{\ryv}}(\xvt, \yv)}\, \mathrm{d}\Hm{m-m_2}(\xvt)\bigg)\,.
%\ee
%(Here, $g(\xv, \yv)$ is $\Hm{m}|_{\sE}$-integrable by our assumption in the formulation of Property~\ref{en:margentropy}, i.e., Condition~\ref{en:caflebesgue} in Theorem~\ref{th:mycaf} is satisfied.)
Thus,  \eqref{eq:marginalentropy} holds.
\section{Proof of Theorem~\ref{th:condproprect}} \label{app:proofcondproprect}
%%%%%%%%%%%%%%%%%%%%%%%%%%%%%%%

\emph{Proof of Property~\mbox{\ref{en:condisrect}}:}
By Theorem~\ref{th:marginals}, the random variable $\ryv$ is $m_2$-rectifiable with Hausdorff density $\theta_{\ryv}^{m_2}$ (given by \eqref{eq:hddensmarginal}) and some support $\sE_2\subseteq \sEt_2$.
Let $\sA_1\subseteq \R^{M_1}$ and $\sA_2\subseteq \R^{M_2}$ be  $\Hm{m_1}$-measurable and $\Hm{m_2}$-measurable, respectively.
Then
\ba\label{eq:conddens1}
& \Pr\{(\rxv, \ryv)\in \sA_1\! \times \! \sA_2\} \notag \\[1mm]
& \rule{5mm}{0mm}\stackrel{\hidewidth \eqref{eq:condprop} \hidewidth}=  \int_{\sA_2}\Pr\{\rxv\in \sA_1\condi \ryv=\; \yv\} \, \mathrm{d}\mu\ryv^{-1}(\yv) \notag \\[1mm]
&  \rule{5mm}{0mm}\stackrel{\hidewidth \eqref{eq:defrvhddensity} \hidewidth}=  \int_{\sA_2}\Pr\{\rxv\in \sA_1\condi \ryv=\; \yv\} \, \theta_{\ryv}^{m_2}(\yv) \, \mathrm{d}\Hm{m_2}|_{\sE_2}(\yv) \notag \\[1mm]
&  \rule{5mm}{0mm}\stackrel{\hidewidth (a) \hidewidth}=  \int_{\sA_2}\Pr\{\rxv\in \sA_1\condi \ryv=\; \yv\} \, \theta_{\ryv}^{m_2}(\yv) \, \mathrm{d}\Hm{m_2}|_{\sEt_2}(\yv)
\ea
where $(a)$ holds  because we can choose $\theta_{\ryv}^{m_2}(\yv)=0$ for  $\yv\in \sE_2^c$.
On the other hand, we have
\ba\label{eq:conddens2}
& \Pr\{(\rxv, \ryv)\in \sA_1\! \times \! \sA_2\} \notag \\[1mm]
&  \rule{3mm}{0mm} \stackrel{\hidewidth \eqref{eq:calcprobina} \hidewidth}= \int_{(\sA_1\times \sA_2) \cap \sE}\theta_{(\rxv, \ryv)}^{m}(\xv, \yv)\, \mathrm{d} \Hm{m}(\xv, \yv) \notag \\
&  \rule{3mm}{0mm}\stackrel{\hidewidth (a) \hidewidth}= \int_{\sA_2\cap \sEt_2} \int_{\sA_1\cap \sE^{(\yv)}}\frac{\theta_{(\rxv, \ryv)}^{m}(\xv, \yv)}{\Jm^{\sE}_{\proj_{\ryv}}(\xv, \yv)}\, \mathrm{d}\Hm{m-m_2}(\xv) \, \mathrm{d}\Hm{m_2}(\yv) \notag \\[1mm]
& \rule{3mm}{0mm} = \int_{\sA_2} \int_{\sA_1\cap \sE^{(\yv)}}\frac{\theta_{(\rxv, \ryv)}^{m}(\xv, \yv)}{\Jm^{\sE}_{\proj_{\ryv}}(\xv, \yv)}\, \mathrm{d}\Hm{m-m_2}(\xv) \,\mathrm{d}\Hm{m_2}|_{\sEt_2}(\yv)
\ea
where in $(a)$ we used \eqref{eq:mycaf} for $g(\xv, \yv)=\theta_{(\rxv, \ryv)}^{m}(\xv, \yv)\geq 0$.
Combining \eqref{eq:conddens1} and \eqref{eq:conddens2}, we obtain that for $\Hm{m_2}|_{\sEt_2}$-almost every $\yv$ and every $\Hm{m_1}$-measurable  set $\sA_1\subseteq \R^{M_1}$
\ba
&\Pr\{\rxv\in \sA_1\condi \ryv=\yv\}\, \theta_{\ryv}^{m_2}(\yv)  \notag \\[1mm]
& \rule{15mm}{0mm} = 
\int_{\sA_1\cap \sE^{(\yv)}}\frac{\theta_{(\rxv, \ryv)}^{m}(\xv, \yv)}{\Jm^{\sE}_{\proj_{\ryv}}(\xv, \yv)}\, \mathrm{d}\Hm{m-m_2}(\xv)\,.\label{eq:proofconddens}
\ea
Because \eqref{eq:proofconddens} holds for  $\Hm{m_2}|_{\sEt_2}$-almost every $\yv$ and $\sE_2\subseteq \sEt_2$, \eqref{eq:proofconddens} also holds for $\Hm{m_2}|_{\sE_2}$-almost every $\yv$.
Furthermore, because $\sE_2$ is a support of $\ryv$, we have $\theta_{\ryv}^{m_2}(\yv)>0$ $\Hm{m_2}|_{\sE_2}$-almost everywhere.
Thus, we obtain for $\Hm{m_2}|_{\sE_2}$-almost every $\yv$ and every $\Hm{m_1}$-measurable set $\sA_1\subseteq \R^{M_1}$
\ba\label{eq:condrddens}
& \Pr\{\rxv\in \sA_1\condi \ryv=\yv\}  \notag \\
& \rule{10mm}{0mm} = 
\int_{\sA_1\cap \sE^{(\yv)}}\frac{\theta_{(\rxv, \ryv)}^{m}(\xv, \yv)}{\Jm^{\sE}_{\proj_{\ryv}}(\xv, \yv)\, \theta_{\ryv}^{m_2}(\yv)}\, \mathrm{d}\Hm{m-m_2}(\xv) \notag \\[1mm]
& \rule{10mm}{0mm} = \int_{\sA_1}\frac{\theta_{(\rxv, \ryv)}^{m}(\xv, \yv)}{\Jm^{\sE}_{\proj_{\ryv}}(\xv, \yv)\, \theta_{\ryv}^{m_2}(\yv)}\, \mathrm{d}\Hm{m-m_2}|_{\sE^{(\yv)}}(\xv)\,.
\ea
Therefore,  $\Pr\{\rxv\in \cdot\condi \ryv=\yv\}\ll \Hm{m-m_2}|_{\sE^{(\yv)}}$.
By Theorem~\ref{th:mycaf},  the set $\sE^{(\yv)}$ is $(m-m_2)$-rectifiable for $\Hm{m_2}$-almost every $\yv$.
%Rev1.8: Avoid conditional random variables
Hence, according to Definition~\ref{def:rectmeasure},
$\Pr\{\rxv\in \cdot\condi \ryv=\yv\}$ is $(m-m_2)$-rectifiable for $\Hm{m_2}|_{\sE_2}$-almost every $\yv$.
%\end{IEEEproof}

\emph{Proof of Property~\mbox{\ref{en:conddensity}}:} 
By \eqref{eq:condrddens}, we have $\frac{\mathrm{d} \Pr\{\rxv\in \cdot\condi \ryv=\yv\}}{\mathrm{d} \Hm{m-m_2}|_{\sE^{(\yv)}}}(\xv) =\frac{\theta_{(\rxv, \ryv)}^{m}(\xv, \yv)}{\Jm^{\sE}_{\proj_{\ryv}}(\xv, \yv)\, \theta_{\ryv}^{m_2}(\yv)}$  for $\Hm{m_2}|_{\sE_2}$-almost every $\yv$. Thus, \eqref{eq:Hausdorff_RN} implies \eqref{eq:conddensity}.
\section{Proof of Theorem~\ref{th:condentropyasexp}} \label{app:proofcondentropyasexp}
%%%%%%%%%%%%%%%%%%%%%%%%%%%%%%%
Starting from \eqref{eq:defcondentropy}, 
we have 
\ba
& \eh{m-m_2}(\rxv \condi \ryv) \notag \\
&  = 
-\int_{\sE_2} \theta_{\ryv}^{m_2}(\yv)\, \int_{\sE^{(\yv)}}\theta_{\Pr\{\rxv\in  \cdot\condi \ryv=\yv\}}^{m-m_2}(\xv)\notag \\
& \rule{10mm}{0mm} \times  \log\theta_{\Pr\{\rxv\in \cdot\condi \ryv=\yv\}}^{m-m_2}(\xv)\, \mathrm{d}\Hm{m-m_2}(\xv) \, \mathrm{d}\Hm{m_2}(\yv) \notag \\
&  \stackrel{\hidewidth \eqref{eq:conddensity} \hidewidth}= - \int_{\sE_2} \theta_{\ryv}^{m_2}(\yv) \int_{\sE^{(\yv)}}\frac{\theta_{(\rxv, \ryv)}^{m}(\xv, \yv)}{\Jm^{\sE}_{\proj_{\ryv}}(\xv, \yv)\, \theta_{\ryv}^{m_2}(\yv)} \notag \\
& \rule{5mm}{0mm} \times \log\bigg(\frac{\theta_{(\rxv, \ryv)}^{m}(\xv, \yv)}{\Jm^{\sE}_{\proj_{\ryv}}(\xv, \yv)\, \theta_{\ryv}^{m_2}(\yv)}\bigg)\, \mathrm{d}\Hm{m-m_2}(\xv) \, \mathrm{d}\Hm{m_2}(\yv)\notag  \\
&  = - \int_{\sE_2}  \int_{\sE^{(\yv)}}\frac{\theta_{(\rxv, \ryv)}^{m}(\xv, \yv)}{\Jm^{\sE}_{\proj_{\ryv}}(\xv, \yv) } \notag \\
& \rule{5mm}{0mm} \times \log\bigg(\frac{\theta_{(\rxv, \ryv)}^{m}(\xv, \yv)}{\Jm^{\sE}_{\proj_{\ryv}}(\xv, \yv)\, \theta_{\ryv}^{m_2}(\yv)}\bigg)\, \mathrm{d}\Hm{m-m_2}(\xv) \,\mathrm{d}\Hm{m_2}(\yv)\notag  \\
& \stackrel{\hidewidth (a) \hidewidth}= - \int_{\sE}\theta_{(\rxv, \ryv)}^{m}(\xv, \yv) \log\bigg(\frac{\theta_{(\rxv, \ryv)}^{m}(\xv, \yv)}{\Jm^{\sE}_{\proj_{\ryv}}(\xv, \yv)\, \theta_{\ryv}^{m_2}(\yv)}\bigg)
% \notag \\ & \rule{60mm}{0mm} \times 
\, \mathrm{d}\Hm{m}(\xv, \yv) \notag \\
%&    = - \int_{\sE}\theta_{(\rxv, \ryv)}^{m}(\xv, \yv) \bigg( \log\bigg(\frac{\theta_{(\rxv, \ryv)}^{m}(\xv, \yv)}{\theta_{\ryv}^{m_2}(\yv)}\bigg) - \log \Jm^{\sE}_{\proj_{\ryv}}(\xv, \yv) \bigg)\, \mathrm{d}\Hm{m}(\xv, \yv) \notag \\
&   \stackrel{\hidewidth \eqref{eq:corexpectation} \hidewidth}= -\E_{(\rxv, \ryv)}\bigg[\log\bigg(\frac{\theta_{(\rxv, \ryv)}^{m}(\rxv, \ryv)}{\theta_{\ryv}^{m_2}(\ryv)}\bigg)\bigg] + \E_{(\rxv, \ryv)}\big[\log \Jm^{\sE}_{\proj_{\ryv}}(\rxv, \ryv)\big] \notag 
\ea
where in $(a)$ we used \eqref{eq:mycaf} with $\sA_1=\R^{M_1}$, $\sA_2=\R^{M_2}$, and 
$
g(\xv, \yv)= \theta_{(\rxv, \ryv)}^{m}(\xv, \yv) \log\Big(\frac{\theta_{(\rxv, \ryv)}^{m}(\xv, \yv)}{\Jm^{\sE}_{\proj_{\ryv}}(\xv, \yv)\, \theta_{\ryv}^{m_2}(\yv)}\Big)
$.
(Here, $g(\xv, \yv)$ is $\Hm{m}|_{\sE}$-integrable by our assumption in Theorem~\ref{th:condentropyasexp} that the right-hand side of \eqref{eq:condentropyasexp} exists and is finite, i.e., Condition~\ref{en:caflebesgue} in Theorem~\ref{th:mycaf} is satisfied.)
%$(b)$ holds because either $\sI$ is finite or we can apply Lebesgue's dominated convergence theorem due to \eqref{eq:fnuniformlybounded} (recall that $\Hm{m_2}(\sE_2)<\infty$ and, thus, a constant $C$ is an integrable function), 
%and $(c)$ holds by the coarea formula~\cite[Th.~3.2.22]{fed69}.
Thus, \eqref{eq:condentropyasexp} holds.

\section{Proof of Theorem~\ref{th:mutinf}} \label{app:proofmutinf}
%%%%%%%%%%%%%%%%%%%%%%%%%%%%%%%

We first note that the product measure  $\mu\rxv^{-1}\times \mu\ryv^{-1}$ can be interpreted as the joint measure induced by the independent random variables $\tilde{\rxv}$ and $\tilde{\ryv}$, where $\tilde{\rxv}$ has the same distribution as $\rxv$ and $\tilde{\ryv}$ has the same distribution as $\ryv$.
Because  $\rxv$ is $m_1$-rectifiable and $\ryv$ is $m_2$-rectifiable, the same holds for $\tilde{\rxv}$ and $\tilde{\ryv}$, respectively.
Furthermore, the Hausdorff densities satisfy $\theta_{\tilde{\rxv}}^{m_1}(\xv)
=\theta_{\rxv}^{m_1}(\xv)$
and $\theta_{\tilde{\ryv}}^{m_2}(\yv)
=\theta_{\ryv}^{m_2}(\yv)$.
By Properties~\ref{en:indepprodrect}--\ref{en:indepprodsupport} in Theorem~\ref{th:prodindeprect}, the joint random variable $(\tilde{\rxv}, \tilde{\ryv})$  is $(m_1+m_2)$-rectifiable with $(m_1+m_2)$-dimensional Hausdorff density
\be\label{eq:densrxtryt}
\theta_{(\tilde{\rxv},\tilde{\ryv})}^{m_1+m_2}(\xv,\yv)
=\theta_{\tilde{\rxv}}^{m_1}(\xv)\theta_{\tilde{\ryv}}^{m_2}(\yv)
=\theta_{\rxv}^{m_1}(\xv)\theta_{\ryv}^{m_2}(\yv) 
\ee
and $\mu(\tilde{\rxv}, \tilde{\ryv})^{-1}\ll \Hm{m_1+m_2}|_{\sE_1\times \sE_2}$.
The rectifiability of $(\tilde{\rxv}, \tilde{\ryv})$ with  $\mu(\tilde{\rxv}, \tilde{\ryv})^{-1}\ll \Hm{m_1+m_2}|_{\sE_1\times \sE_2}$ implies that the measure $\mu\rxv^{-1}\times \mu\ryv^{-1}$ is $(m_1+m_2)$-rectifiable and  
\be\label{eq:abscondprodmeasure}
 \mu\rxv^{-1}\times \mu\ryv^{-1} \ll \Hm{m_1+m_2}|_{\sE_1\times \sE_2}\,.
\ee

\emph{Proof of Part \ref{en_mutinfcasecomp} (case $m=m_1+m_2$):}
%We claim that $\Pr\{(\rxv, \ryv)\in (\sE_1\times \sE_2)^c\}=0$.
%Indeed, this holds because $(\sE_1\times \sE_2)^c = (\sE_1^c\times \R^{M_2}) \cup (\R^{M_1}\times \sE_2^c)$ and hence 
%$\Pr\{(\rxv, \ryv)\in (\sE_1\times \sE_2)^c\} \leq \Pr\{\rxv\in \sE_1^c\}+\Pr\{\ryv\in \sE_2^c\}=0$.
For any $\Hm{m}$-measurable set $\sA\subseteq \R^{M_1+M_2}$, we have
\ba\label{eq:jointacindep}
& \mu(\rxv, \ryv)^{-1}(\sA) \notag \\
%& = \Pr\{(\rxv, \ryv)\in \sA\cap (\sE_1\times \sE_2)\} \notag \\
& \rule{6mm}{0mm}\stackrel{\hidewidth \eqref {eq:calcprobina} \hidewidth}= \int_{\sA } \theta_{(\rxv, \ryv)}^m(\xv, \yv)\, \mathrm{d}\Hm{m}|_{\sE}(\xv, \yv)  \notag \\[1mm]
& \rule{6mm}{0mm}\stackrel{\hidewidth (a)\hidewidth }= \int_{\sA } \theta_{(\rxv, \ryv)}^m(\xv, \yv)\, \mathrm{d}\Hm{m}|_{\sE_1\times \sE_2}(\xv, \yv)  \notag \\
& \rule{6mm}{0mm}\stackrel{\hidewidth (b) \hidewidth}= \int_{ \sA } \frac{\theta_{(\rxv, \ryv)}^m(\xv, \yv)}{\theta_{\rxv}^{m_1}(\xv)\theta_{\ryv}^{m_2}(\yv)} \theta_{\rxv}^{m_1}(\xv)\theta_{\ryv}^{m_2}(\yv) 
%\notag \\ & \rule{48mm}{0mm} \times 
\, \mathrm{d}\Hm{m}|_{\sE_1\times \sE_2}(\xv, \yv)  \notag \\[1mm]
& \rule{6mm}{0mm}\stackrel{\hidewidth \eqref{eq:densrxtryt} \hidewidth}= \int_{ \sA } \frac{\theta_{(\rxv, \ryv)}^m(\xv, \yv)}{\theta_{\rxv}^{m_1}(\xv)\theta_{\ryv}^{m_2}(\yv)} \theta_{(\tilde{\rxv},\tilde{\ryv})}^{m}(\xv,\yv)  \, \mathrm{d}\Hm{m}|_{\sE_1\times \sE_2}(\xv, \yv)  \notag \\[1mm]
& \rule{6mm}{0mm}\stackrel{\hidewidth (c) \hidewidth}= \int_{\sA } \frac{\theta_{(\rxv, \ryv)}^m(\xv, \yv)}{\theta_{\rxv}^{m_1}(\xv)\theta_{\ryv}^{m_2}(\yv)}   \, \mathrm{d}\big(\mu\rxv^{-1}\times \mu\ryv^{-1}\big)(\xv, \yv)\,. %\notag \\[1mm]
%& \stackrel{\hidewidth (c) \hidewidth}= \int_{\sA} \frac{\theta_{(\rxv, \ryv)}^m(\xv, \yv)}{\theta_{\rxv}^{m_1}(\xv)\theta_{\ryv}^{m_2}(\yv)}   \, \mathrm{d}\big(\mu\rxv^{-1}\times \mu\ryv^{-1}\big)(\xv, \yv)\,.
\ea
Here, $(a)$ holds because $\sE\subseteq \sE_1\times \sE_2$ and because we can choose  $\theta_{(\rxv, \ryv)}^m(\xv, \yv)=0$ on $\sE^c$,
 $(b)$ holds because $\theta_{\rxv}^{m_1}(\xv)\theta_{\ryv}^{m_2}(\yv)>0$  $\Hm{m}$-almost everywhere on $\sE_1\times \sE_2$, 
%$(b)$ holds by~\eqref{eq:densrxtryt}, 
and $(c)$ holds because, by~\eqref{eq:defrvhddensity}, $\theta_{(\tilde{\rxv},\tilde{\ryv})}^{m}=\frac{\mathrm{d}(\mu\rxv^{-1}\times \mu\ryv^{-1})}{\mathrm{d}\Hm{m}|_{\sE_1\times \sE_2}}$ $\Hm{m}|_{\sE_1\times \sE_2}$-almost everywhere. 
%and $(c)$ holds because, by Property~\ref{en:cordensnull} in Corollary~\ref{cor:propsrectrandvar}, $\theta_{(\rxv, \ryv)}^m(\xv, \yv)=0$ $\Hm{m}$-almost everywhere on $\sE^c$.
By \eqref{eq:jointacindep}, we obtain that $\mu(\rxv, \ryv)^{-1}\ll  \mu\rxv^{-1}\times \mu\ryv^{-1}$ with Radon-Nikodym derivative
\be\label{eq:quotmutinf}
\frac{\mathrm{d}\mu(\rxv, \ryv)^{-1}}{\mathrm{d}\big(\mu\rxv^{-1}\times \mu\ryv^{-1}\big)} (\xv, \yv)
= \frac{\theta_{(\rxv, \ryv)}^m(\xv, \yv)}{\theta_{\rxv}^{m_1}(\xv)\theta_{\ryv}^{m_2}(\yv)}\,.
\ee
Inserting \eqref{eq:quotmutinf} into \eqref{eq:gyptheorem1} yields
\ba \label{eq:provemutint1}
 I(\rxv; \ryv)
% \notag \\ & \rule{3mm}{0mm}
& = \int_{\R^{M_1+M_2}}\log \bigg(\frac{\theta_{(\rxv, \ryv)}^m(\xv, \yv)}{\theta_{\rxv}^{m_1}(\xv)\theta_{\ryv}^{m_2}(\yv)}\bigg) \, \mathrm{d}\mu(\rxv, \ryv)^{-1}(\xv, \yv) \notag \\[1mm]
& \stackrel{\hidewidth \eqref{eq:defrvhddensity} \hidewidth}= \int_{\sE}\theta_{(\rxv, \ryv)}^m(\xv, \yv)\log \bigg(\frac{\theta_{(\rxv, \ryv)}^m(\xv, \yv)}{\theta_{\rxv}^{m_1}(\xv)\theta_{\ryv}^{m_2}(\yv)}\bigg) \, \mathrm{d}\Hm{m}(\xv, \yv)
\ea
which is  \eqref{eq:mutinfasdens}.
Furthermore, we can rewrite \eqref{eq:provemutint1} as 
\ba
I(\rxv; \ryv) 
\, & \stackrel{\hidewidth \eqref{eq:corexpectation} \hidewidth}=  \, \E_{(\rxv, \ryv)}\bigg[\log \bigg(\frac{\theta_{(\rxv, \ryv)}^m(\rxv, \ryv)}{\theta_{\rxv}^{m_1}(\rxv)\theta_{\ryv}^{m_2}(\ryv)}\bigg)\bigg] \notag \\[1mm]
&  = \, \E_{(\rxv, \ryv)}[\log \theta_{(\rxv,  \ryv)}^{m}(\rxv, \ryv) ] - \E_{(\rxv, \ryv)}[\log \theta_{\rxv}^{m_1}(\rxv) ] \notag \\
& \rule{42mm}{0mm} - \E_{(\rxv, \ryv)}[\log \theta_{\ryv}^{m_2}(\ryv) ] \notag \\
& \stackrel{\hidewidth \eqref{eq:defjointentropyexp} \hidewidth}=  - \eh{m}(\rxv, \ryv) - \E_{\rxv}[\log \theta_{\rxv}^{m_1}(\rxv) ] - \E_{\ryv}[\log \theta_{\ryv}^{m_2}(\ryv) ] \notag \\
& \stackrel{\hidewidth \eqref{eq:defmdimentropy} \hidewidth}=   -\eh{m}(\rxv,\ryv) + \eh{m_1}(\rxv)+\eh{m_2}(\ryv) \label{eq:mutinfjointtocond}
\ea
which is \eqref{eq:mutinfasjoint}.
Finally, we obtain the first expression in~\eqref{eq:mutinfascondent} by inserting \eqref{eq:chainrulecomp} into \eqref{eq:mutinfjointtocond}.
The second expression in~\eqref{eq:mutinfascondent} is obtained by symmetry.

\emph{Proof of Part \ref{en_mutinfcasenoncomp}  (case $m<m_1+m_2$):}
We first show that $\mu(\rxv, \ryv)^{-1}\centernot\ll\mu\rxv^{-1}\times \mu\ryv^{-1}$.
To this end, we show that the assumption $\mu(\rxv, \ryv)^{-1}\ll  \mu\rxv^{-1}\times \mu\ryv^{-1}$ leads to a contradiction.
Using \eqref{eq:abscondprodmeasure}, we have $\mu(\rxv, \ryv)^{-1}\ll \mu\rxv^{-1}\times \mu\ryv^{-1} \ll \Hm{m_1+m_2}|_{\sE_1\times \sE_2}$.
By Property~\ref{en:rectsetsdimm} in Lemma~\ref{lem:proprectset} and because $\sE$ is an $m$-rectifiable set and $m_1+m_2>m$, we obtain $\Hm{m_1+m_2}(\sE)=0$. 
This implies  $\Hm{m_1+m_2}|_{\sE_1\times \sE_2}(\sE)=0$.
On the other hand, by~\eqref{eq:probineisone},  $\mu(\rxv, \ryv)^{-1}(\sE)=1$.
Thus, we have a contradiction to $\mu(\rxv, \ryv)^{-1}\ll \Hm{m_1+m_2}|_{\sE_1\times \sE_2}$.
Hence, $\mu(\rxv, \ryv)^{-1}\centernot\ll\mu\rxv^{-1}\times \mu\ryv^{-1}$ and, by \eqref{eq:gyptheorem2}, $I(\rxv;\ryv)=\infty$.

\section{Proof of Lemma~\ref{lem:ourentasinf}} \label{app:ourentasinfproofs}
%%%%%%%%%%%%%%%%%%%%%%%%%%%%%%%
%Rev1.5: Use Csiszars generalized entropy in place of KL-divergence
%We will use two alternative characterizations of the Kullback-Leibler divergence $D_{\text{KL}}(\mu\Vert \nu)$ between probability measures $\mu$ and $\nu$ on $\R^M$.
%Usually,  the Kullback-Leibler divergence is defined by \cite{KL51}
%\be\label{eq:KLasRN}
%D_{\text{KL}}(\mu\Vert \nu)\triangleq 
%\begin{cases}
%\displaystyle\int_{\R^M} \log \bigg(\frac{\mathrm{d}\mu}{\mathrm{d}\nu}(x)\bigg)\, \mathrm{d}\mu(x) & \text{ if } \mu\ll \nu \\
%\infty &  \text{ else. }
%\end{cases}
%\ee
%By the Gelfand-Yaglom-Perez theorem \cite[Lem.~5.2.3]{Gray1990Entropy}, this definition coincides with the following:
Let $\mathfrak{P}$ denote the set of all finite, measurable partitions of $\R^M$, i.e., for $\mathfrak{Q}=\{\sA_1, \dots, \sA_N\}\in \mathfrak{P}$ the sets $\sA_i$ are mutually disjoint, measurable, and satisfy $\bigcup_{i=1}^N \sA_i=\R^M$.
Using the interpretation of $\eh{m}(\rxv)$  as a generalized entropy with respect to the Hausdorff measure $\Hm{m}|_{\sE}$ (cf.\ Remark~\ref{rem:genent}), we obtain by \cite[eq.\ (1.8)]{cs73}
\be\label{eq:triventsup}
\eh{m}(\rxv) = - \sup_{\mathfrak{Q}\in \mathfrak{P}} \sum_{\sA\in \mathfrak{Q}} \mu\rxv^{-1}(\sA) \log \bigg(\frac{\mu\rxv^{-1}(\sA)}{\Hm{m}|_{\sE}(\sA)}\bigg)\,.
\ee
Because $\mu\rxv^{-1}(\sE^c)=0$ and $\Hm{m}|_{\sE}(\sE^c)=0$, we have for all $\mathfrak{Q}\in \mathfrak{P}$
\ba%\label{eq:triventsupcape}
& \sum_{\sA\in \mathfrak{Q}} \mu\rxv^{-1}(\sA) \log \bigg(\frac{\mu\rxv^{-1}(\sA)}{\Hm{m}|_{\sE}(\sA)}\bigg) \notag \\
& \rule{10mm}{0mm} =   \sum_{\sA\in \mathfrak{Q}} \mu\rxv^{-1}(\sA\cap \sE) \log \bigg(\frac{\mu\rxv^{-1}(\sA\cap \sE)}{\Hm{m}|_{\sE}(\sA\cap \sE)}\bigg) \notag \\
&  \rule{10mm}{0mm}= \sum_{\sA'\in \widetilde{\mathfrak{Q}}} \mu\rxv^{-1}(\sA') \log \bigg(\frac{\mu\rxv^{-1}(\sA')}{\Hm{m}|_{\sE}(\sA')}\bigg) \label{eq:changepartitionset}
\ea
where $\widetilde{\mathfrak{Q}}\triangleq \{\sA\cap \sE: \sA\in \mathfrak{Q}\}\in \mathfrak{P}_{m,\infty}^{(\sE)}$.
Hence,  for every $\mathfrak{Q}\in \mathfrak{P}$ there exists a $\widetilde{\mathfrak{Q}}\in \mathfrak{P}_{m,\infty}^{(\sE)}$ such that \eqref{eq:changepartitionset} holds.
%\be\notag
%\sum_{\sA\in \mathfrak{Q}} \mu\rxv^{-1}(\sA) \log \bigg(\frac{\mu\rxv^{-1}(\sA)}{\Hm{m}|_{\sE}(\sA)}\bigg)
%= \sum_{\sA'\in \widetilde{\mathfrak{Q}}} \mu\rxv^{-1}(\sA') \log \bigg(\frac{\mu\rxv^{-1}(\sA')}{\Hm{m}|_{\sE}(\sA')}\bigg)\,.
%\ee
Thus, the supremum in \eqref{eq:triventsup} does not change if we replace $ \mathfrak{P}$ by $\mathfrak{P}_{m,\infty}^{(\sE)}$, i.e., we obtain further
\ba 
\eh{m}(\rxv) & = - \!\!\sup_{\mathfrak{Q}\in \mathfrak{P}_{m,\infty}^{(\sE)}} \sum_{\sA\in \mathfrak{Q}} \mu\rxv^{-1}(\sA) \log \bigg(\frac{\mu\rxv^{-1}(\sA)}{\Hm{m}|_{\sE}(\sA)}\bigg) \notag \\
& = \!\!\inf_{\mathfrak{Q}\in \mathfrak{P}_{m,\infty}^{(\sE)}}\!\bigg(\!  -\sum_{\sA\in \mathfrak{Q}} \mu\rxv^{-1}(\sA) \log \bigg(\frac{\mu\rxv^{-1}(\sA)}{\Hm{m}|_{\sE}(\sA)}\bigg) \bigg) \label{eq:ourentasinffirstproof}\\
& = \!\!\inf_{\mathfrak{Q}\in \mathfrak{P}_{m,\infty}^{(\sE)}}\! \bigg( \! -\sum_{\sA\in \mathfrak{Q}} \mu\rxv^{-1}(\sA) \log   \mu\rxv^{-1}(\sA) \notag \\
&  \rule{24mm}{0mm} + \sum_{\sA\in \mathfrak{Q}} \mu\rxv^{-1}(\sA) \log \Hm{m}|_{\sE}(\sA)\bigg)\notag \\
& = \!\!\inf_{\mathfrak{Q}\in \mathfrak{P}_{m,\infty}^{(\sE)}} \!\! \bigg(H([\rxv]_{\mathfrak{Q}}) 
% \notag \\&  \rule{24mm}{0mm} 
+\! \sum_{\sA\in \mathfrak{Q}} \mu\rxv^{-1}(\sA) \log \Hm{m}|_{\sE}(\sA)\bigg)\,. \label{eq:ourentasinfproof}
\ea
Here, \eqref{eq:ourentasinffirstproof} is \eqref{eq:ourentasinffirst} and \eqref{eq:ourentasinfproof} is \eqref{eq:ourentasinf}.

%%%%%%%%%%%%%%%%%%%%%%%%%%%%%%%
\section{Proof of Theorem~\ref{th:sourcelowerupperbound}} \label{app:sourcecodingproofs}
%%%%%%%%%%%%%%%%%%%%%%%%%%%%%%%

%We finally proceed to the proof of Theorem~\ref{th:sourcelowerupperbound}. 
%We first prove the lower bound \eqref{eq:sourcelowerbound}.
\subsection{Proof of Lower Bound \mbox{\eqref{eq:sourcelowerbound}}}
Let ${\mathfrak{Q}}\in \mathfrak{P}_{m,\delta}^{(\sE)}$ be an $(m, \delta)$-partition of $\sE$ according to Definition~\ref{def:mdpartition}, i.e., ${\mathfrak{Q}}= \{\sA_1, \dots, \sA_N\}$ where $\bigcup_{i=1}^N\sA_i=\sE$, $\sA_i\cap \sA_j=\emptyset$, and  $\Hm{m}(\sA_i)\leq \delta$ for all $i,j\in \{1, \dots, N\}$, $i\neq j$.
Note that ${\mathfrak{Q}}$ also belongs to $\mathfrak{P}_{m,\infty}^{(\sE)}$.
Then, starting from \eqref{eq:ourentasinf}, we obtain 
\ba%\label{boundentbydiscent}
 \eh{m}(\rxv) & =  \inf_{\mathfrak{Q}'\in \mathfrak{P}_{m,\infty}^{(\sE)}} \! \bigg(H([\rxv]_{\mathfrak{Q}'}) 
 \notag \\[-3mm]
 &  \rule{25mm}{0mm}
+ \sum_{\sA\in \mathfrak{Q}'} \mu\rxv^{-1}(\sA) \log  \Hm{m}|_{\sE}(\sA) \bigg) \notag \\
&   \leq H([\rxv]_{\mathfrak{Q}})+ \sum_{i=1}^N \mu\rxv^{-1}(\sA_i) \log  \Hm{m}|_{\sE}(\sA_i) \notag \\
&   \stackrel{\hidewidth (a) \hidewidth}\leq H([\rxv]_{\mathfrak{Q}})+ \sum_{i=1}^N \mu\rxv^{-1}(\sA_i) \log \delta  \notag \\
&   \stackrel{\hidewidth (b) \hidewidth}= H([\rxv]_{\mathfrak{Q}})+ \log \delta \notag
\ea
where $(a)$ holds because $\Hm{m}|_{\sE}(\sA_i)\leq \delta$ and $(b)$ holds because $\sum_{i=1}^N \mu\rxv^{-1}(\sA_i)=\mu\rxv^{-1}(\sE)=1$.
Multiplying by $\ld e$, we have equivalently
\be\label{boundentbydiscent}
(\eh{m}(\rxv) - \log \delta) \ld e
 \leq  H([\rxv]_{\mathfrak{Q}}) \ld e\,.
\ee
By \eqref{eq:sourcebounddiscrete}, we have
\be\label{eq:applysourcebounddisc}
H([\rxv]_{\mathfrak{Q}}) \ld e\leq L^*([\rxv]_{\mathfrak{Q}})\,.
\ee
Combining \eqref{boundentbydiscent} and \eqref{eq:applysourcebounddisc}, we obtain
\be \notag 
(\eh{m}(\rxv) - \log \delta) \ld e \leq  L^*([\rxv]_{\mathfrak{Q}}) 
\ee
which implies \eqref{eq:sourcelowerbound}.
%\end{IEEEproof}
\subsection{Proof of Upper Bound  \eqref{eq:sourceupperboundspecial}}
We first state a preliminary result.
\begin{lemma}\label{lem:splitmakessmall}
Let $\rxv$ be an $m$-rectifiable random variable, i.e., $\mu\rxv^{-1}\ll \Hm{m}|_{\sE}$ for an $m$-rectifiable set $\sE\subseteq \R^M$,  with $m\in  \{1, \dots, M\}$ and   $\Hm{m}(\sE)<\infty$.
Furthermore, let $\mathfrak{Q}=\{\sA_1, \dots, \sA_N\}\in \mathfrak{P}_{m,\infty}^{(\sE)}$, % satisfy $\Hm{m}|_{\sE}(\sA_i)\neq 0$ for $i\in \{1, \dots, N\}$.
where each $\sA_i$ is constructed as the union of disjoint sets $\sA_{i,1}, \dots, \sA_{i, k_i}$, i.e., $\sA_i=\bigcup_{j=1}^{k_i}\sA_{i,j}$ with $\sA_{i,j_1}\cap \sA_{i,j_2}=\emptyset$ for $j_1\neq j_2$. 
%Assume that $\Hm{m}(\sA_{1,i})=\Hm{m}(\sA_1)/k$ for $i=1, \dots, k$, i.e., $\sA_1$ is partitioned into sets of equal measure.
Finally, let $\widetilde{\mathfrak{Q}}\triangleq \{\sA_{1,1}, \dots, \sA_{1, k_1}, \dots, \sA_{N,1}, \dots, \sA_{N, k_N}\}$.
Then
\ba
& -\sum_{\sA\in \mathfrak{Q}} \mu\rxv^{-1}(\sA) \log \bigg(\frac{\mu\rxv^{-1}(\sA)}{\Hm{m}|_{\sE}(\sA)}\bigg) \notag \\
& \rule{10mm}{0mm}\geq -\sum_{\sA\in \widetilde{\mathfrak{Q}}} \mu\rxv^{-1}(\sA) \log \bigg(\frac{\mu\rxv^{-1}(\sA)}{\Hm{m}|_{\sE}(\sA)}\bigg)\,.\label{eq:splitmakessmall}
\ea
\end{lemma}
\begin{IEEEproof}
The inequality~\eqref{eq:splitmakessmall} can be written as
\ba
&-\sum_{i=1}^N \mu\rxv^{-1}(\sA_i) \log \bigg(\frac{\mu\rxv^{-1}(\sA_i)}{\Hm{m}|_{\sE}(\sA_i)}\bigg)  \notag \\
& \rule{10mm}{0mm}
\geq -\sum_{i=1}^N \sum_{j=1}^{k_i} \mu\rxv^{-1}(\sA_{i,j}) \log \bigg(\frac{\mu\rxv^{-1}(\sA_{i,j})}{\Hm{m}|_{\sE}(\sA_{i,j})}\bigg)\,. \notag 
\ea
Therefore, it suffices to show that
\ba %\label{eq:equalparttoshow}
&\mu\rxv^{-1}(\sA_i) \log \bigg(\frac{\mu\rxv^{-1}(\sA_i)}{\Hm{m}|_{\sE}(\sA_i)}\bigg) \notag \\
& \rule{10mm}{0mm}
\leq \sum_{j=1}^{k_i}\mu\rxv^{-1}(\sA_{i,j}) \log \bigg(\frac{\mu\rxv^{-1}(\sA_{i,j})}{\Hm{m}|_{\sE}(\sA_{i,j})}\bigg) \notag
\ea
for $i \in \{1, \dots, N\}$.
This latter inequality follows from the log sum inequality \cite[Th.~2.7.1]{Cover91}.
\end{IEEEproof}

We now proceed to the proof of \eqref{eq:sourceupperboundspecial}.
By \eqref{eq:ourentasinffirst},   for each $\varepsilon'>0$, there exists a partition $\mathfrak{Q}=\{\sA_1, \dots, \sA_N\}\in \mathfrak{P}_{m,\infty}^{(\sE)}$ such that
\be\label{eq:boundhmconcretetilde}
\eh{m}(\rxv) > -\sum_{\sA\in\mathfrak{Q}} \mu\rxv^{-1}(\sA) \log \bigg(\frac{\mu\rxv^{-1}(\sA)}{\Hm{m}|_{\sE}(\sA)}\bigg) 
-\varepsilon'\,.
\ee
%Let $\mathfrak{Q}=\{\sA_1, \dots, \sA_N\}\in \mathfrak{P}_{m,\infty}^{(\sE)}$ satisfy \eqref{eq:boundhmconcrete} and $\Hm{m}|_{\sE}(\sA_i)\neq 0$ (which exists due to Lemma~\ref{lem:boundhmconcrete}).
Let us choose, in particular, $\varepsilon'\triangleq \frac{\varepsilon}{2 \ld e}$.
 %satisfying
%\be\label{eq:boundaijdownup}
%e^{-\varepsilon'}\delta_{\varepsilon} \leq \Hm{m}|_{\sE}(\sA_{i,j})\leq \delta_{\varepsilon}
%\ee
%for some $\delta_{\varepsilon}>0$.
We define 
\be\label{eq:defdeltaepsilon}
\delta_{\varepsilon}\triangleq \big(1-e^{-\varepsilon'}\big)\!\!\! \min_{\substack{\sA_i\in \mathfrak{Q} \\  \Hm{m}|_{\sE}(\sA_i)\neq 0  }}\!\!\! \Hm{m}|_{\sE}(\sA_i) >0\,.
\ee
%which is nonzero.
%For $\delta\in (0, \delta_{\varepsilon})$, we will partition each set $\sA_i$ into disjoint subsets $\sA_{i,j}$ such that the resulting partition $\mathfrak{Q}_{\delta}$ of $\sE$ containing all sets $\sA_{i,j}$, satisfies  $\mathfrak{Q}_{\delta}\in \mathfrak{P}_{m,\delta}^{(\sE)}$. 
Choosing some $\delta\in (0, \delta_{\varepsilon})$, we furthermore define 
\be \notag 
J_{i,\delta}\triangleq \frac{\Hm{m}|_{\sE}(\sA_i)}{\delta}
\ee
 and 
\be
M_{i,\delta}\triangleq \begin{cases}
\lceil J_{i,\delta} \rceil & \text{if }\Hm{m}|_{\sE}(\sA_i)\neq 0\\
1 & \text{if } \Hm{m}|_{\sE}(\sA_i)=0\,.
\end{cases}\notag 
\ee
Let us partition each set $\sA_i\in \mathfrak{Q}$ into $M_{i,\delta}$ disjoint  subsets $\sA_{i,j}$ of equal Hausdorff measure,%
\footnote{Because $\Hm{m}$ is a nonatomic measure, we can always find subsets of arbitrary but smaller measure (see \cite[Sec.~2.5]{Fry04}).}  
i.e.,
\be\notag
\Hm{m}|_{\sE}(\sA_{i,j})= \frac{\Hm{m}|_{\sE}(\sA_i)}{M_{i,\delta}}
\ee
and
\be\notag
\bigcup_{j=1}^{M_{i,\delta}}\sA_{i,j}= \sA_i\,.
\ee
For  $\sA_i\in \mathfrak{Q}$ such that $\Hm{m}|_{\sE}(\sA_i)= 0$, we have $M_{i,\delta}=1$, and thus this partition degenerates to $\sA_{i,1}=\sA_i$, which implies $\Hm{m}|_{\sE}(\sA_{i,1})=0$.
For  $\sA_i\in \mathfrak{Q}$ such that $\Hm{m}|_{\sE}(\sA_i)\neq 0$, we have $M_{i,\delta}=\lceil J_{i,\delta} \rceil$, and thus
\be \label{eq:boundsizeaij}
\Hm{m}|_{\sE}(\sA_{i,j}) = \frac{\Hm{m}|_{\sE}(\sA_i)}{\lceil J_{i,\delta} \rceil}= \frac{J_{i,\delta}}{ \lceil J_{i,\delta} \rceil} \, \delta \leq \delta\,.
\ee
In either case we have $\Hm{m}|_{\sE}(\sA_{i,j})\leq \delta$.

Let us denote by $\mathfrak{Q}_{\delta}$ the partition of $\sE$ containing all the sets $\sA_{i,j}$.
Then $\Hm{m}|_{\sE}(\sA_{i,j})\leq \delta$ implies $\mathfrak{Q}_{\delta}\in \mathfrak{P}_{m,\delta}^{(\sE)}$.
Furthermore,   for  $\sA_{i,j}\in \mathfrak{Q}_{\delta}$ satisfying $\Hm{m}|_{\sE}(\sA_{i,j})\neq 0$, 
\ba\label{eq:boundaijbyjid}
\Hm{m}|_{\sE}(\sA_{i,j}) \,
%& = \frac{\Hm{m}|_{\sE}(\sA_i)}{\lceil J_{i,\delta} \rceil} \notag \\
& \stackrel{\hidewidth \eqref{eq:boundsizeaij} \hidewidth}=\, \frac{J_{i,\delta}}{\lceil J_{i,\delta}\rceil}\, \delta \notag \\[1mm]
& = \,\frac{\lceil J_{i,\delta} \rceil-\big(\lceil J_{i,\delta} \rceil- J_{i,\delta}\big)}{\lceil J_{i,\delta} \rceil}\, \delta \notag \\
& =\, \bigg(1- \frac{\lceil J_{i,\delta} \rceil- J_{i,\delta}}{\lceil J_{i,\delta} \rceil}\bigg)\delta \notag \\
& \stackrel{\hidewidth (a) \hidewidth}> \, \bigg(1- \frac{1}{\lceil J_{i,\delta} \rceil}\bigg)\delta 
\ea
where $(a)$ holds because $\lceil J_{i,\delta} \rceil- J_{i,\delta}<1$.
Furthermore, we can  bound $\lceil J_{i,\delta} \rceil$ as (note that $\Hm{m}|_{\sE}(\sA_{i,j})\neq 0$ implies $\Hm{m}|_{\sE}(\sA_i)\neq 0$)
\ba
\lceil J_{i,\delta} \rceil \geq J_{i,\delta}
& =\frac{\Hm{m}|_{\sE}(\sA_i)}{\delta} \notag \\[1mm]
& > \frac{\Hm{m}|_{\sE}(\sA_i)}{\delta_{\varepsilon}} \notag \\
& \stackrel{ \hidewidth \eqref{eq:defdeltaepsilon} \hidewidth }= \frac{\Hm{m}|_{\sE}(\sA_i)}{(1-e^{-\varepsilon'})\min\limits_{\substack{i'\in \{1,\dots, N\}\\\Hm{m}|_{\sE}(\sA_{i'}) \neq 0 }}\Hm{m}|_{\sE}(\sA_{i'})} \notag \\[-1mm]
& \geq \frac{1}{1-e^{-\varepsilon'}}\,. \label{eq:boundjid}
\ea
Inserting \eqref{eq:boundjid} into \eqref{eq:boundaijbyjid}, we obtain   for all sets $\sA_{i,j}\in \mathfrak{Q}_{\delta}$ satisfying $\Hm{m}|_{\sE}(\sA_{i,j})\neq 0$
\be\label{eq:boundhmofaijfinal}
\Hm{m}|_{\sE}(\sA_{i,j})
 > \Bigg(1- \frac{1}{\frac{1}{1-e^{-\varepsilon'}}}\Bigg)\delta 
 = e^{-\varepsilon'}\delta\,.
\ee
%Hence, also the first inequality in \eqref{eq:boundaijdownup} is satisfied.
%Denoting by $\mathfrak{Q}_{\delta}$ the partition of $\sE$ containing all sets $\sA_{i,j}$,
Combining our results yields
\ba\label{eq:boundourentbydisc}
\eh{m}(\rxv) \; & \stackrel{\hidewidth \eqref{eq:boundhmconcretetilde} \hidewidth}>\;  -\sum_{\sA\in \mathfrak{Q}} \mu\rxv^{-1}(\sA) \log \bigg(\frac{\mu\rxv^{-1}(\sA)}{\Hm{m}|_{\sE}(\sA)}\bigg) 
-\varepsilon' \notag \\
& \stackrel{\hidewidth \eqref{eq:splitmakessmall} \hidewidth}\geq\;  -\sum_{\sA\in \mathfrak{Q}_{\delta}} \mu\rxv^{-1}(\sA) \log \bigg(\frac{\mu\rxv^{-1}(\sA)}{\Hm{m}|_{\sE}(\sA)}\bigg) 
-\varepsilon'\notag \\
&  \stackrel{\hidewidth  (a) \hidewidth}\geq\;  -\sum_{\substack{\sA\in \mathfrak{Q}_{\delta} \\ \hidewidth \Hm{m}|_{\sE}(\sA)\neq 0 \hidewidth}} \mu\rxv^{-1}(\sA) \log \bigg(\frac{\mu\rxv^{-1}(\sA)}{\Hm{m}|_{\sE}(\sA)}\bigg) 
-\varepsilon'\notag \\
& \stackrel{\hidewidth \eqref{eq:boundhmofaijfinal} \hidewidth}>\;  -\sum_{\substack{\sA\in \mathfrak{Q}_{\delta} \\ \hidewidth \Hm{m}|_{\sE}(\sA)\neq 0 \hidewidth}} \mu\rxv^{-1}(\sA) \log \bigg(\frac{\mu\rxv^{-1}(\sA)}{e^{-\varepsilon'}\delta}\bigg) 
-\varepsilon' \notag \\
& \stackrel{\hidewidth (b) \hidewidth}=\;  -\sum_{ \sA\in \mathfrak{Q}_{\delta}  } \mu\rxv^{-1}(\sA) \log \bigg(\frac{\mu\rxv^{-1}(\sA)}{e^{-\varepsilon'}\delta}\bigg) 
-\varepsilon' \notag \\
%& =\;  -\sum_{\sA\in \mathfrak{Q}_{\delta}} \mu\rxv^{-1}(\sA) \log \big(\mu\rxv^{-1}(\sA)\big) \notag \\
%& \rule{20mm}{0mm} + (\log \delta - \varepsilon' ) \sum_{\sA\in \mathfrak{Q}_{\delta}} \mu\rxv^{-1}(\sA)  
  %- \varepsilon' \notag \\
& \stackrel{\hidewidth (c) \hidewidth}= \, -\!\sum_{\sA\in \mathfrak{Q}_{\delta}} \mu\rxv^{-1}(\sA) \log \big(\mu\rxv^{-1}(\sA)\big) % \notag \\[-2mm]
%& \rule{50mm}{0mm}
+ \log \delta
-2\varepsilon' \notag \\[1mm]
& =\;  H([\rxv]_{\mathfrak{Q}_{\delta}}) + \log \delta
-2\varepsilon' \notag \\[1mm]
& \stackrel{\hidewidth (d) \hidewidth}>\;  \frac{L^*([\rxv]_{\mathfrak{Q}_{\delta}})-1}{\ld e} + \log \delta
-2\varepsilon'
\ea
where  $(a)$ and $(b)$ hold because, by $\mu\rxv^{-1}\ll \Hm{m}|_{\sE}$, $\Hm{m}|_{\sE}(\sA)= 0$ implies $\mu\rxv^{-1}(\sA)= 0$ and thus the additional restriction $\Hm{m}|_{\sE}(\sA)\neq 0$ removes only summands that are zero, 
$(c)$ holds because $\mathfrak{Q}_{\delta}$ is a partition of $\sE$ and thus $\sum_{\sA\in \mathfrak{Q}_{\delta}} \mu\rxv^{-1}(\sA)  =\mu\rxv^{-1}(\sE)  =1$,
and $(d)$ holds by the second inequality in \eqref{eq:sourcebounddiscrete}.
%By the second inequality in \eqref{eq:boundaijdownup}, we have $\mathfrak{Q}_{\delta}\in \mathfrak{P}_{m,\delta_{\varepsilon}}^{(\sE)}$ and 
Finally, rewriting \eqref{eq:boundourentbydisc} gives (recall $\varepsilon'= \frac{\varepsilon}{2 \ld e}$)
\be \notag 
L^*([\rxv]_{\mathfrak{Q}_{\delta}}) < \eh{m}(\rxv)\ld e- \log \delta \ld e +1
+ \varepsilon
\ee
which is \eqref{eq:sourceupperboundspecial}.

%

%%%%%%%%%%%%%%%%%%%%%%%%%%%%%%%
\section{Proof of Lemma~\ref{lem:rslbtoinfty}} \label{app:rslbtoinfty}
%%%%%%%%%%%%%%%%%%%%%%%%%%%%%%%

Because $R_{\text{SLB}}(D,s)=\eh{m}(\rxv)- (sD +\log \gamma(s))$, where $\eh{m}(\rxv)$ is finite and does not depend on $s$, it is sufficient to show $\lim_{s\to \infty}\big(sD +\log \gamma(s)\big) =\infty$.
For $\yv\in \R^M$, we define the set of all $\xv$ whose distortion relative to $\yv$ is less than $D/2$,
\be\notag
\sC(\yv)\triangleq \bigg\{\xv\in \R^M: d(\xv, \yv)<\frac{D}{2}\bigg\}\,.
\ee 
We obtain
\ba
sD & +\log \gamma(s) \notag \\
& \stackrel{\hidewidth \eqref{eq:defyvt} \hidewidth}=\; sD + \log
\bigg(\sup_{\yv\in \R^M} \int_{\sE} e^{-s d(\xv,\yv)}\, \mathrm{d}\Hm{m}(\xv)\bigg) \notag \\
& \stackrel{\hidewidth (a) \hidewidth}=\; \sup_{\yv\in \R^M} \bigg( sD + \log
\bigg( \int_{\sE} e^{-s d(\xv,\yv)}\, \mathrm{d}\Hm{m}(\xv)\bigg) \bigg)\notag \\
& =\;\sup_{\yv\in \R^M} \log
\bigg( e^{sD}\int_{\sE} e^{-s d(\xv,\yv)}\, \mathrm{d}\Hm{m}(\xv)\bigg) \notag \\
& =\;\sup_{\yv\in \R^M} \log
\bigg(\int_{\sE} e^{s (D- d(\xv,\yv))}\, \mathrm{d}\Hm{m}(\xv)\bigg) \notag \\
& \stackrel{\hidewidth (b) \hidewidth}\geq\;\sup_{\yv\in \R^M} \log
\bigg(\int_{\sE\cap \sC(\yv)} e^{s (D- d(\xv,\yv))}\, \mathrm{d}\Hm{m}(\xv)\bigg) \notag \\
& \stackrel{\hidewidth (c) \hidewidth}\geq\;\sup_{\yv\in \R^M} \log
\bigg(\int_{\sE\cap \sC(\yv)} e^{s D/2}\, \mathrm{d}\Hm{m}(\xv)\bigg) \notag \\
& =\;\sup_{\yv\in \R^M} \log
\big(e^{s D/2}\Hm{m}\big(\sE\cap \sC(\yv)\big)\big) \notag \\[1mm]
& =\; s \, \frac{D}{2} + \sup_{\yv\in \R^M} \log \Hm{m}\big(\sE\cap \sC(\yv)\big) \label{eq:boundshlb}
\ea
where $(a)$ holds because $\log$ is a monotonically increasing function, 
$(b)$ holds because ${e^{s (D- d(\xv,\yv))}>0}$,
and $(c)$ holds because $d(\xv,\yv)<D/2$ for all $\xv\in \sC(\yv)$.
Because $\mu\rxv^{-1}(\sE)=1$ (see \eqref{eq:probineisone}), the absolute continuity $\mu\rxv^{-1}\ll \Hm{m}|_{\sE}$ implies $\Hm{m}(\sE)>0$. 
Thus, there  exists a $\yvb\in \R^{M}$ such that $\delta\triangleq\Hm{m}\big(\sE\cap \sC(\yvb)\big)>0$.
Clearly, this implies $\sup_{\yv\in \R^M} \log \Hm{m}\big(\sE\cap \sC(\yv)\big)\geq\log \delta$, and
hence, by~\eqref{eq:boundshlb}, 
\be  \label{eq:boundshlbybar}
sD +\log \gamma(s)  \geq s \, \frac{D}{2} + \log \delta\,.
\ee
For fixed but arbitrary $K>0$ and all $s\geq\frac{2\, (K -\log \delta)}{D}$, we have $s\frac{D}{2} + \log \delta\geq K$, and thus \eqref{eq:boundshlbybar} implies
\be \notag %\label{eq:boundshlbybar}
sD +\log \gamma(s)  \geq K\,.
\ee
Since $K$ can be chosen arbitrarily large, this shows that
$\lim_{s\to \infty}\big(sD +\log \gamma(s)\big) =\infty$.

%%%%%%%%%%%%%%%%%%%%%%%%%%%%%%%
\section{Proof of Theorem~\ref{th:rdupperbound}} \label{app:rdupperbound}
%%%%%%%%%%%%%%%%%%%%%%%%%%%%%%%

Consider the source $\rxv$ on $\R^2$ as specified in Theorem~\ref{th:rdupperbound}.
The main idea of the proof is to construct a specific source code  and calculate its rate and expected distortion. 
We can then use the  source coding theorem~\cite[Th.~11.4.1]{Gray1990Entropy} to conclude that the calculated rate is an upper bound on the RD function.

To this end, recall that a $(k,n)$ source code for a sequence $\rxv_{1:k}\in (\R^2)^k$ of $k$ independent realizations of $\rxv$
consists of  an encoding function $f\colon (\R^2)^k\to \{1, \dots, n\}$ and a decoding function $g\colon \{1, \dots, n\}\to (\R^2)^k$.
The rate of this code is defined as $R_{f,g}\triangleq(\log n)/k$
and the expected distortion is given by
\be\notag
D_{f,g}=\E_{\rxv_{1:k}}\big[ \lVert \rxv_{1:k}-g(f(\rxv_{1:k})) \rVert^2 \big]\,.
\ee
By the source coding theorem~\cite[Th.~11.4.1]{Gray1990Entropy}, every $(k,n)$ code  with expected distortion $D_{f,g}$ must have a rate 
greater than or equal to $R(D_{f,g})$.
In particular, this has to hold for the special case $k=1$.
The rate of these $(1,n)$ codes reduces to $R_{f,g}=\log n$, and the expected distortion is given by
\be\label{eq:dist1ncode}
D_{f,g}=\E_{\rxv}\big[\lVert \rxv-g(f(\rxv))\rVert^2 \big]\,.
\ee
Thus, the implication of the source coding theorem is that for a $(1,n)$ code with expected distortion $D_{f,g}$, we have
\be \label{eq:bound1ncode}
\log n \geq R(D_{f,g})\,.
\ee

We directly design the composed function $q\triangleq g\circ f$.
Because $\rxv$ has probability zero outside $\sS_1$, we only have to define $q$ on the unit circle. 
Furthermore, because $f$ maps  $\xv$ to one of at most $n$ distinct values,  $q = g\circ f$ can also attain at most $n$ distinct values.
We define $q$ to map each circle segment defined by an angle interval $\big[i\frac{2\pi}{n}, (i+1)\frac{2\pi}{n}\big)$, $i\in \{0, \dots, n-1\}$, onto one associated ``center'' point, which is not constrained to lie on the unit circle. 
To this end, we only have to consider the circle segment defined by $\{\xv = (\cos \phi\; \sin \phi)^{\trans}: \phi \in [-\pi/n, \pi/n)\}$ since the problem is invariant under rotations.
Because of  symmetry, we choose the ``center'' associated with this segment to be some point $(x_1\; 0)^{\trans}$, i.e., $q(\xv)=(x_1\; 0)^{\trans}$ for all $\xv = (\cos \phi\; \sin \phi)^{\trans}$ with $\phi \in [-\pi/n, \pi/n)$.
According to \eqref{eq:dist1ncode}, the expected distortion is then obtained as
\ba\label{eq:calcexpdistex}
D_{q}& =
\E_{\rxv}\big[\lVert \rxv-q(\rxv)\rVert^2 \big] \notag \\[1mm]
& = \int_0^{2\pi} \frac{1}{2\pi} \bigg\lVert
\binom{\cos \phi}{\sin \phi} - q \bigg(\binom{\cos \phi}{\sin \phi}\bigg)\bigg\rVert^2 \, \mathrm{d}\phi \notag \\[1mm]
& = \frac{n}{2\pi} \int_{-\pi/n}^{\pi/n}    \bigg\lVert
\binom{\cos \phi}{\sin \phi}
- \binom{x_1}{0}\bigg\rVert^2 \, \mathrm{d}\phi \notag \\
& = \frac{n}{2\pi} \int_{-\pi/n}^{\pi/n}   
\big((\cos \phi -x_1)^2 + \sin^2 \phi \big) \, \mathrm{d}\phi  \notag \\
& = \frac{n}{2\pi}  \int_{-\pi/n}^{\pi/n} 
\big(1 + x_1^2 - 2x_1\cos \phi\big) \, \mathrm{d}\phi  \notag \\
%& = 1 + x_1^2 - \frac{n x_1}{\pi} \int_{-\pi/n}^{\pi/n} \cos \phi \, \mathrm{d}\phi \\
& = 1 + x_1^2 - \frac{2 n x_1}{\pi} 
\sin \frac{\pi}{n}\,.
\ea
Minimizing the expected  distortion with respect to $x_1$ gives the optimum value of $x_1$ as
\be\label{eq:optx1}
x_1^* = \frac{n}{\pi} 
\sin \frac{\pi}{n}\,.
\ee
The corresponding quantization function will be denoted by $q^*$.
Inserting~\eqref{eq:optx1} into \eqref{eq:calcexpdistex}  yields $\bar{D}_n$ in \eqref{eq:defdbarn}:
\ba
D_{q^*}& =
\E_{\rxv}\big[\lVert \rxv-q^*(\rxv)\rVert^2 \big] \notag \\
& = 
1 + \bigg(\frac{n}{\pi} 
\sin \frac{\pi}{n}\bigg)^2 
- 2\bigg(\frac{ n }{\pi } \sin  \frac{\pi}{n}\bigg)^2 \notag  \\
& = 1 - \bigg(\frac{n }{\pi } 
\sin  \frac{\pi}{n}\bigg)^2 \notag \\
& = \bar{D}_n\,. \notag 
\ea
Thus, we found a $(1,n)$ code with expected distortion $D_{q^*}=\bar{D}_n$.
Hence, by~\eqref{eq:bound1ncode}, we have $\log n \geq R(D_{q^*})=R(\bar{D}_n)$, which is \eqref{eq:rdboundatdbar}.

%%%%%%%%%%%%%%%%%%%%%%%%%%%%%%%
\section*{Acknowledgment} 
%%%%%%%%%%%%%%%%%%%%%%%%%%%%%%%%
%
%We wish to thank.
The authors wish to thank  the anonymous reviewers for numerous
 insightful comments and, 
in particular, for suggesting reference \cite{cs73}.

%%%%%%%%%%%%%%%%%%%%%%%%%%%%%%%%%
%References
%%%%%%%%%%%%%%%%%%%%%%%%%%%%%%%%%

%\renewcommand{\baselinestretch}{1.07}\small\normalsize

%\bibliography{referencesrect}
%\bibliographystyle{IEEEtran}

\end{document}